\documentclass[apj]{emulateapj}

\usepackage{color}
\usepackage{comment}
\usepackage{natbib,aas_macros}
\newcommand{\OII}{[O\,\textsc{ii}]}
\newcommand{\Ha}{H$\alpha$}
\newcommand{\Hb}{H$\beta$}
\newcommand{\Hc}{H$\gamma$}
\newcommand{\Hd}{H$\delta$}

\newcommand{\NII}{[N\,\textsc{ii}]}

\newcommand{\OIII}{[O\,\textsc{iii}]}

\newcommand{\NeIII}{[Ne\,\textsc{iii}]}

\newcommand{\TeOII}{$T_\mathrm{e}$(O\,\textsc{ii})}
\newcommand{\TeOIII}{$T_\mathrm{e}$(O\,\textsc{iii})}
\newcommand{\neOII}{$n_\mathrm{e}$(O\,\textsc{ii})}

\newcommand{\ergscm}{erg\,s$^{-1}$\,cm$^{-2}$}

\newcommand{\ergscmHz}{erg\,s$^{-1}$\,cm$^{-2}$\,Hz$^{-1}$}
\newcommand{\metal}{$12$+$\log ({\rm O/H})$}
\newcommand{\HI}{H\,\textsc{i}}
\newcommand{\HII}{H\,\textsc{ii}}
\newcommand{\mass}{$\log$(M$_{\star}$/M$_{\odot}$)}
\newcommand{\sfr}{$\log$(SFR/M$_{\odot}$\,yr$^{-1}$)}
\newcommand{\EW}{$\textrm{EW}_0$}
\newcommand{\EBV}{$E$($B$$-$$V$)}

\newcommand{\HSCEMPGa}{HSC J1429$-$0110} 
\newcommand{\HSCEMPGb}{HSC J2314$+$0154} 
\newcommand{\HSCEMPGc}{HSC J1142$-$0038} 
\newcommand{\HSCEMPGd}{HSC J1631$+$4426} 
\newcommand{\SDSSEMPGa}{SDSS J0002$+$1715} 
\newcommand{\SDSSEMPGb}{SDSS J1642$+$2233} 
\newcommand{\SDSSEMPGc}{SDSS J2115$-$1734} 
\newcommand{\SDSSEMPGd}{SDSS J2253$+$1116} 
\newcommand{\SDSSEMPGe}{SDSS J2310$-$0211}  
\newcommand{\SDSSEMPGf}{SDSS J2327$-$0200} 

\newcommand*{\mod}[1]{\textcolor{black}{#1}}

\shorttitle{
Subaru EMPG Survey with ML
}
\shortauthors{Kojima et al.}

\begin{document}
\title{
Extremely Metal-Poor Representatives Explored by the Subaru Survey (EMPRESS). I.\\
A Successful Machine Learning Selection of Metal-Poor Galaxies and \\
the Discovery of a Galaxy with $M_*<10^6 M_\odot$ and $0.016 Z_\odot$
}
\author{
Takashi Kojima\altaffilmark{1,2}, 
Masami Ouchi\altaffilmark{3,1,4},
Michael Rauch\altaffilmark{5},
Yoshiaki Ono\altaffilmark{1},
Kimihiko Nakajima\altaffilmark{3},
Yuki Isobe\altaffilmark{1,2},
\\ 
Seiji Fujimoto\altaffilmark{6,7},
Yuichi Harikane\altaffilmark{3,8,1},
Takuya Hashimoto\altaffilmark{9},
Masao Hayashi\altaffilmark{3},
Yutaka Komiyama\altaffilmark{3},\\
Haruka Kusakabe\altaffilmark{10},
Ji Hoon Kim\altaffilmark{11,12},
Chien-Hsiu Lee\altaffilmark{13}, 
Shiro Mukae\altaffilmark{1,14},
Tohru Nagao\altaffilmark{15},\\
Masato Onodera\altaffilmark{11,16},
Takatoshi Shibuya\altaffilmark{17},
Yuma Sugahara\altaffilmark{18,3,1,2},
Masayuki Umemura\altaffilmark{19},
Kiyoto Yabe\altaffilmark{4}
}
\email{E-mail: tkojima@icrr.u-tokyo.ac.jp}
\altaffiltext{1}{%
Institute for Cosmic Ray Research, The University of Tokyo,
5-1-5 Kashiwanoha, Kashiwa, Chiba 277-8582, Japan
}
\altaffiltext{2}{%
Department of Physics, Graduate School of Science, The University of Tokyo, 7-3-1 Hongo, Bunkyo, Tokyo 113-0033, Japan
}
\altaffiltext{3}{%
National Astronomical Observatory of Japan, 2-21-1 Osawa, Mitaka, Tokyo 181-8588, Japan
}
\altaffiltext{4}{%
Kavli Institute for the Physics and Mathematics of the Universe (WPI), 
University of Tokyo, Kashiwa, Chiba 277-8583, Japan
}
\altaffiltext{5}{%
Carnegie Observatories, 813 Santa Barbara Street, Pasadena, CA 91101, USA
}
\altaffiltext{6}{%
Cosmic Dawn Center (DAWN), Copenhagen, Denmark
}
\altaffiltext{7}{%
Niels Bohr Institute, University of Copenhagen, Lyngbyvej 2, DK2100 Copenhagen, Denmark
}
\altaffiltext{8}{%
Department of Physics and Astronomy, University College London, Gower Street, London WC1E 6BT, UK
}
\altaffiltext{9}{%
Tomonaga Center for the History of the Universe (TCHoU), Faculty of Pure and Applied Sciences, University of Tsukuba, Tsukuba, Ibaraki 305-8571, Japan
}
\altaffiltext{10}{%
Observatoire de Gen\`{e}ve, Universit\'e de Gen\`{e}ve, 51 Ch. des Maillettes, 1290 Versoix, Switzerland
}
\altaffiltext{11}{%
Subaru Telescope, National Astronomical Observatory of Japan, National Institutes of Natural Sciences (NINS), 650 North Aohoku Place, Hilo, HI 96720, USA
}
\altaffiltext{12}{%
Metaspace, 36 Nonhyeon-ro, Gangnam-gu, Seoul 06312, Republic of Korea
}
\altaffiltext{13}{%
NSF's National Optical Infrared Astronomy Research Laboratory, Tucson, AZ 85719, USA
}
\altaffiltext{14}{%
Department of Astronomy, Graduate School of Science, The University of Tokyo, 7-3-1 Hongo, Bunkyo, Tokyo, 113-0033, Japan
}
\altaffiltext{15}{%
Research Center for Space and Cosmic Evolution, Ehime University, 2-5 Bunkyo-cho, Matsuyama, Ehime 790-8577, Japan
}
\altaffiltext{16}{%
Department of Astronomical Science, SOKENDAI (The Graduate University for Advanced Studies), Osawa 2-21-1, Mitaka, Tokyo, 181-8588, Japan
}
\altaffiltext{17}{%
Kitami Institute of Technology, 165 Koen-cho, Kitami, Hokkaido 090-8507, Japan
}
\altaffiltext{18}{%
Waseda Research Institute for Science and Engineering, Faculty of Science and Engineering, Waseda University, 3-4-1, Okubo, Shinjuku, Tokyo 169-8555, Japan
}
\altaffiltext{19}{%
Center for Computational Sciences, University of Tsukuba, Tsukuba, Ibaraki 305-8577, Japan.
}

\altaffiltext{\dag}{Partly based on data obtained with the Subaru Telescope. The Subaru Telescope is operated by the National Astronomical Observatory of Japan.}
\altaffiltext{\ddag}{The data presented herein were partly obtained at the W. M. Keck Observatory, which is operated as a scientific partnership among the California Institute of Technology, the University of California and the National Aeronautics and Space Administration. The Observatory was made possible by the generous financial support of the W. M. Keck Foundation.}
\altaffiltext{\#}{This paper includes data gathered with the 6.5 meter Magellan Telescopes located at Las Campanas Observatory, Chile.}

\begin{abstract}
We have initiated a new survey for local extremely metal-poor galaxies (EMPGs) with Subaru/Hyper Suprime-Cam (HSC) large-area ($\sim 500$ deg$^2$) optical images reaching a $5\sigma$ limit of $\sim 26$ magnitude, about 100 times deeper than the Sloan Digital Sky Survey (SDSS). To select $Z/Z_{\odot}<0.1$ EMPGs from $\sim$ 40 million sources detected in the Subaru images, we first develop a machine-learning (ML) classifier based on a deep neural network algorithm with a training data set consisting of optical photometry of galaxy, star, and QSO models. We test our ML classifier with SDSS objects having spectroscopic metallicity measurements, and confirm that our ML classifier accomplishes 86\%-completeness and 46\%-purity EMPG classifications with photometric data. Applying our ML classifier to the photometric data of the Subaru sources as well as faint SDSS objects with no spectroscopic data, we obtain 27 and 86 EMPG candidates from the Subaru and SDSS photometric data, respectively. We conduct optical follow-up spectroscopy for 10 out of our EMPG candidates with Magellan/LDSS-3$+$MagE, Keck/DEIMOS, and Subaru/FOCAS, and find that the 10 EMPG candidates are star-forming galaxies at $z=0.007-0.03$ with large {\Hb} equivalent widths of 104--265 {\AA}, stellar masses of {\mass}$=$5.0--7.1, and high specific star-formation rates of $\sim$300\,Gyr$^{-1}$, which are similar to those of early galaxies at $z\gtrsim6$ reported recently. 
We spectroscopically confirm that 3 out of 10 candidates are truly EMPGs with $Z/Z_{\odot}<0.1$, one of which is {\HSCEMPGd}, the most metal-poor galaxy with $Z/Z_{\odot}=0.016$ reported ever.
\end{abstract}


\section{INTRODUCTION} \label{sec:intro}
The early universe is dominated by a large number of young, low-mass, metal-poor galaxies.
Theoretically, first galaxies are formed at $z\sim10$--$20$ from gas already metal-enriched by Pop-III (i.e., metal-free) stars \citep[e.g.,][]{Bromm2011}. 
According to hydrodynamical simulation \citep[e.g.,][]{Wise2012}, first galaxies are created in dark matter (DM) mini halos with $\sim10^{8}M_{\odot}$ and have low stellar masses ($M_{\star}\sim10^{4}$--$10^{6}M_{\odot}$), low metallicities ($Z\sim0.1$--$1\% Z_{\odot}$), and high specific star-formation rates (${\rm sSFR}\sim100$ Gyr$^{-1}$) at $z\sim10$.
The typical stellar mass ($M_{\star}\sim10^{4}$--$10^{6}M_{\odot}$) is remarkably small, comparable to those of star clusters.
Such cluster-like galaxies are undergoing an early stage of the galaxy formation, which is characterized by an intensive star formation.
One of ultimate goals of modern astronomy is to understand the early-stage galaxy formation by probing the cluster-like galaxies.
The cluster-like galaxies are also the key galaxy population, which are building blocks of the galaxy formation hierarchy.

Recent observations \citep[e.g.,][]{Stark2014a} have reported that low-mass, young galaxies of {\mass}$\sim$6--9 at $z\sim2$ show strong emission lines with very high equivalent widths (EWs), $\sim$1000 {\AA}, for {\OIII}+{\Hb} lines.
Such very high EWs suggest an intensive star formation predicted by the hydrodynamical simulation of the first galaxies \citep[e.g.,][]{Wise2012}.
Stellar synthesis and photoionization models \citep{Inoue2011} also demonstrate that the rest-frame equivalent width of {\Ha} line, {\EW}({\Ha}), can reach $\sim$1,000--3,000 {\AA} for stellar ages of $\lesssim$100 Myr.
The association between the first galaxies and local, low-mass galaxies is partly investigated with 
cosmological hydrodynamic zoom-in simulations by \citet{Jeon2017}. 
\citet{Jeon2017} suggest that the Local Group (LG) ultra-faint dwarf galaxies (UDGs) began as young, low-mass, star-forming galaxies (SFGs) in the past time and have been quenched during the epoch of the cosmic re-ionization.
Thus, the LG UDGs themselves are not analogs of the high-z galaxies because they are already quenched and old.
Contrary to the LG UDGs, low-mass, young, SFGs discovered in the local Universe are undergoing an intensive star-forming phase and can be regarded as analogs of high-z galaxies.

In the last decade, metal-poor galaxies with a large {\OIII}$\lambda$5007 rest-frame equivalent width, {\EW}({\OIII}$\lambda$5007), have been discovered by the broad-band excess technique in the data of Sloan Digital Sky Survey \citep[SDSS;][]{York2000}. 
For example, \citet{Cardamone2009} have reported metal-poor, actively star-forming galaxies at $z\sim0.3$ in the SDSS data, which have been named  ``green pea galaxies'' (GPs) after their compact size and intrinsically green color caused by the large {\EW}({\OIII}$\lambda$5007) up to $\sim$1500 {\AA}. 
\citet{Yang2017b} have also discovered metal-poor, highly star-forming galaxies at $z\sim0.04$ in the SDSS data  selected with the $g$-band excess with the very large {\EW}({\OIII}$\lambda$5007)$\sim$500--2500 {\AA}.
The galaxies found by \citet{Yang2017b} have been nicknamed ``blueberry galaxies'' (BBs).

Typical metallicities of these GPs/BBs are {\metal}=8.0$\pm$0.3, which fall into a \textit{moderate} metallicity range compared to extremely metal-poor galaxies (EMPGs) such as J0811$+$4730 \citep[][]{Izotov2018a}, SBS0335$-$052 \citep[e.g.,][]{Izotov2009}, AGC198691 \citep[][]{Hirschauer2016}, J1234$+$3901 \citep[][]{Izotov2019a}, LittleCub \citep[][]{Hsyu2017}, DDO68 \citep[][]{Pustilnik2005, Annibali2019}, IZw18 \citep[e.g.,][]{Izotov1998a, Thuan2005a}, and LeoP \citep[][]{Skillman2013} in the range of {\metal}$\sim$7.0--7.2.
Stellar synthesis and photoionization models \citep{Inoue2011} suggest that the {\EW}({\OIII}$\lambda$5007) takes maximum values around {\metal} $\sim$ 8.0.
Thus, galaxies selected with a single broadband excess such as GPs/BBs may be somewhat biased towards a large {\EW}({\OIII}$\lambda$5007), i.e., a \textit{moderate} metallicity of {\metal} $\sim$ 8.0.

\begin{figure}
\epsscale{1.1}
\plotone{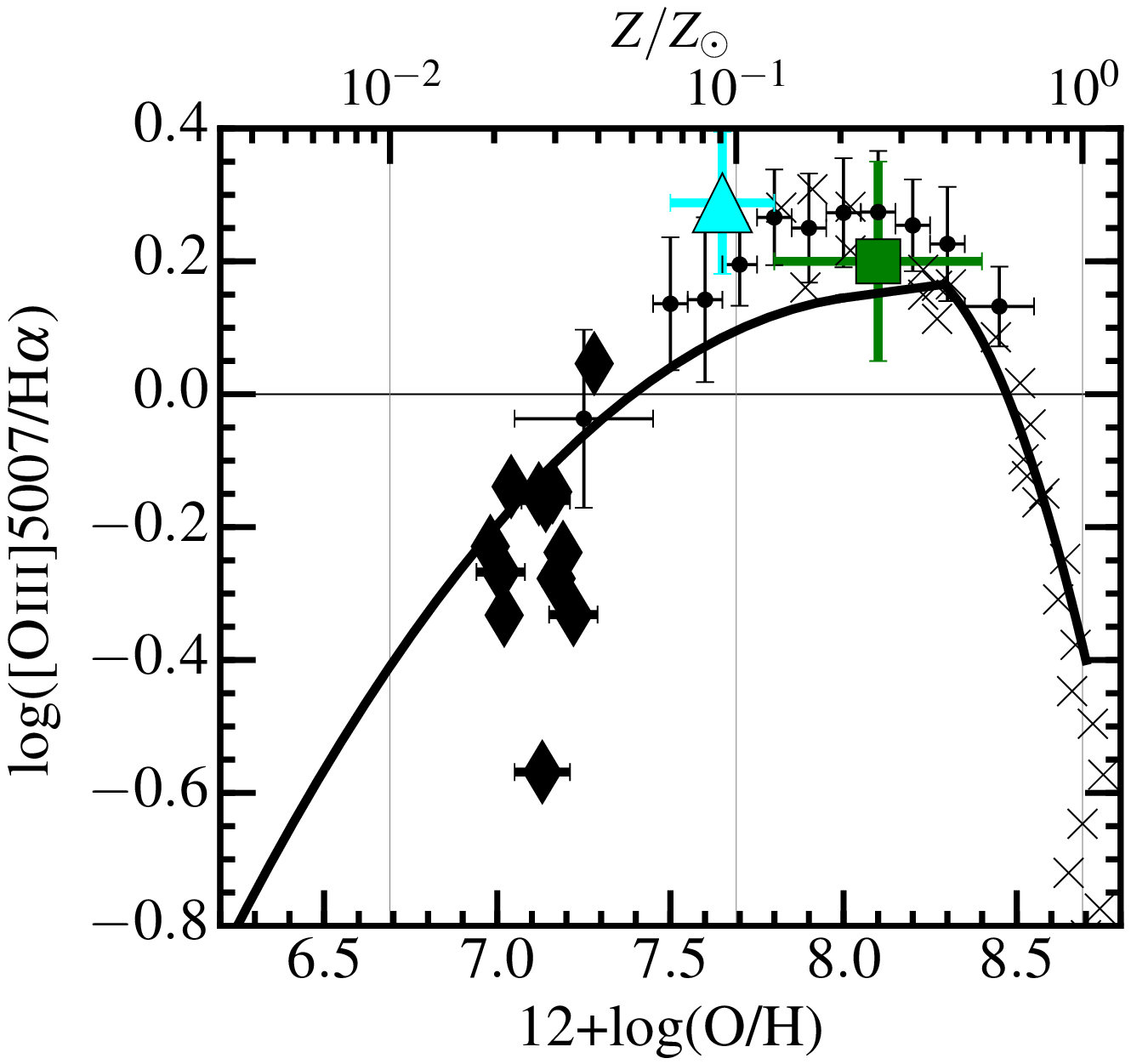}
\caption{[\textsc{Oiii}]/H$\alpha$ ratio as a function of gas-phase metallicity. The cross marks \citep{Andrews2013} and dots \citep{Nagao2006} represent the average of local star-forming galaxies. 
We also show the typical values of SDSS GPs \citep[green square;][]{Cardamone2009,Amorin2010} and BBs \citep[cyan triangle;][]{Yang2017b}.
The solid line is a theoretical prediction \citep{Inoue2011}. 
The diamonds are representative metal-poor galaxies, J0811$+$4730 \citep[][]{Izotov2018a}, SBS0335$-$052 \citep[e.g.,][]{Izotov2009}, AGC198691 \citep[][]{Hirschauer2016}, J1234$+$3901 \citep[][]{Izotov2019a}, LittleCub \citep[][]{Hsyu2017}, DDO68 \citep[][]{Pustilnik2005, Annibali2019}, IZw18 \citep[e.g.,][]{Izotov1998a, Thuan2005a}, and LeoP \citep[][]{Skillman2013} in the range of {\metal}$\sim$7.0--7.2.
The GPs and BBs show \textit{moderate} metallicities around {\metal}$\sim$8.0, which correspond to a high [\textsc{Oiii}]/H$\alpha$ ratio of $\sim$2, while the representative EMPGs have relatively low [\textsc{Oiii}]/H$\alpha$ ratio of $\sim$0.3--1.0.
\label{fig:O3metal}}
\end{figure}
\begin{figure}
\epsscale{1.1}
\plotone{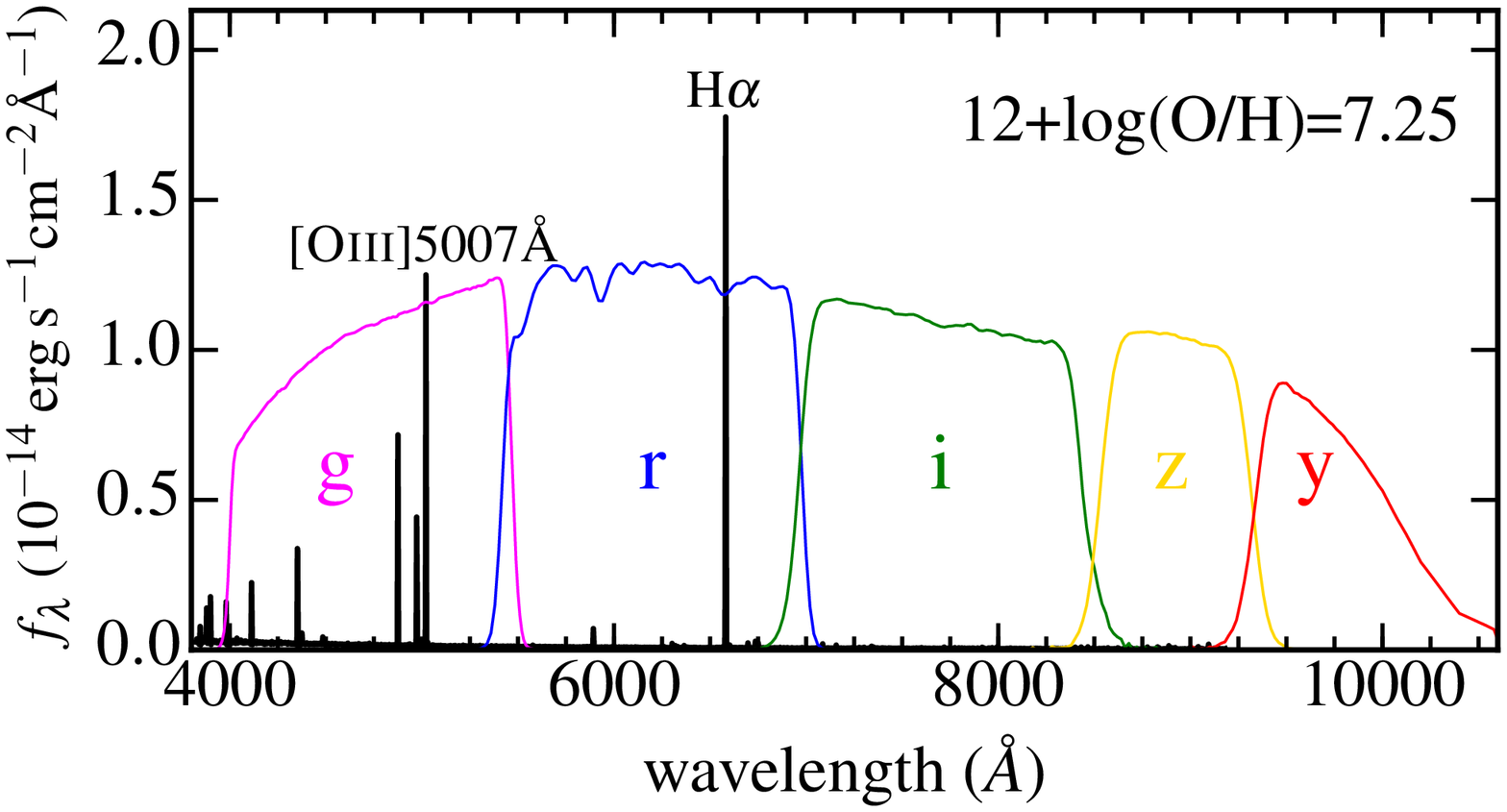}
\plotone{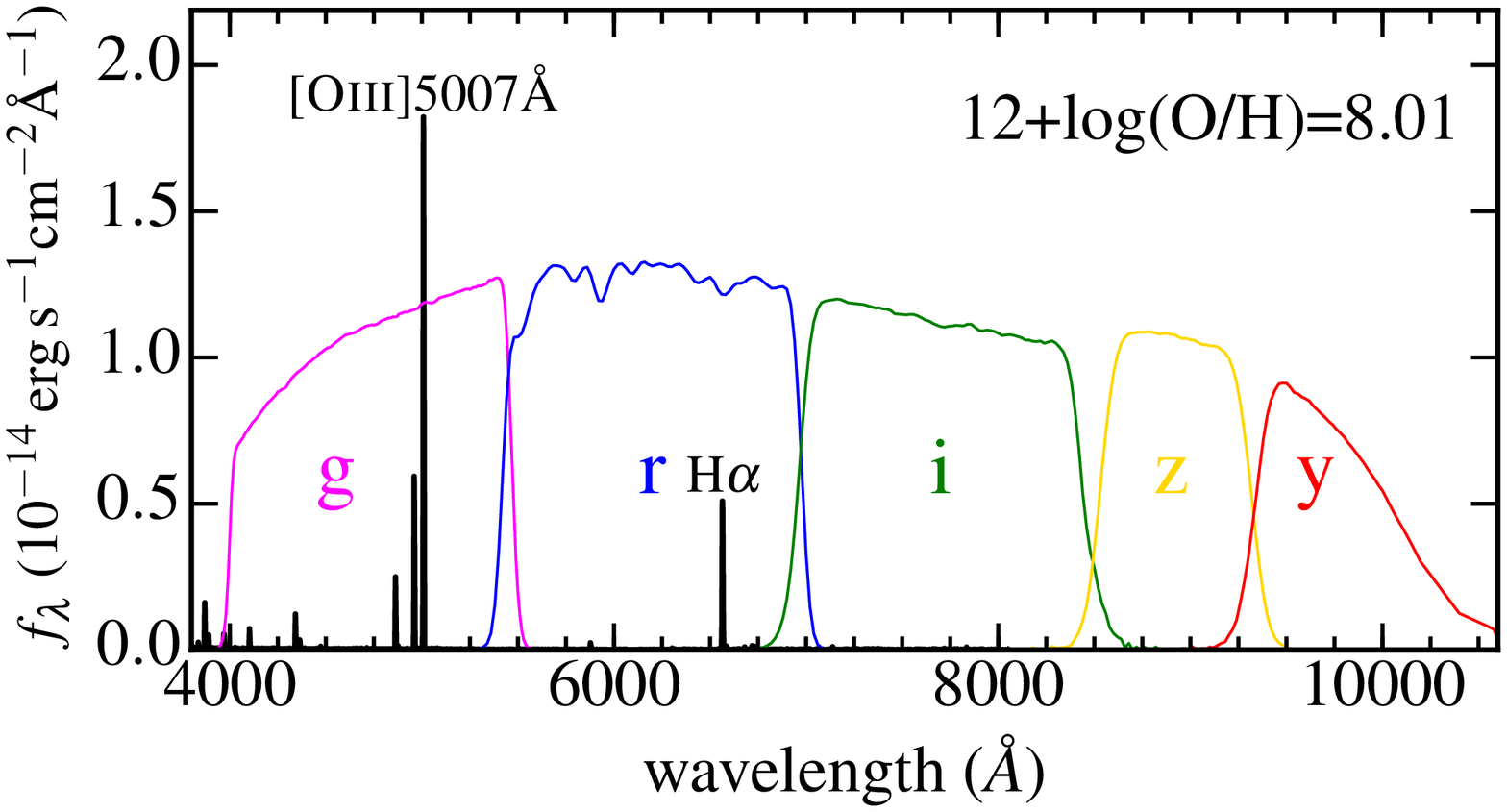}
\caption{Top: Spectrum example of an EMPG with a very low metallicity of 12+log(O/H)=7.25 \citep{Kniazev2003}. 
Bottom: Same as top panel, but for a GP with a moderate metallicity of 12+log(O/H)=8.01 \citep{Jaskot2013}.
We show the GP spectrum in the rest-frame for an easy comparison with the EMPG spectrum.
The color curves are throughput curves of HSC $grizy$-band filters for reference.
In the optical spectrum of this typical EMPG (top panel), {\Ha} is the strongest line. 
\label{fig:spectrum_ref}}
\end{figure}

As shown in Figure \ref{fig:O3metal}, the models of \citet{Inoue2011} also exhibit a peak of {\OIII}$\lambda$5007/{\Ha} flux ratio at around {\metal} $\sim$ 8.0.
The {\OIII}$\lambda$5007/{\Ha} ratio monotonically decreases with decreasing metallicity in the range of {\metal} $<$ 8.0 simply because the oxygen element becomes deficient.
Indeed, as shown in Figure \ref{fig:O3metal}, representative metal-poor galaxies \citep[e.g.,][{\metal}$\sim$7.0--7.2]{Thuan2005a, Izotov2009, Skillman2013, Hirschauer2016, Izotov2018a} have a ratio of {\OIII}$\lambda$5007/{\Ha}=0.4--1.0.
The {\OIII}$\lambda$5007 line is no longer the strongest emission line in an optical spectrum of an EMPG as demonstrated in Figure \ref{fig:spectrum_ref}.
Thus, optical spectra of the EMPGs are characterized by multiple strong emission lines such as hydrogen Balmer lines and {\OIII}$\lambda\lambda$4959,5007.
The strong emission lines of EMPGs cause $g$- and $r$-band excesses at $z\lesssim0.03$.

Recently, \citet{Hsyu2018} and \citet{Senchyna2019a} have started metal-poor galaxy surveys with the SDSS data, where they have selected objects that show $g$- and $r$-band excesses.
Unfortunately, the EMPGs have similar colors to those of other types of objects (i.e., blue stars, transient objects) on a classical color-color diagram of $r-i$ vs. $g-r$ \citep[][]{Hsyu2018, Senchyna2019a}.
In this paper, we aim to use machine learning (ML) techniques to find an EMPG selection criteria that is not limited by the linear relationships in color-color space, which is the most common selection criteria used in previous works.

In this study, we target EMPGs with strong emission lines in the local universe ($z\lesssim0.03$), which may be a local analog of a high-$z$ star-forming galaxy.
Because such galaxies are intrinsically faint and rare in the local universe, wide-field, deep imaging data are necessary.
However, the SDSS data are not deep enough to discover EMPGs with {\mass}$<$6, which are possible candidates of the most metal-deficient galaxies inferred from the mass-metallicity relation \citep[e.g.,][]{Andrews2013,Wuyts2012b}.
Deeper, wide-field imaging data surveys have since been conducted to discover faint EMPGs that are undetected by SDSS-like surveys.
In March 2014, the Subaru telescope has started a large-area ($\sim$1,400 deg$^2$), deep survey with Hyper Suprime-Cam \citep[HSC;][]{Miyazaki2012, Miyazaki2018, Komiyama2018, Kawanomoto2018, Furusawa2018}, called HSC Subaru Strategic Program \citep[HSC SSP;][]{Aihara2018a}.

Based on the HSC-SSP data, we have initiated a new survey for local EMPGs, which has been named ``Extremely Metal-Poor Representatives Explored by the Subaru Survey'' (EMPRESS).
The final goal of our research is to discover faint EMPGs by exploiting the Subaru HSC-SSP data, whose $i$-band limiting magnitude ($i_{\rm lim}$$\sim$26 mag) is $\sim$5 mag deeper than the one of the SDSS data ($i_{\rm lim}$$\sim$21 mag).
We also use the SDSS data to complement brighter EMPGs in this study. 
This paper is the first paper from our EMPRESS program, which explores EMPGs based on the S17A and S18A data releases of HSC SSP.
This first paper will be followed by other papers in which we investigate details of elemental abundances, physical states of inter-stellar medium, size and morphology, stellar population of our EMPGs.
We plan to continue updating the EMPG sample with the future HSC-SSP data release and up-coming follow-up spectroscopy.

The outline of this paper is as follows.
In Section \ref{sec:data}, we explain the Subaru HSC-SSP data as well as how we construct a source catalog from the HSC-SSP data.
We also make a source catalog from SDSS photometry data to complement our EMPG sample.
Section \ref{sec:classifier} explains our new selection technique that we develop with the ML and shows results of a test of our ML selection.
Section \ref{sec:selection} explains the selection of EMPG candidates from the source catalogs.
We describe our optical spectroscopy carried out for our EMPG candidates in Section \ref{sec:spectroscopy}.
Section \ref{sec:reduction} explains the reduction and calibration processes of our spectroscopy data.
In Section \ref{sec:analysis}, we estimate emission line fluxes and galaxy properties such as stellar mass, star-formation rate, and metallicity. 
We show the results of our spectroscopy and compare our EMPG sample with other low-$z$ galaxy samples in the literature in Section \ref{sec:results}.
Then we summarize our results in Section \ref{sec:summary}.

Throughout this paper, magnitudes are on the AB system \citep{Oke1983}. 
We adopt the following cosmological parameters, $(h, \Omega_m, \Omega_{\Lambda}) = (0.7, 0.3, 0.7)$.
The definition of the solar metallicity is given by {\metal}=8.69 \citep{Asplund2009}.
We also define an EMPG as a galaxy with {\metal}$<$7.69 (i.e., $Z/Z_{\odot}$$<$0.1) in this paper, which is almost the same as in previous metal-poor galaxy studies \citep[e.g.,][]{Kunth2000, Izotov2012, Guseva2017}.
In this paper, we try to select EMPGs with a large {\EW}(H$\alpha$) (e.g., $\gtrsim$800 {\AA}) because our motivation is to discover local counterparts of high-$z$, low-mass galaxies whose specific star-formation rate (sSFR) is expected to be high \citep[$\gtrsim$10 Gyr$^{-1}$, e.g.,][]{Ono2010b, Stark2017, Elmegreen2017b, Harikane2018b}.

\section{DATA} \label{sec:data}
\begin{figure*}
\epsscale{1.1}
\plotone{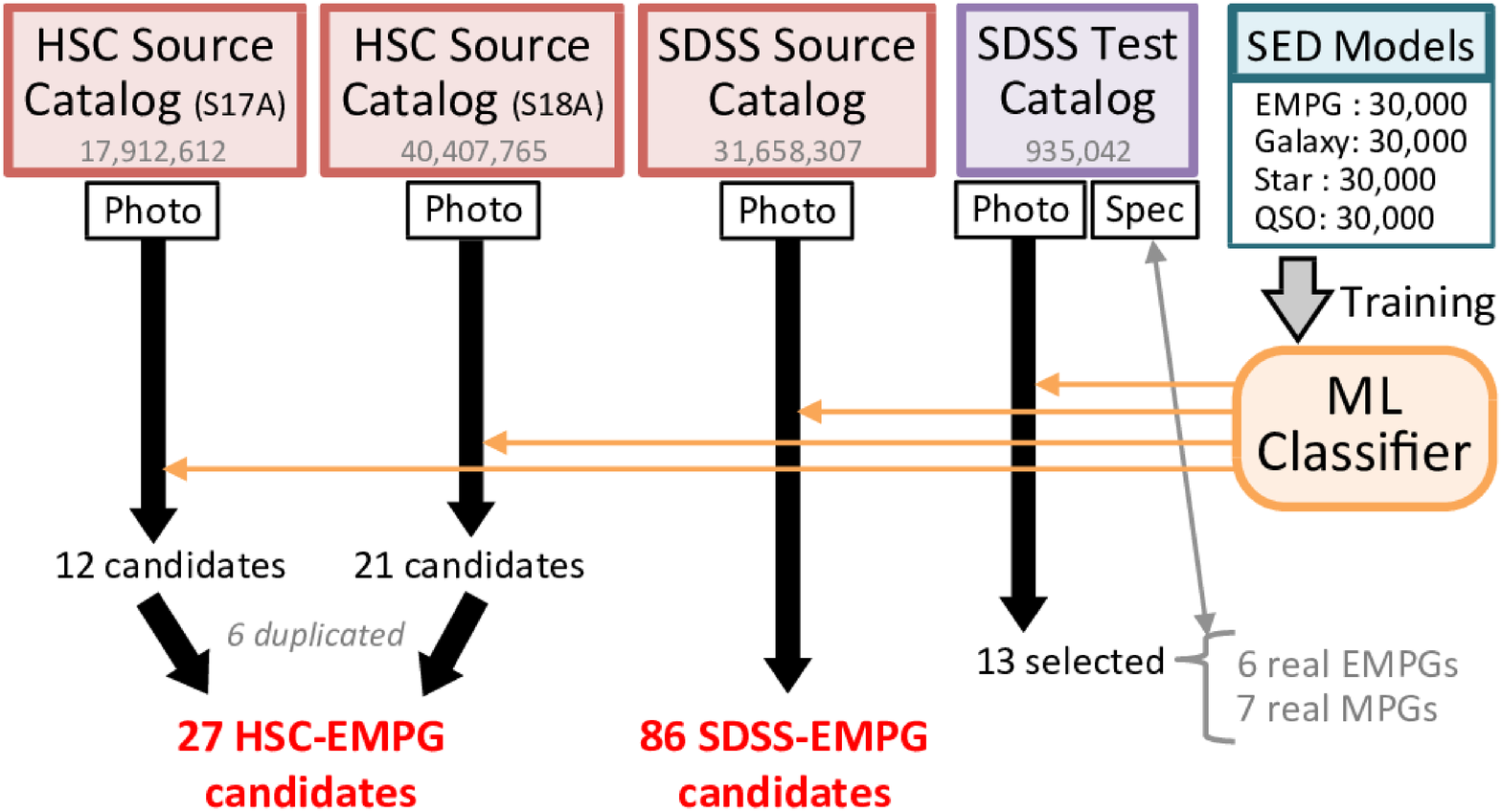}
\caption{Picture of our selection flow. 
We select our EMPG candidates from the HSC source catalogs (Section \ref{subsec:HSCcatalog}) and SDSS  source catalog (Section \ref{subsec:SDSScatalog}), which consist of photometric data.
To test our ML classifier, we use the SDSS test catalog (Section \ref{subsec:SDSStestcatalog}), which is composed of photometry$+$spectroscopy data.
Our ML classifier (Section \ref{subsec:classifier}) is trained by spectral energy distribution (SED) models of galaxies, stars, and QSOs (Section \ref{subsec:model}).
We do not use the existing observational data in the training because we target very faint EMPGs that no previous survey could discover.
Part of details are omitted in this flow for simplicity. See the details in each sections.
\label{fig:flow}}
\end{figure*}

We explain HSC-SSP imaging data used in this study in Section \ref{subsec:HSCdata}.
We construct source catalogs from the HSC-SSP and SDSS data in Sections \ref{subsec:HSCcatalog} and \ref{subsec:SDSScatalog}, respectively.
These source catalogs and following selection processes are summarized in Figure \ref{fig:flow}.

\subsection{HSC-SSP Imaging Data} \label{subsec:HSCdata}
%
We use the HSC-SSP internal data of S17A and S18A data releases, which are taken from 2014 March to 2017 May and from 2014 March to 2018 Jan, respectively.
The internal S17A+S18A data are explained in the second data release (DR2) paper of HSC SSP \citep[][]{Aihara2019}.
The HSC-SSP survey data are taken in three layers of Wide, Deep, and UltraDeep with 5 broad-band filters of $grizy$.
In the HSC-SSP S17A and S18A data releases, images were reduced with the HSC pipeline, {\tt hscPipe} v5.4 and v6.7 \citep[][]{Bosch2018}, respectively, with codes of the Large Synoptic Survey Telescope (LSST) software pipeline \citep[][]{Ivezic2008, Axelrod2010, Juric2015, Ivezic2019}.
The pipeline conducts the bias subtraction, flat fielding, image stacking, astrometry and zero-point magnitude calibration, source detection, and magnitude measurement.
The {\tt hscPipe} v6.7 (S18A) uses the global background subtraction, a lower detection threshold, a new artifact rejection algorithm, the different co-add weighting between old and new $i$/$r$-band filters, and the updated way of the point-spread-function (PSF) estimation \citep[detailed in][]{Aihara2019}.
These pipeline differences slightly change the detection and magnitude measurements, which may affect our classification results.
Indeed, as we will explain later in Section \ref{subsec:HSC_EMPG}, we find that part of EMPG candidates are selected only in either of S17A or S18A data, which is caused by the different {\tt hscPipe} versions between S17A and S18A.
To maximize the EMPG sample size, we use both S17A and S18A data in this paper.
The details of the observations, data reduction, and detection and photometric catalog are described in \citet{Aihara2019} and \citet{Bosch2018}.
We use {\tt cmodel} magnitudes corrected for Milky-Way dust extinction to estimate the total magnitudes of a source.
The {\tt cmodel} magnitudes are deblended by fitting profiles to multiple sources on an image.
Thus, even if a certain source is overlapped with other sources, the {\tt cmodel} magnitude represents a total magnitude of the source, which is almost free from its overlapping sources.
See the detailed algorithm of the {\tt cmodel} photometry in \citet{Bosch2018}.

\subsection{HSC source catalog} \label{subsec:HSCcatalog}
We explain how we construct an HSC source catalog, from which we select EMPG candidates.
We select sources from the HSC-SSP Wide field data.
We use isolated or cleanly deblended sources that fall within $griz$-band images.
We also require that none of the pixels in their footprints are interpolated, none of the central 3 $\times$ 3 pixels are saturated, none of the central 3 $\times$ 3 pixels suffer from cosmic rays, and there are no bad pixels in their footprints. 
Then we exclude sources whose {\tt cmodel} magnitude or centroid position measurements are flagged as problematic by hscPipe. 
We exclude sources close to a bright star \citep{Coupon2018, Aihara2019}.
We require the detection in the $griz$-band images.
Here we select objects whose photometric measurements are brighter than 5$\sigma$ limiting magnitudes, $g<26.5$, $r<26.0$, $i<25.8$, and $z<25.2$ mag, which are estimated by \citet[][]{Ono2018} with 1.5-arcsec diameter circular apertures.
Note again that we use {\tt cmodel} photometry to select EMPG candidates. 
In this study, we do not use $y$-band photometry because the $y$-band limiting magnitude is shallower \citep[$y=24.5$ mag; ][]{Ono2018} than the other 4 bands and the $y$-band imaging has not yet been completed  in part of the survey area that we use in this study. 
We also require that the photometric measurement errors are less than 0.1 mag in $griz$ bands.
Here the photometric measurement errors are given by {\tt hscPipe}.
Finally, we obtain 17,912,612 and 40,407,765 sources in total from the HSC-SSP S17A and S18A data, respectively.
The effective areas are 205.82 and 508.84 deg$^2$ in the HSC-SSP S17A and S18A data, respectively.
Note again that there is overlap between S17A and S18A data (see also Sections \ref{subsec:HSCdata} and \ref{subsec:HSC_EMPG}).
Table \ref{table:flags} summarizes the selection criteria that we apply to make the HSC source catalog.

\begin{deluxetable*}{lccl}
\tablecolumns{4}
\tabletypesize{\scriptsize}
\tablecaption{Selection criteria in our source catalog construction. %
\label{table:flags}}
\tablehead{%
\colhead{Parameter} &
\colhead{Value} &
\colhead{Band} &
\colhead{Comment} \\
\colhead{} &
\colhead{} &
\colhead{} &
\colhead{} \\
\colhead{(1)} &
\colhead{(2)} &
\colhead{(3)} &
\colhead{(4)} 
}
\startdata
\verb|isprimary| & True & --- & Object is a primary one with no deblended children. \\
\verb|detect_ispatchinner| & True & --- &  Object falls on the inner region of a coadd patch. \\
\verb|detect_istractinner| & True & --- & Object falls on the inner region of a coadd tract. \\
\verb|pixelflags_edge| & False & $griz$ & Object locates within images. \\
\verb|pixelflags_interpolatedcenter| & False & $griz$ & None of the central $3 \times 3$ pixels of an object is interpolated. \\
\verb|pixelflags_saturatedcenter| & False & $griz$ &  None of the central $3 \times 3$ pixels of an object is saturated. \\
\verb|pixelflags_crcenter| & False & $griz$ & None of the central $3 \times 3$ pixels of an object is masked as cosmic ray. \\
\verb|pixelflags_bad| & False & $griz$ & None of the pixels in the footprint of an object is labelled as bad. \\
\verb|cmodel_flag| & False & $griz$ & Cmodel flux measurement has no problem. \\
\verb|merge_peak| & True & $griz$  & Detected in $griz$ bands.\\
\verb|mask_bright_objectcenter|\tablenotemark{a} & False & $griz$ & No bright stars near an object.\\
\verb|cmodel_mag| & $<26.5$ & $g$ & $g$-band cmodel magnitudes are smaller than the 5$\sigma$ limiting magnitude. \\
                               & $<26.0$ & $r$ & $r$-band cmodel magnitudes are smaller than the 5$\sigma$ limiting magnitude. \\
                               & $<25.8$ & $i$ & $i$-band cmodel magnitudes are smaller than the 5$\sigma$ limiting magnitude. \\
                               & $<25.2$ & $z$ & $z$-band cmodel magnitudes are smaller than the 5$\sigma$ limiting magnitude. \\
\verb|cmodel_magsigma| & $<0.1$ & $griz$ & Errors of $griz$-bamd cmodel magnitudes are less than 0.1 mag. 
\enddata
\tablecomments{
{$^a$} Only used in the S18A catalog because we find many contaminants in S18A data caused by nearby bright stars, while we do not in S17A.
}
\end{deluxetable*}

\subsection{SDSS source catalog} \label{subsec:SDSScatalog}
We construct a SDSS source catalog from the 13th release \citep[DR13;][]{Albareti2017} of the SDSS photometry data. 
Although the SDSS data are $\sim$5 mag shallower ($i_{\rm lim}$$\sim$21 mag) than HSC-SSP data ($i_{\rm lim}$$\sim$26 mag), we expect to select apparently bright, but intrinsically faint EMPGs from the SDSS data in the local volume (i.e., $z\lesssim0.001$).
Such apparently bright, but intrinsically faint EMPGs can be discovered because a total unique area of the SDSS (14,555 deg$^{2}$) data is $\sim$30 times larger than the S18A HSC-SSP data ($509$ deg$^{2}$).
It should also be noted that 99.0\% of total SDSS sources have not yet been observed in spectroscopy\footnote{\texttt{https://www.sdss.org/dr13/scope/}}.
Here we select objects whose photometric measurements are brighter than SDSS limiting magnitudes, $u<22.0$, $g<22.2$, $r<22.2$, $i<21.3$, and $z<21.3$ mag\footnote{Magnitudes reaching 95\% completeness, which are listed in \texttt{https://www.sdss.org/dr13/scope/}}.
We only obtain objects whose magnitude measurement errors are $<$0.1 mag in $ugriz$ bands.
Here we use {\tt Modelmag} for the SDSS data.
Among flags in the {\tt PhotoObjALL} catalog, we require that a {\tt clean} flag is ``1'' (i.e., {\it True}) to remove objects with photometry measurement issues.
The {\tt clean} flag\footnote{Details are described in\\ \texttt{http://www.sdss.org/dr13/algorithms/photo\_flags\_recommend/}} eliminates the duplication, deblending/interpolation problem, suspicious detection, and detection at the edge of an image.
We also remove objects with a {\it True} comic-ray flag and/or a {\it True} blended flag, which often mimics a broad-band excess in photometry. 
We reject relatively large objects with a ninety-percent petrosian radius greater than 10 arcsec to best eliminate contamination of HII regions in nearby spiral galaxies. 
Finally, we derive 31,658,307 sources in total from the SDSS DR13 photometry data.
The total unique area of SDSS DR13 data is 14,555 deg$^2$.

\section{CONSTRUCTION OF CLASSIFIER} \label{sec:classifier}
In this section, we construct a classifier based on ML, which will be applied to the HSC-SSP and SDSS source catalogs to select EMPGs.
We target galaxies that have a metallicity of {\metal}$=$6.69--7.69 (i.e., 1--10\% solar metallicity) with a rest-frame {\Ha} equivalent width of {\EW}({\Ha})$>$800 {\AA}.
The basic idea of our selection technique is to build an object classifier that separates EMPG candidates from other types of objects, such as non-EMPG galaxies\footnote{We define a non-EMPG galaxy to have a metallicity 12+log(O/H) larger than 7.69.}, stars, and QSOs.
We construct the object classifier with a deep neural network \citep[DNN;][]{Lecun2015}.
In Section \ref{subsec:targets}, we discuss typical colors of EMPGs to show how we determine the ranges of metallicity, equivalent width, and redshift of EMPGs that we target in this study.
Section \ref{subsec:classifier} explains how we construct our ML classifier that distinguishes EMPGs from non-EMPG galaxies, stars, and QSOs.
In Section \ref{subsec:test}, we test our ML classifier with the SDSS photometry+spectroscopy data (i.e., data of SDSS objects that are detected in photometry and observed in spectroscopy) to check whether our ML classifier successfully selects EMPGs.
We refer to a catalog made from the SDSS photometry+spectroscopy data as a SDSS test catalog in this paper.

\subsection{EMPG Colors} \label{subsec:targets}
We examine typical colors of EMPGs in the literature.
This paper only focuses on EMPGs at $z\lesssim0.03$, where the {\OIII}+{\Hb} and {\Ha} lines fall on the $g$-band and $r$-band, respectively.

\begin{figure}
\epsscale{1.1}
\plotone{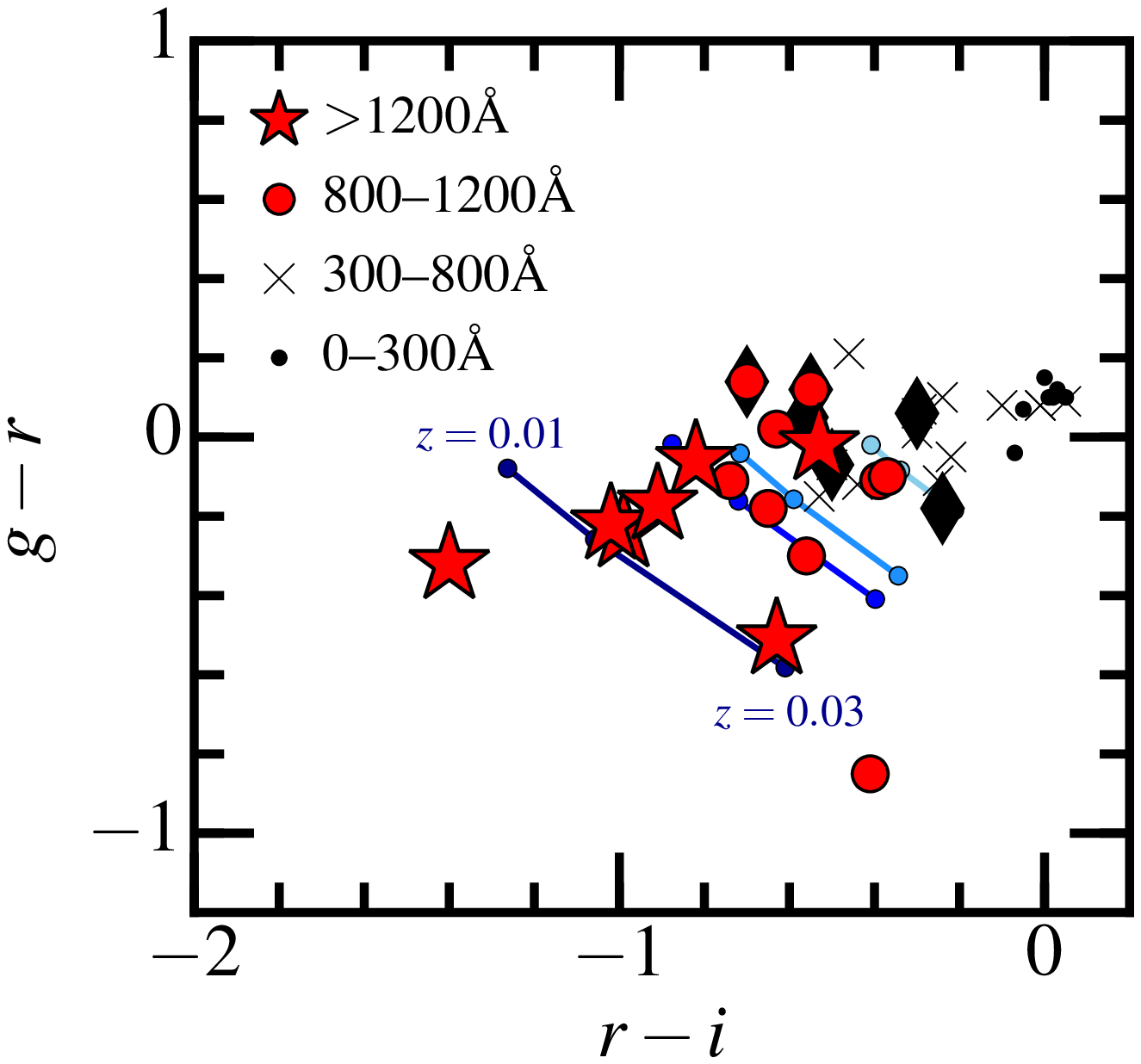}
\caption{Color-color diagram of $g-r$ vs. $r-i$ for previously reported metal-poor galaxies with  {\metal}$<$7.69 at $z<0.03$.
The red stars, red circles, black crosses, and black dots represent SDSS metal-poor galaxies with {\EW}({\Ha})$>$1,200 {\AA}, {\EW}({\Ha})$=$800--1,200 {\AA}, {\EW}({\Ha})$=$300--800 {\AA}, and {\EW}({\Ha})$=$0--300 {\AA}, respectively.
The diamonds show EMPGs at $z<0.03$ with {\metal}$\sim$7.0--7.2, AGC198691 \citep[][]{Hirschauer2016}, LittleCub \citep[][]{Hsyu2017}, DDO68 \citep[][]{Pustilnik2005, Annibali2019}, IZw18 \citep[e.g.,][]{Izotov1998a, Thuan2005a}, and LeoP \citep[][]{Skillman2013}, which have SDSS photometry data.
The four blue solid lines present the \textsc{beagle} model calculations with {\EW}({\Ha})$\sim$2,500, 1,500, 1,000, and 500 {\AA} (from dark blue to light blue) under the assumption of {\metal}=7.50.
On the blue solid lines, redshifts are indicated with dots ($z=0.01$, $0.02$, and $0.03$ from upper left to lower right).
The models are calculated in the same manner as the EMPG models in Section \ref{subsec:model}.
The SDSS metal-poor galaxies with a larger {\EW}({\Ha}) show smaller $r-i$ values due to the strong {\Ha}-line contribution in an $r$-band magnitude, which are consistent with the \textsc{beagle} model calculations.
\label{fig:ref_color}}
\end{figure}

We compile SDSS metal-poor galaxies at $z<0.03$ with {\metal}$<$7.69 from the literature \citep{Kunth2000, Kniazev2003, Guseva2007, Izotov2007, Izotov2009, Pustilnik2010, Izotov2012, Pilyugin2012, SanchezAlmeida2016, Guseva2017}.
Figure \ref{fig:ref_color} shows these SDSS metal-poor galaxies on the $r-i$ vs. $g-r$ diagram, whose {\EW}(\Ha) values are in the ranges of 0---300, 300--800, 800--1,200, and $>$1,200 {\AA}.
In Figure \ref{fig:ref_color}, metal-poor galaxies with a higher {\EW}(\Ha) have a smaller $r-i$ value with $g-r\sim0$ due to the $g$- and $r$-band excesses caused by strong nebular emission lines (top panel of Figure \ref{fig:spectrum_ref}).
This trend is also supported by the stellar synthesis and photoionization models as shown with solid lines in Figure \ref{fig:ref_color}. 
These $g$- and $r$-band excesses are typical for EMPGs with strong emission lines, which basically enables us to separate EMPGs from other types of objects (e.g., galaxies, stars, and QSOs) only with photometric data.
In addition, as described in Section \ref{sec:intro}, EMPGs with strong emission lines are expected to be local analogs of high-$z$ SFGs because high-$z$ SFGs have a high sSFR, which corresponds to high emission-line equivalent widths, and a low metallicity.

\begin{figure}
\epsscale{1.1}
\plotone{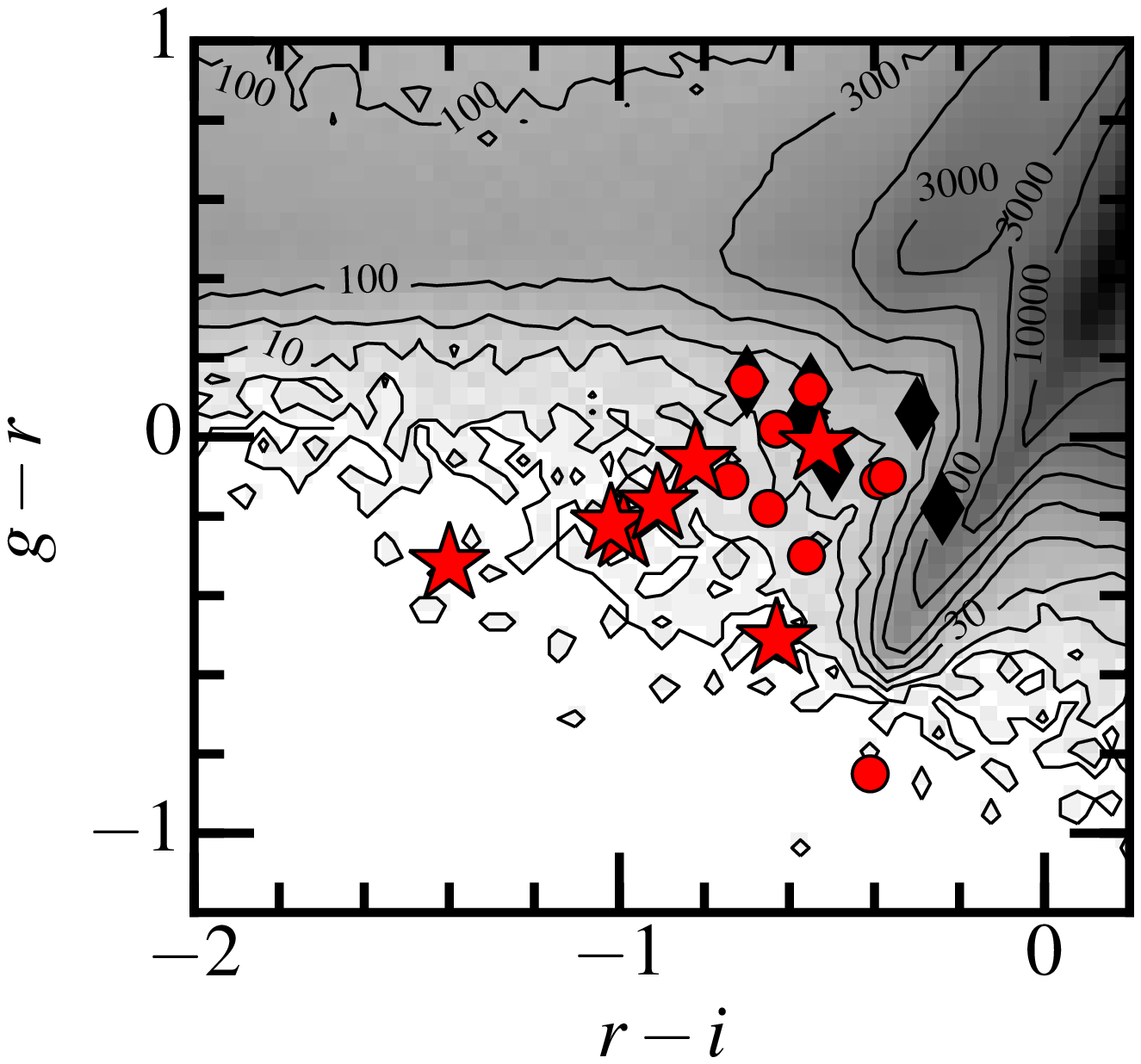}
\caption{The same as Figure \ref{fig:ref_color}, but only with the SDSS EMPGs with {\EW}(\Ha)$>$800 {\AA}.
The black mesh and contours represent a two-dimensional histogram of the SDSS source catalog (discussed later in Section\ref{subsec:SDSS_EMPG}) .
The contours indicate the number of sources ($N$=1, 3, 10, 30,..., 10000) in each bin with a size of $\Delta m$=0.04 mag. 
On this color-color diagram, the EMPGs overlap largely with many SDSS sources, most of which are contaminants such as stars.
\label{fig:ref_sources}}
\end{figure}

As described in Section \ref{sec:intro}, there are many contaminants in EMPG candidates selected with the classical color-color selection.
Figure \ref{fig:ref_sources} shows the SDSS EMPGs with {\EW}(\Ha)$>$800 {\AA} on the $r-i$ vs. $g-r$ diagram as well as the SDSS source catalog created in Section \ref{subsec:SDSScatalog}.
Figure \ref{fig:ref_sources} demonstrates that the position of the EMPGs are overlapped by many sources on the $r-i$ vs. $g-r$ diagram.
With the visual inspection, we find that most of the overlapping sources are contaminants such as stars and artifacts. 
Thus, we suggest that the classical color-color diagram is not effective for selecting EMPGs.

\begin{figure}
\epsscale{1.1}
\plotone{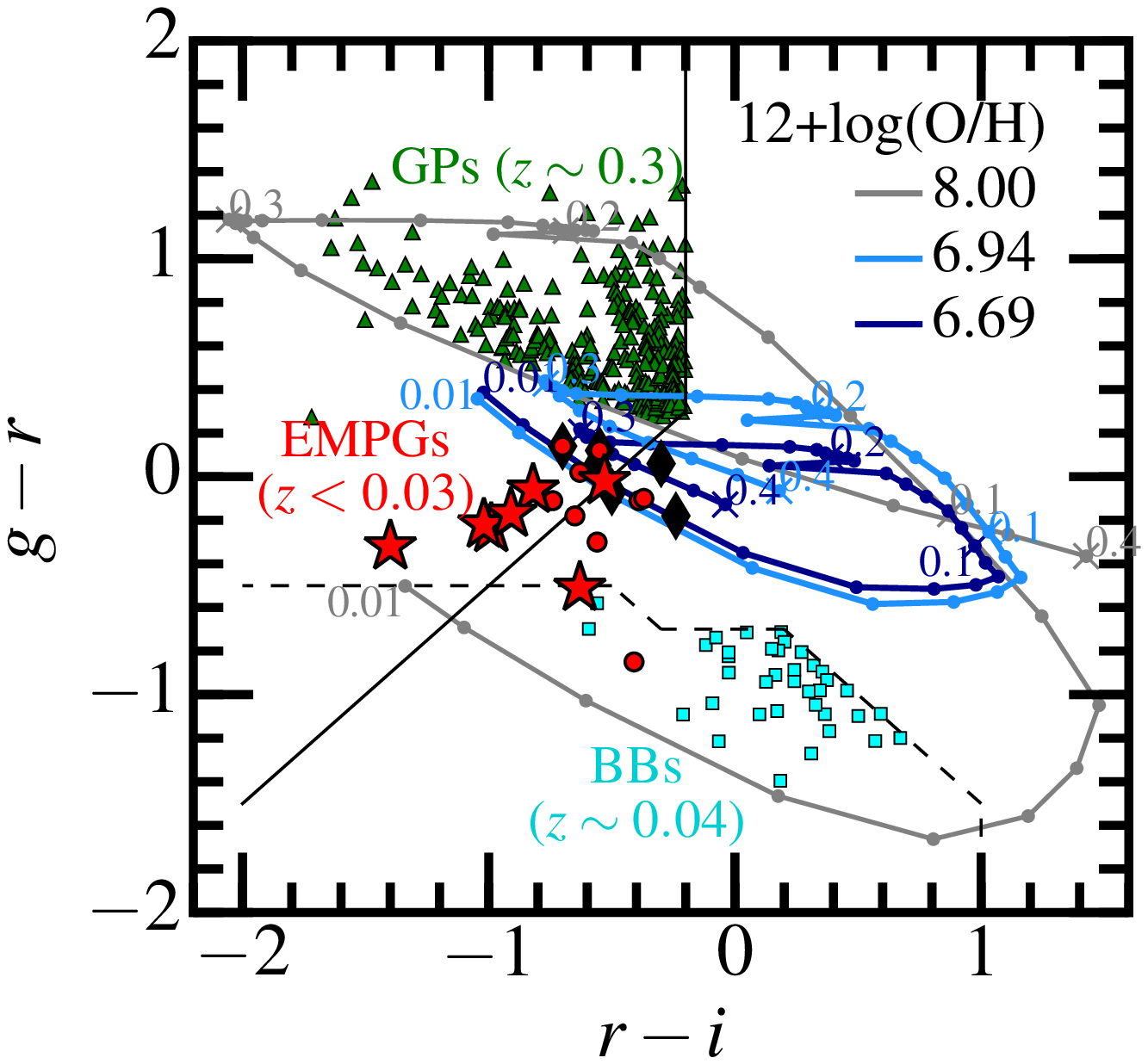}
\caption{Model calculation of $g-r$ and $r-i$ colors for galaxies with {\EW}({\Ha})$\sim$2,000 {\AA} that have {\metal}=6.69 (dark blue line; EMPGs), {\metal}=6.94 (light blue line; EMPGs), and {\metal}=8.00 (gray line; GPs).
Dots and crosses are placed in every step of $\Delta z$=0.01 and 0.1, respectively.
The green triangles and cyan squares are GPs at $z\sim0.3$ \citep{Cardamone2009} and BBs at $z\sim0.04$ \citep{Yang2017b}, respectively. 
The black solid and dotted lines are boundaries used to select GPs at $z\sim0.3$ \citep[][]{Cardamone2009} and BBs at $z\sim0.04$ \citep{Yang2017b}, respectively.
EMPGs derived from the literature are shown with the red stars, red circles, and black diamonds (see Figure \ref{fig:ref_color}).
\label{fig:model_color}}
\end{figure}

To contrast an effectiveness of a selection with the $g$- and $r$-band excesses, we compare known EMPGs and EMPG models with the GP/BB selections on the $r-i$ vs. $g-r$ diagram.
Figure \ref{fig:model_color} demonstrates that the EMPGs, GPs and BBs little overlap on the $r-i$ vs. $g-r$ diagram.
The EMPGs show colors of $r-i\sim-1$ and $g-r\sim0$, which suggest the $g$- and $r$-band excesses.
The solid and dotted lines are the selection criteria of the GPs \citep[][]{Cardamone2009} and BBs \citep{Yang2017b}, respectively.
We also show galaxy models of {\metal}$=$8.00 (GPs/BBs) as well as {\metal}$=$6.69 and 6.94 (EMPGs) in Figure \ref{fig:model_color} (see model details in Section \ref{subsec:model}).
The {\metal}$=$6.69 and 6.94 models are of lower metallicity than the lowest metallicity galaxy currently known, 12+log(O/H)=6.98 \citep[J0811$+$4730,][]{Izotov2018a}. 
However, we include these models with the expectation that such low metallicity systems can be found with the deeper HSC-SSP data.
The EMPG models with {\metal}$=$6.69 and 6.94 overlap little with the GP and BB selections, which basically means that the GP/BB selection criteria are not the best to select EMPGs with {\metal}$\lesssim$7.0. 
The basic idea of $g$- and $r$-band excesses seems to satisfy a necessary condition needed to select EMPGs with {\metal}$\lesssim$7.0.
The next step is to reduce as many contaminants as possible, as explained in Figure \ref{fig:ref_sources}.

\subsection{ML Classifier} \label{subsec:classifier}
We construct an object classifier with DNN that efficiently separates EMPGs from other types of objects  by reducing as many contaminants as possible.
The DNN, which is one of the most utilized artificial neural networks, is used to solve a classification problem.
The DNN is composed of multiple layers of neural networks and optimized with training data to best separate multiple types of objects.
We train the classifier with SED models because only a limited number of metal-poor galaxies have been identified fainter than 21 mag \citep{Izotov2018a, Izotov2019a}, which means that the existing metal-poor galaxy sample is too small to make an ML training sample.
In the following sections, we explain the merits of DNN (Section \ref{subsec:merits}), how the ML classifier is constructed (Section \ref{subsec:structure}) and how the training samples are generated from models (Section \ref{subsec:model}). 

\subsubsection{Merits} \label{subsec:merits}
\begin{figure}
\epsscale{1.2}
\plotone{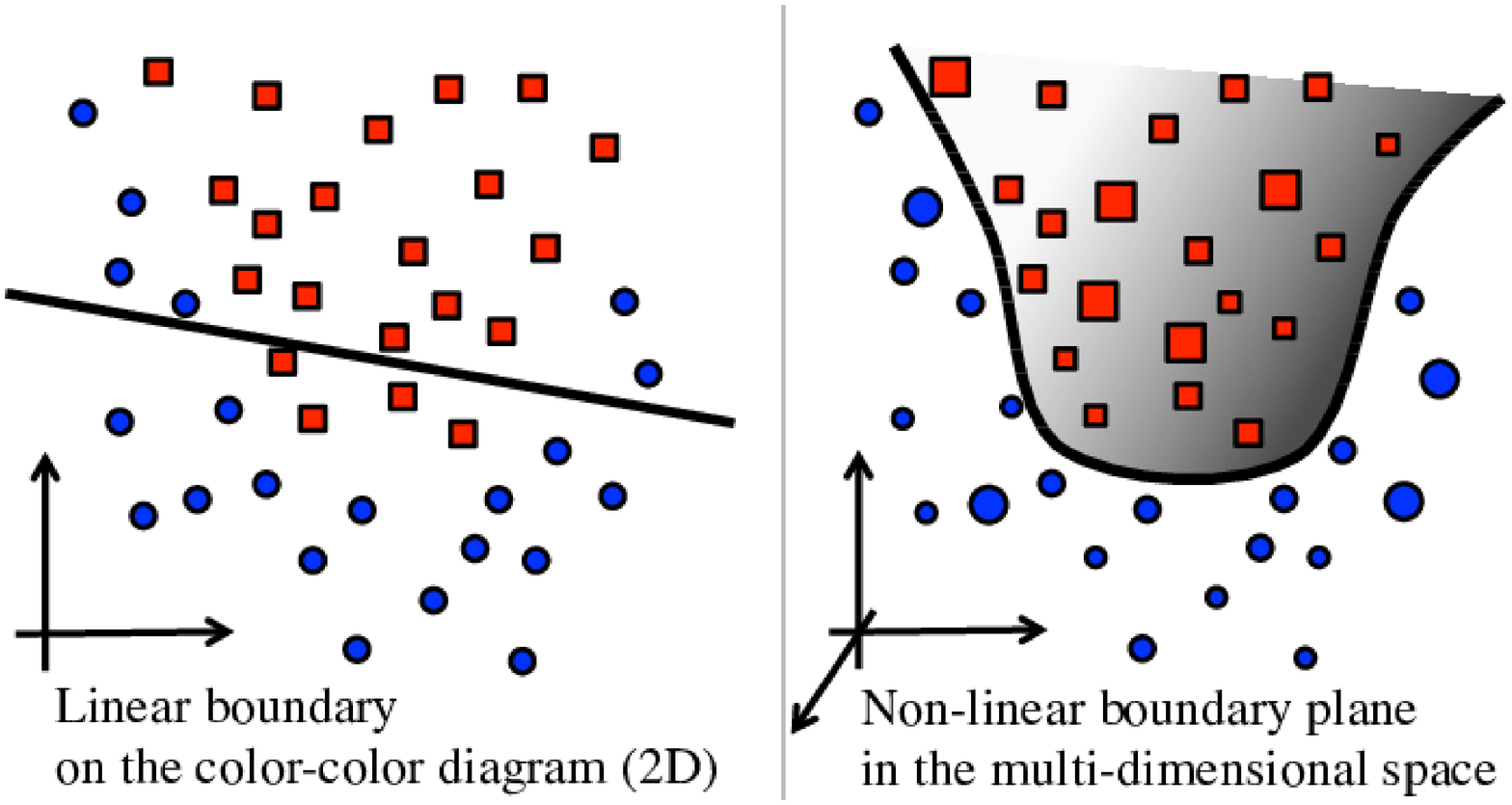}
\caption{Schematic illustrations of two methods of the object classification. 
Left---The object classification on the color-color diagram with a linear boundary.
Two different types of objects are presented with the squares and circles.
Right---The object classification in the multi-dimensional space (e.g., 5 dimensional space of $grizy$-band magnitudes) with a non-linear boundary.
The DNN classification corresponds to the right illustration, while the left panel demonstrates a classical color-color selection.
The size of circles and squares represents a distance on the line of sight.
\label{fig:merit}}
\end{figure}

There are four merits of the use of the DNN as shown below.

i) We can select objects in the multi-dimensional photometry space (e.g., in a $grizy$ 5-dimensional space), while classical selections use a combination of simple linear projections onto a 2-dimensional color-color diagram.
As discussed in Section \ref{subsec:targets}, EMPGs are overlapped by many sources on a color-color diagram of $g-r$ vs. $r-i$ (Figure \ref{fig:ref_sources}) for instance.
In principle, if we use criteria in a multi-dimensional space, we can eliminate such overlapping sources more efficiently. 

ii) We can use a non-linear boundary that separates object types.
The DNN can determine a non-linear boundary thanks to a non-linear function, called an activation function, which is used in the DNN structure (see Section \ref{subsec:structure}). 
Although classical selections try to separate object types with a straight line on a color-color diagram, such a simple, straight line does not always separate different types of objects well.
The use of a non-linear boundary usually reduces the contamination and increases the completeness.

iii) A boundary is optimized by the DNN algorithm, where as the classical boundaries are not optimized mathematically.
The DNN enables the objective determination of the boundaries.
Figure \ref{fig:merit} is a schematic illustration of merits i) to iii).

iv) The DNN selection is very fast. 
Indeed, in principle, we are able to select EMPG candidates by fitting with SED models of galaxies, stars, and QSOs in a wide range of parameters.
However, such SED fitting takes much longer time than the DNN.
Our DNN classifier requires only several minutes to train itself and classify sources once we produce SED models of galaxies, stars, and QSOs.

Although there are many other ML algorithms such as a support vector machine (SVM), we begin with the DNN due to the four merits described above.
We focus on the use of DNN in this paper because our purpose is to construct an EMPG sample, not to find the most efficient selection method.
In Section \ref{subsec:identification}, we spectroscopically confirm that the success rate of our ML classifier is over $\gtrsim50\%$, which is high enough to construct an EMPG sample.
Thus, the comparison between the DNN and other ML techniques is out of the scope of this paper.

\subsubsection{Structure} \label{subsec:structure}
We construct an object classifier that distinguishes four object types of EMPGs, non-EMPG galaxies, stars, and QSOs.
For every source input, the classifier calculates probabilities of the four types and chooses only one type whose probability is the highest in the four.
In our calculation, we use an open-source software library for ML, {\tt Tensorflow}\footnote{https://www.tensorflow.org}.
Its detailed structure and training process are explained below. 

\begin{figure*}
\epsscale{1.2}
\plotone{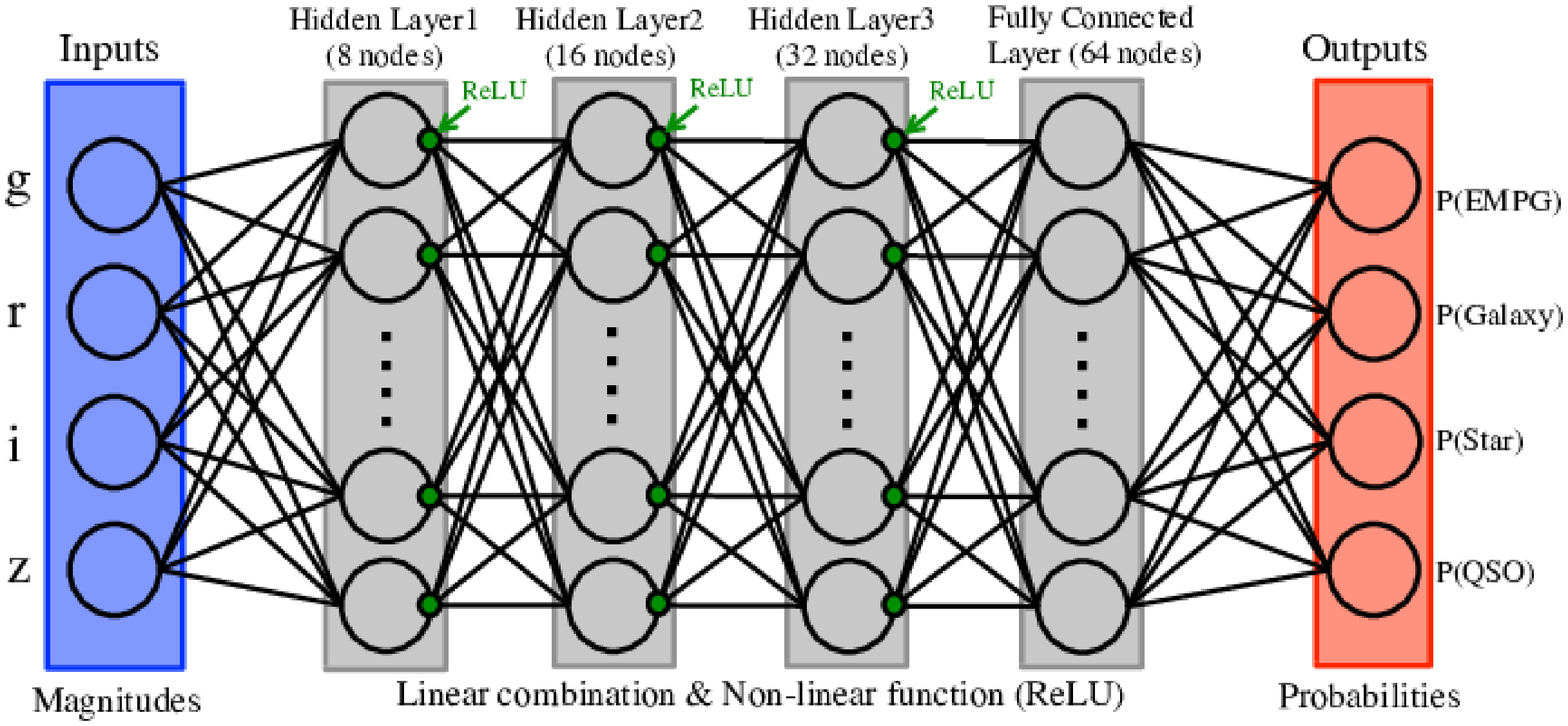}
\caption{Schematic illustration of the structure of our ML classifier based on the DNN.
The nodes (blank circles) and branches (solid lines) represent a linear combination.
The green circles are ReLU activation functions.
\label{fig:structure}}
\end{figure*}

The object classifier is constructed with the DNN that consists of three hidden layers and one fully connected layer.
Figure \ref{fig:structure} is an schematic illustration of the structure of our classifier.
The three hidden layers and one fully connected layer have 8/16/32 and 64 nodes, respectively.
As Figure \ref{fig:structure} shows, these nodes are connected with branches, which represent a linear-combination calculation. 
Each node in the hidden layers is followed by an activation function called rectified linear unit \citep[ReLU;][]{Nair2010}.
The activation function, ReLU, is a non-linear function, which is essential to construct a deep-layer structure.
In the fully connected layer, 10\% of the nodes are dropped at random to avoid over-fitting.

Inputs of our classifier are four (five) photometric magnitudes of HSC $griz$ bands (SDSS $ugriz$ bands).
We do not use the HSC $y$-band photometry, which is shallower than the other bands, to reach as faint magnitudes as possible.
After calculations, the classifier outputs four probabilities of the EMPG, non-EMPG galaxy, star, and QSO and chooses only one type whose probability is the highest in the four.
Here we obtain probabilities with the {\tt softmax} function.
The {\tt softmax} function is a mathematical function that normalizes a vector with an exponential function.

The structure of the neural network is optimized so that the sum of output errors is minimized. 
We optimize our classifier with a training sample, in which object types are already known beforehand.
The optimization process is usually called ``training''.
We use the cross entropy and Adam optimizer \citep{Kingma2014}, which are built in the {\tt Tensorflow} software, to calculate and minimize the errors in the training, respectively.
To train our classifier, we prepare a training sample with the SED models, which will be detailed in Section \ref{subsec:model}.
Then the training sample is divided into two independent data sets.
Here, 80\% of the training sample is used as training data and the other 20\% as check data.
We use the training data to train the neural network, while the check data are prepared to check whether the classifier successfully identifies and separates EMPGs from other objects types.
In every training step, 100 models are randomly chosen from the training sample and used to train the neural network.
The one training step is defined by a training with the 100 models.
This training step is repeated 10,000 times.

\begin{figure}
\epsscale{1.0}
\plotone{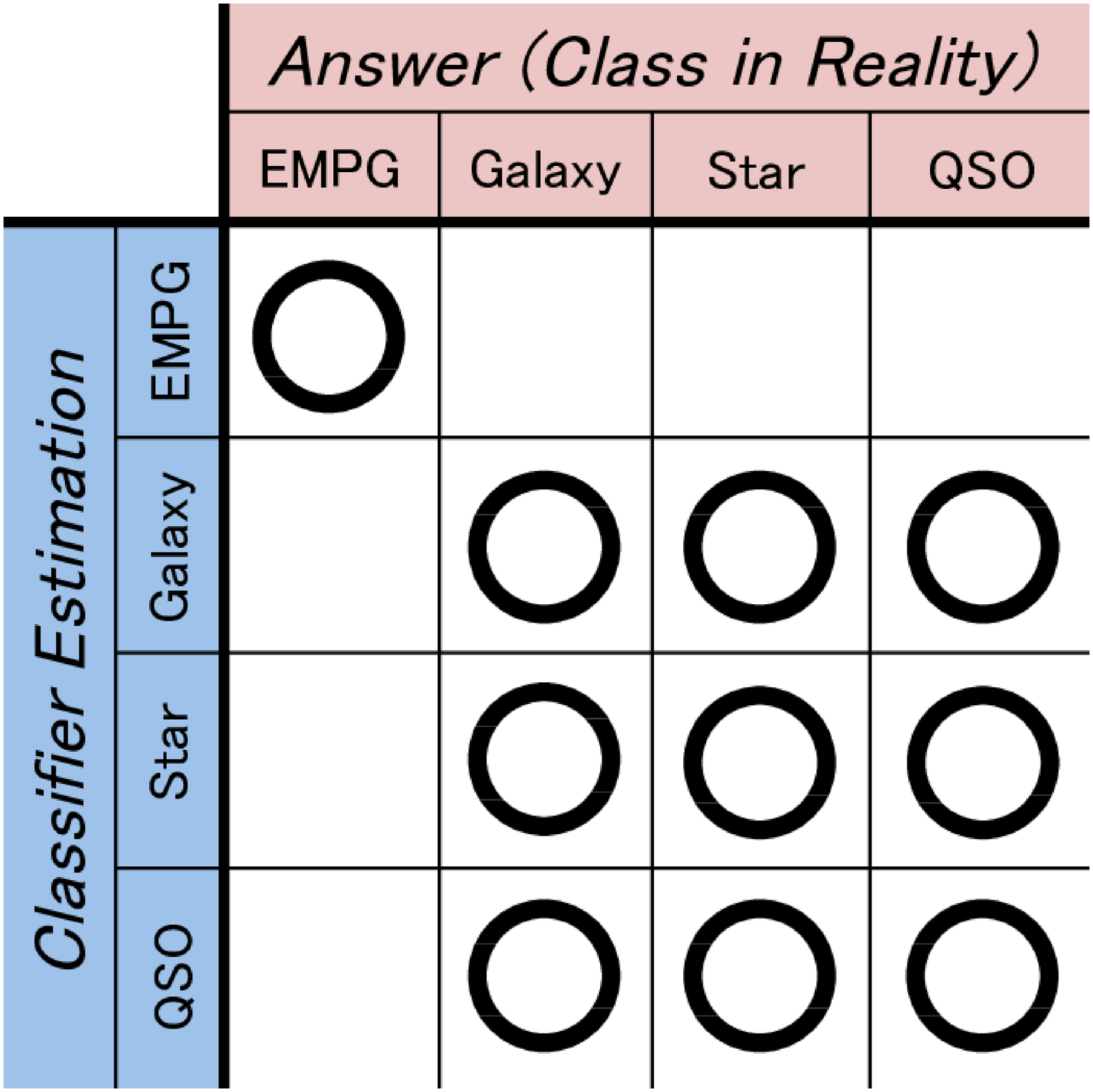}
\caption{Matrix that explains the success cases defined in this paper.
The columns and rows correspond to answers in reality and estimations made by the classifier, respectively.
We mark success cases with circles.
Note that we ignore mistakes in the classification between galaxies, stars, and QSOs because we only aim to select EMPGs.
\label{fig:success}}
\end{figure}

Success cases are defined by the true-positive EMPGs (i.e., a real EMPG classified as an EMPG) and its true-negative (i.e., a real galaxy/star/QSO classified as a galaxy/star/QSO) as summarized in Figure \ref{fig:success}.
Here we only focus on whether an object is an EMPG or not.
In other words, we ignore mistakes in the classification between galaxies, stars, and QSOs.
Our EMPG classification is not affected by this ignorance.
We define a success rate by the number of success cases over the number of total classification. 
In the calculation, we repeat the training step 10,000 times, until which the success rate exceeds 99.5\% constantly.

\subsubsection{Training Sample} \label{subsec:model}
We prepare the training sample, which is used to train ML classifier explained in Section \ref{subsec:structure}.
The training sample consists of photometric magnitudes calculated from models of EMPGs, non-EMPG galaxies, stars, and QSOs.
The photometric magnitudes are calculated from the SED models by convoluting the SEDs with the throughput curves of the HSC broadband filters \citep[][]{Kawanomoto2018} or the SDSS broadband filters \citep[][]{Fukugita1996}.
Below, we detail the models of the EMPG, non-EMPG galaxy, star, and QSO.

\begin{figure}
\epsscale{1.15}
\plotone{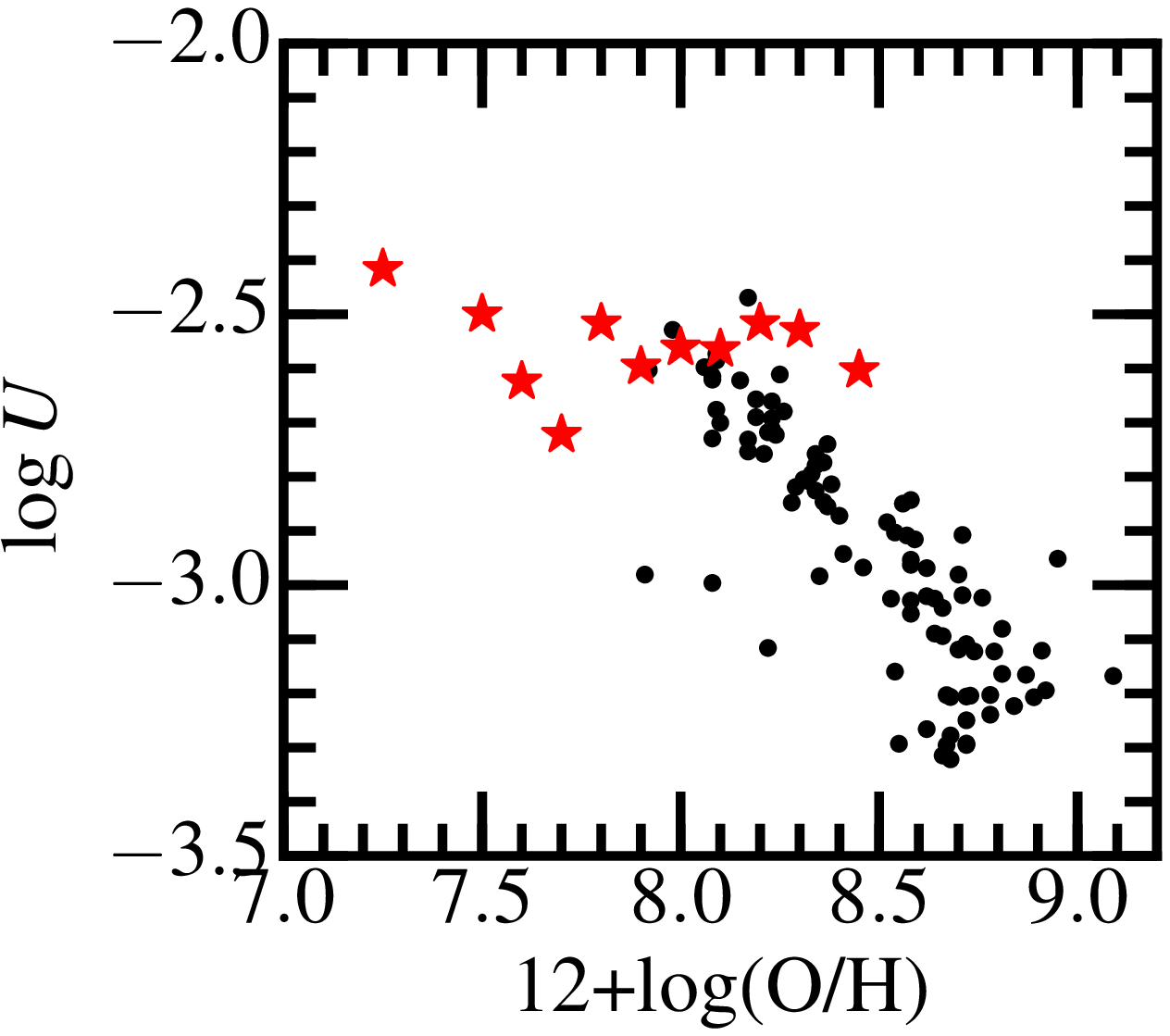}
\caption{Ionization parameters of typical local galaxies as a function of metallicity.
Red stars represent averages of the local metal-poor galaxies \citet[][sample A+B]{Nagao2006}.
Black dots are obtained from the SDSS composite spectra of \citet{Andrews2013}.
Metallicities are based on the electron temperature measurements.
Ionization parameters are calculated assuming the photoionization model of \citet{Kewley2002}.
This figure suggests that galaxies with {\metal}$=$7.0--7.5 have $\log\,U\sim-2.5$.
\label{fig:qZ}}
\end{figure}

\begin{figure}
\epsscale{1.15}
\plotone{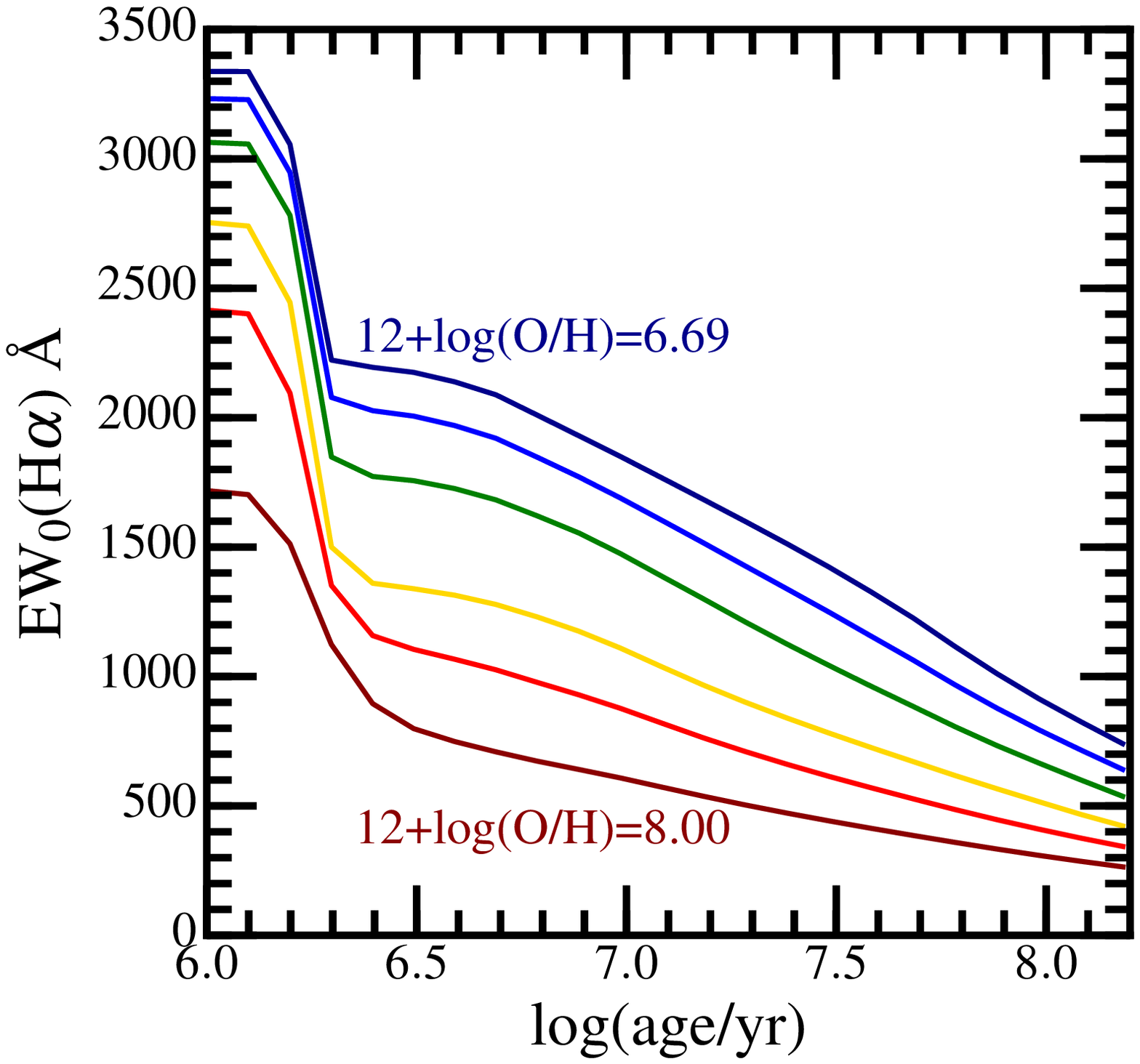}
\caption{{\EW}({\Ha}) values as a function of age. 
The colors of the solid lines correspond to metallicities, {\metal}=6.69, 6.94, 7.19, 7.50, 7.69, and 8.00 from dark blue to dark red.
These relations are provided by the \textsc{beagle} models under the assumption of the constant star-formation.
\label{fig:age}}
\end{figure}

\underline{\bf 1) EMPG model:}
We generate EMPG SEDs with the SED interpretation code, \textsc{beagle} \citep{Chevallard2016}.
The \textsc{beagle} code calculates both the stellar continuum and the nebular emission (line + continuum) in a self-consistent manner, using the stellar population synthesis code \citep[][]{Bruzual2003} and the photoionization code, \textsc{cloudy} \citep[][]{Ferland2013}.
The \textsc{beagle} codes use the \textsc{cloudy} photoionization models produced by \citet{Gutkin2016}, where the photoionization calculations are stopped when the electron density falls below 1\% of the hydrogen density or if the temperature falls below 100 K.
In the \textsc{cloudy} photoionization models, we assume the solar carbon-to-oxygen abundance ratio (C/O) and the metallicity-dependent nitrogen-to-oxygen abundance ratio (N/O) given in \citet{Gutkin2016}.
The assumption of the C/O and N/O ratios does not affect our model photometry because carbon and nitrogen lines are very faint.
In the \textsc{beagle} calculation, we change five parameters of stellar mass, maximum stellar age (called just ``age'' hereafter), gas-phase metallicity, ionization parameter, $\log\,U$, and redshift as shown below.
\begin{itemize}
\item {\mass}=(4.0, 4.5, 5.0, 5.5, 6.0, 6.5, 7.0, 7.5, 8.0, 8.5, 9.0)
\item log(age/yr)=(6.00, 6.25, 6.50, 6.60, 6.70, 6.80, 6.90, 7.00, 7.25, 7.50, 7.75)
\item {\metal}=(6.69, 7.19, 7.69)
\item $\log\,U$=($-2.7$, $-2.5$, $-2.3$)
\item redshift=(0.01, 0.02)
\end{itemize}
These stellar mass, age, and gas-phase metallicity cover typical values of known EMPGs \citep[e.g.,][{\metal}$\sim$7.0--7.2]{Thuan2005a, Izotov2009, Skillman2013, Hirschauer2016, Izotov2018a} and/or theoretical predictions \citep[e.g.,][]{Wise2012}.
A stellar metallicity is matched to a gas-phase metallicity here.
The ionization parameter is defined by a ratio of hydrogen ionizing-photon flux, $S_{H^0}$ and hydrogen gas density, $n_{H}$, normalized by speed of light, $c$,
\begin{eqnarray}
  \log\,U\equiv \frac{S_{H^0}}{c\,n_{H}}. \label{eq:logU}
\end{eqnarray}
We choose ionization parameters of $\log\,U$=($-2.7$, $-2.5$, $-2.3$), which are typical values for metal-poor galaxies as demonstrated in Figure \ref{fig:qZ}.
The constant star-formation history is assumed in the model.
Here, we also assume no dust attenuation because we target very metal-poor galaxies, where the dust production is insufficient.
Indeed, representative metal-poor galaxies \citep[e.g.,][{\metal}$\sim$7.0--7.2]{Thuan2005a, Izotov2009, Skillman2013, Hirschauer2016, Izotov2018a} show a negligibly small dust attenuation with a color excess of {\EBV}$\sim$0.
The \citet{Chabrier2003} stellar initial mass function (IMF) is applied in the \textsc{beagle} code \citep{Chevallard2016}.
In total, we generate 2,178 (=11$\times$11$\times$3$\times$3$\times$2) SEDs with the parameters described above.
For each SED, the \textsc{beagle} code also calculates the photometric magnitudes with response curves of the HSC and SDSS filters, as well as emission line equivalent widths.
From the 2,178 model SEDs, we only select 1,397 models that satisfy $i<26$ mag and {\EW}({\Ha})$>$1,000 {\AA}.
The $26$ mag corresponds to about an $i$-band limiting magnitude of the HSC imaging data.
The {\EW}({\Ha})$>$1,000 {\AA} corresponds to an age $\lesssim$ 10--100 Myr in this metallicity range, {\metal}$<$7.69 as shown in Figure \ref{fig:age}.
Although we aim to select EMPGs with {\EW}({\Ha})$>$800 {\AA} in this paper, we here limit EMPG models with {\EW}({\Ha})$>$1,000 {\AA}.
This is because, if we limit EMPG models with {\EW}({\Ha})$>$800 {\AA}, we tend to obtain galaxies with {\EW}({\Ha})$<$800 {\AA} due to the 0.1 mag photometry error (explained below), as well as the increasing number of contamination.
Then, to take the magnitude errors of 0.1 mag (see Sections \ref{subsec:HSCcatalog} and \ref{subsec:SDSScatalog}) into consideration in models, we generate random numbers under the assumption of the normal distribution with $\sigma=0.1$ and add them to the photometric magnitudes.
Here we generate 30 sets of random numbers for each model\footnote{We remove random numbers beyond $\sigma=0.1$ from models as we eliminate sources with $\sigma>0.1$ in the source catalogs (Sections \ref{subsec:HSCcatalog} and \ref{subsec:SDSScatalog}). We continue to generate random numbers until the total number becomes 30.}. 
Thus, we obtain a total of 41,910 (=1,397$\times$30) models including magnitude errors.
We do not use models that satisfy $0.02<z\leq0.03$ because we find that a contamination rate increases in that case.

\underline{\bf 2) Galaxy model (non-EMPG):}
We introduce two types of non-EMPG galaxies: normal SFGs and GPs.
First, we generate SEDs of normal SFGs with the \textsc{beagle} code similarly to the EMPG models.
In the calculation, we change five parameters of stellar mass, age, metallicity,  redshift, and $V$-band dust-attenuation optical depth ($\tau_{\rm V}$) as shown below, assuming a bursty star-formation history.
\begin{itemize}
\item {\mass}=8.0, 9.0, 10.0, 11.0
\item log(age/yr)=8.5, 9.0, 9.5, 10.0
\item {\metal}=8.19, 8.69, 8.89
\item redshift=0.02, 0.04, 0.06, 0.08, 0.10
\item log($\tau_{\rm V}$)=0.0, 2.0
\end{itemize}
These stellar mass, age, metallicity, and $V$-band dust-attenuation optical depth are selected from typical values of local SFGs.
We fix an ionization parameter to $\log\,U$=$-3.0$, which is a value representative of local galaxies as demonstrated in Figure \ref{fig:qZ}.
In total, we generate 480 SEDs with the parameters described above.
The photometric magnitudes are calculated in the same manner as the EMPG models.
From the 480 models, we only select models that satisfy $i<26$ mag.
After the $i$-band magnitude selection, 471 models remain.
We introduce magnitude errors of 0.1 mag similarly to the EMPG models, generating 100 sets of random numbers for the 471 models.
Then we have 47,100 normal-SFG models in total including magnitude errors.

Second, we also create GP SEDs with the \textsc{beagle} code with following 4 parameters.
\begin{itemize}
\item {\mass}=7.0, 7.5, 8.0, 8.5, 9.0, 9.5, 10.0
\item log(age/yr)=6.0, 6.1, 6.2,..., 8.0
\item $\log\,U$=($-3.0$, $-2.5$, $-2.0$)
\item redshift=0.08, 0.09, 0.10,..., 0.40
\end{itemize}
Metallicity is fixed at {\metal}=8.00, which is typical value of GPs (see Figure \ref{fig:O3metal}).
We also assume dust free in the GP models.
We obtain 14,553(=7$\times$21$\times$3$\times$21) models.
From the 14,553 models, we use 3,234 models that satisfy $i<26$ mag.
We also introduce 0.1-mag errors in magnitude  as described above, generating 10 sets of random numbers for 3,234 models.
Then we have 32,340 GP models in total.

We combine the 47,100 normal-SFG models and 32,340 GP models into 79,440 models of non-EMPG galaxies.
We do not include galaxies over $z=0.1$ in the non-EMPG galaxy models because galaxies at $z>0.1$ have quite different colors from those of low-$z$ EMPGs.
The inclusion of high-$z$ galaxy models does not change our result.

\underline{\bf 3) Stellar model:}
We use stellar SED models of \citet{Castelli2004}, where 53 types of stars are modeled from O-type to M-type.
For each stellar type, SEDs are calculated in a metallicity range of $\log(Z/Z_{\odot})=(-2.5, -2.0, -1.5, -1.0, -0.5, \pm0.0, +0.2, +0.5)$.
Thus, we obtain 424 (=53$\times$8) model SEDs in total.
These stellar model SEDs are obtained from a STScI web site\footnote{\texttt{ftp://ftp.stsci.edu/cdbs/grid/ck04models}}.
Assuming the HSC and SDSS filters, we calculate $u-i$, $g-i$, $r-i$, and $z-i$ colors from the 424 model SEDs.
Then, we determine $i$-band magnitudes, selecting 10 values in the range of  $i=15$--$26$ mag at regular intervals.
Multiplying the 424 sets of $u-i$, $g-i$, $r-i$, and $z-i$ colors and the 10 $i$-band magnitudes, we generate 4,240 (=424$\times$10) sets of stellar models with photometric magnitudes.
In addition, we also introduce magnitude errors (0.1 mag) similarly to the EMPG models, obtaining 42,400 (=4,240$\times$10) stellar models in total.

\underline{\bf 4) QSO model:}
We use a composite spectrum of QSOs at $1<z<2.1$ observed by the X-SHOOTER spectrograph installed on Very Large Telescope (VLT) \citep{Selsing2016}.
This composite spectrum covers the wide wavelength range of 1,000--11,000 {\AA} in the rest-frame.
From this composite spectrum, we generate mock spectra by varying three parameters as follows: the power law index ($\alpha$) of an intrinsic near ultraviolet (NUV) slope $f_{\lambda}\propto \lambda^{\alpha}$, the $V$-band dust-attenuation optical depth, and the redshift.
\begin{itemize}
\item $\alpha$=$-2.0$, $-1.5$, $-1.0$, $-0.5$
\item $\tau_{\rm V}$=0.0, 0.5, 1.0
\item redshift=0.1, 0.2, 0.3, ..., 3.0.
\end{itemize}

The intrinsic NUV slope and $V$-band dust-attenuation optical depth of typical QSOs are well covered by the parameters above \citep[e.g.,][]{Telfer2002, Selsing2016}.
Then we get 360(=4$\times$3$\times$30) QSO model SEDs in total.
Similarly to the stellar models, we calculate $u-i$,$g-i$, $r-i$, and $z-i$ colors from the 360 model SEDs.
Here, we take 10 values of $i$-band magnitude in the range of $i=15$--$26$ at regular intervals, obtaining 3,600 models.
In addition, we also introduce magnitude errors (0.1 mag) similarly to the EMPG models, obtaining 36,000 (=3,600$\times$10) stellar models in total.\\

\begin{figure}
\epsscale{1.1}
\plotone{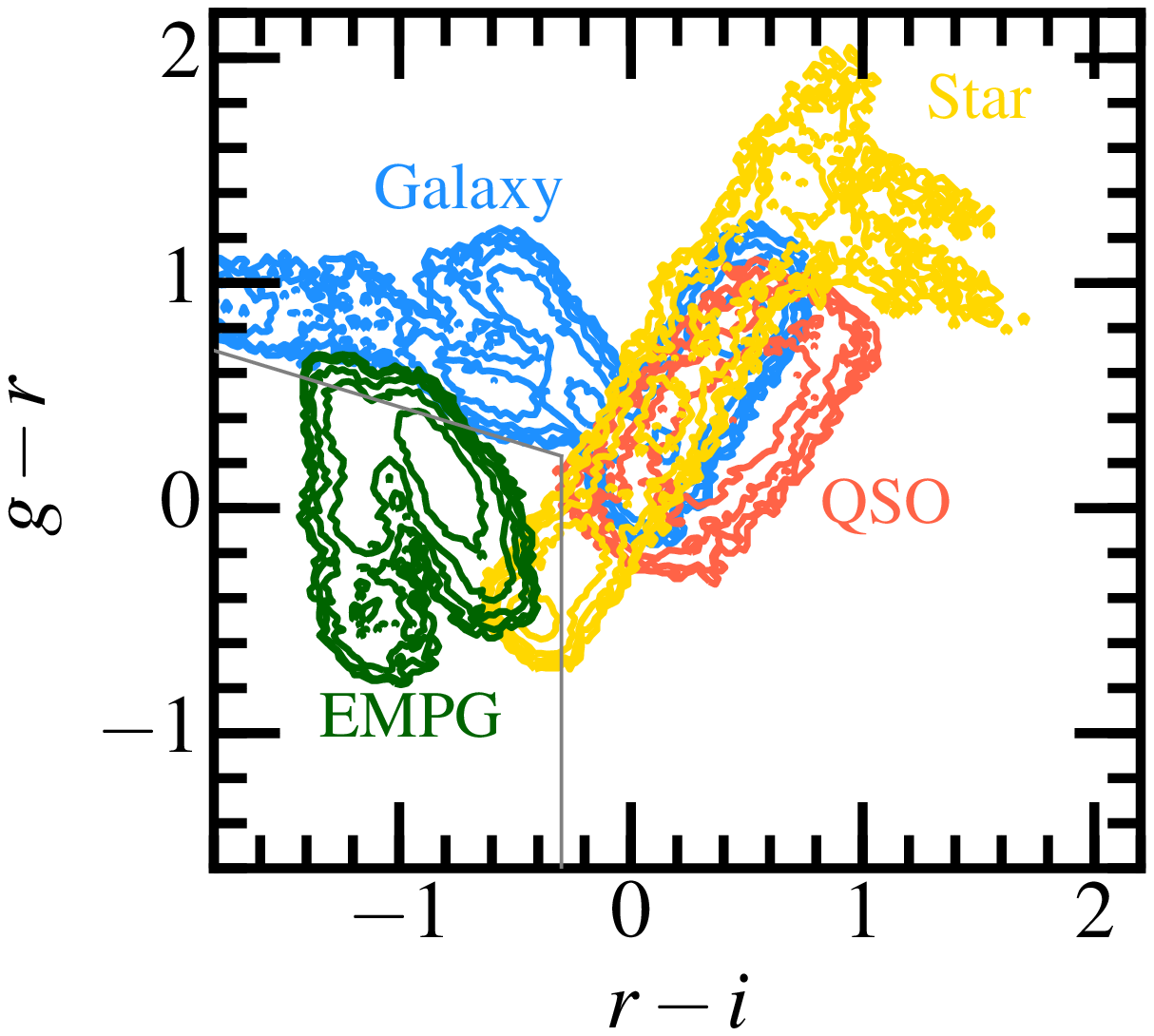}
\plotone{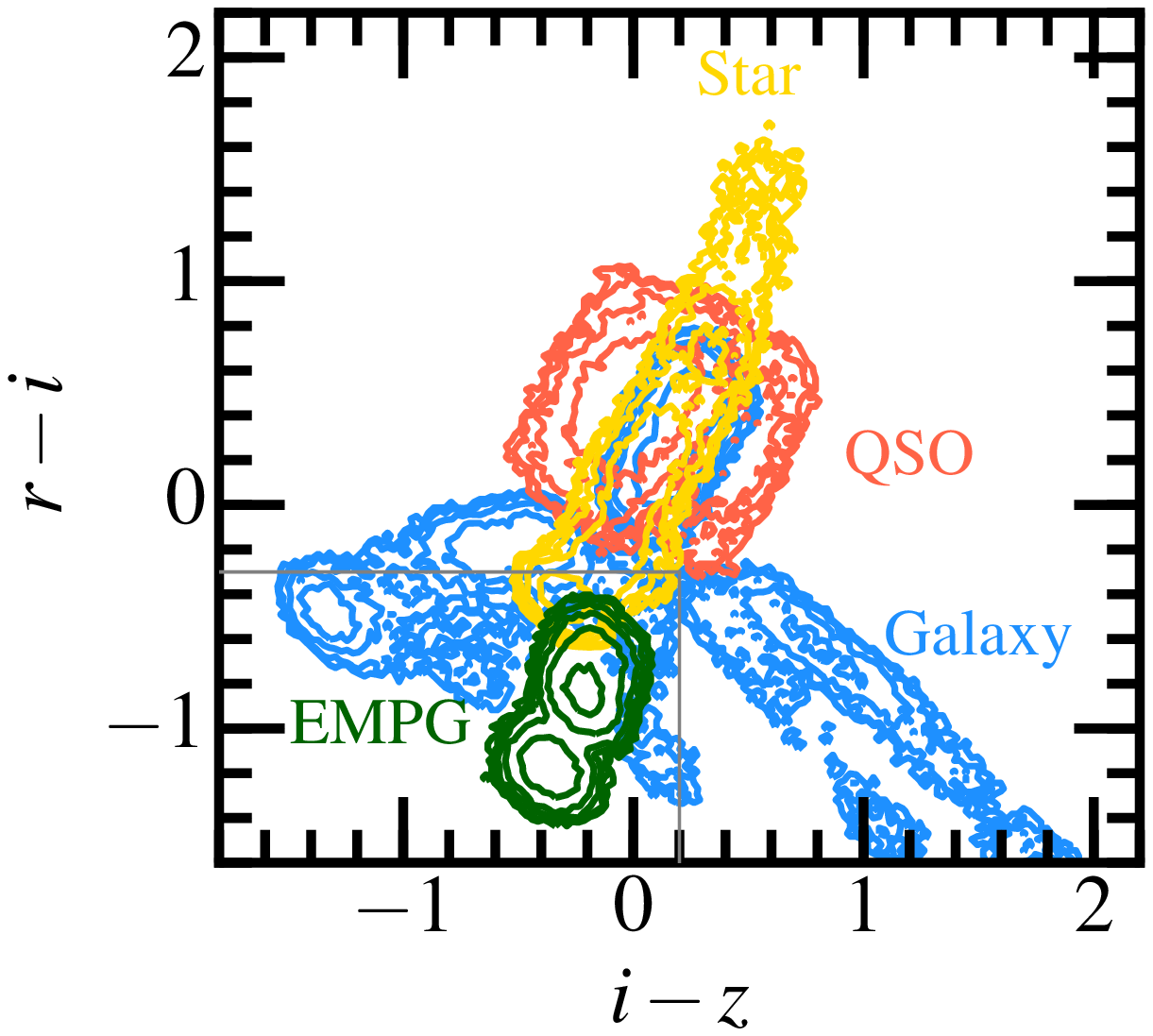}
\caption{Top: Model templates of EMPGs (green), non-EMPG galaxies (blue), stars (yellow), and QSOs (red) on the color-color diagrams of $g-r$ vs. $r-i$.
The contours show the number of these models ($N$=1, 3, 10 ,30, 100, 300) in each bin with a size of $\Delta m$=0.025 mag. 
Bottom: Same as the top panel, but for $r-i$ vs. $i-z$.
The gray lines present the Equations (\ref{eq:color_a})--(\ref{eq:color_c}) described in Section \ref{subsec:HSC_EMPG}.
\label{fig:mock}}
\end{figure}

In this section, we have generated 41,910, 79,440, 42,400, and 36,000 models for the EMPG, non-EMPG galaxy, star, and QSO, respectively. 
Selecting 30,000 models from each of EMPG, non-EMPG galaxy, star, and QSO, we obtain a training sample composed of 120,000 (=30,000$\times$4) models in total.
Figure \ref{fig:mock} shows models of EMPGs, non-EMPG galaxies, stars, and QSOs on the projected color-color diagrams of $g-r$ vs. $r-i$ and $r-i$ vs. $i-z$.
Here, EMPGs are overlapped with non-EMPG galaxies and stars on these projected color-color diagrams, which potentially causes contamination in the EMPG selection.

\subsection{Test with SDSS Data} \label{subsec:test}
Before we apply our ML classifier to the HSC-SSP and SDSS source catalogs, we test if our classifier successfully distinguishes EMPGs from other types of objects (non-EMPG galaxy, star, or QSO).
We carry out the test with SDSS data whose sources are detected in photometry and observed in spectroscopy.
Such data set is a good test sample because we can easily check object types (galaxy, star, or QSO) and metallicities in their spectra.
We can also see if a source satisfies the EMPG condition of 12+log(O/H)$<$7.69.
We do not expect to discover unconfirmed EMPGs in the SDSS test catalog because SDSS sources with the spectroscopic confirmation have been intensively investigated by many authors \citep[e.g.,][]{SanchezAlmeida2016, Guseva2017}.
This step is important towards discovering unconfirmed EMPGs in the HSC data, whose limiting magnitude is $\gtrsim$5 mag deeper than SDSS, potentially pushing to lower metallicities than what has been discovered so far through SDSS.
Here we explain how we create a SDSS test catalog in Section \ref{subsec:SDSStestcatalog}, and the test results  are described in Section \ref{subsec:tests}.

\subsubsection{SDSS Test Catalog} \label{subsec:SDSStestcatalog}
We construct a {\it SDSS test catalog} from the SDSS DR13 data. 
The SDSS DR13 data is based on the SDSS-I through SDSS-IV data, which contain the extragalactic spectroscopy data from the Sloan Digital Sky Survey \citep[SDSS-I,][]{York2000}, the Sloan Digital Sky Survey II \citep[SDSS-II,][]{Abazajian2009}, the Baryon Oscillation Spectroscopic Survey \citep[BOSS,][]{Dawson2013}, and extended Baryon Oscillation Spectroscopic Survey \citep[eBOSS,][]{Dawson2016}.
In the SDSS data, typical wavelength ranges are 3,800--9,200 {\AA} or 3,650-10,400 {\AA} and a typical spectral resolution is $R$=1,500--2,500.
We only select objects cross-matched with the photometry catalog of {\tt PhotoObjAll} and the spectroscopy catalogs of {\tt SpecObjAll}, {\tt galSpecExtra}, and {\tt galSpecLine}.
Then we construct the SDSS {\it test} catalog in the same way as the SDSS {\it source} catalog described in  Section \ref{subsec:SDSScatalog}.
The SDSS test catalog is composed of 935,042 sources (579,961 galaxies, 327,421 stars, and 27,660 QSOs) in total.
The spectroscopic effective area of the SDSS data is 9,376 deg$^2$.


\subsubsection{Tests} \label{subsec:tests}
Applying our ML classifier  (Section \ref{subsec:classifier}) to the SDSS test catalog (Section \ref{subsec:SDSStestcatalog}), we obtained thirteen EMPG candidates from the SDSS test catalog.
We have checked their object classes (galaxy, star, and QSO) that are given in the {\tt SpecObjAll} catalog based on spectroscopy.
Based on the images, spectra and the object classes, we identify all of the thirteen candidates as galaxies.
We find that these thirteen galaxies are well-studied with spectroscopic data in the literature.

For the thirteen galaxies, we obtain redshift, {\EW}(\Ha), and {\EW}(\Hb) values from {\tt SpecObjAll} and {\tt galSpecLine} catalogs.
Metallicities of the thirteen galaxies are derived from the literature \citep{Kunth2000, Kniazev2003, Guseva2007, Izotov2007, Izotov2009, Pustilnik2010, Izotov2012, Pilyugin2012, SanchezAlmeida2016, Guseva2017}.
These metallicities are calculated based on electron temperature measurements.
We find that six out of the thirteen galaxies satisfy the EMPG condition, {\metal}$<$7.69.
Although the other seven galaxies do not fulfill the EMPG definition, they still have low metallicities in the range of {\metal}$\sim$7.8--8.5.
As we have expected, all of the thirteen galaxies show large {\EW}(\Ha) values (750--1,700 {\AA}).
We summarize their object classes, redshifts, {\EW}(\Ha), {\EW}(\Hb), and metallicities in Table \ref{table:propsdss}.

The success rate, or a purity, of our EMPG selection is $46$\% (6/13) for the SDSS test catalog.
It is worth noting that the other $54$\% galaxies (7/13) also show a low metallicity as described above.
In the local metal-poor galaxy sample obtained from the literature in Section \ref{subsec:targets}, we find 7 EMPGs, which are also included in the SDSS test catalog and have a high equivalent width, {\EW}({\Ha})$>$800{\AA}.
Very high EWs are necessary to be selected by the $gr$-band excess technique.
In other words, we have successfully selected the 6 EMPGs as above out of the 7 known high-EW EMPGs in the SDSS test catalog, which suggests that our selection reaches $86$\% (6/7) completeness.
Thus, we conclude that our selection method has successfully selected EMPGs and EMPG-like galaxies from the SDSS test catalog.

\begin{deluxetable*}{lccccccc}
\tablecolumns{8}
\tabletypesize{\scriptsize}
\tablecaption{Parameters of EMPG candidates selected in the SDSS test %
\label{table:propsdss}}
\tablehead{%
\colhead{\#} &
\colhead{ID} &
\colhead{class} &
\colhead{redshift} &
\colhead{{\EW}(H$\alpha$)} &
\colhead{{\EW}(H$\beta$)} &
\colhead{12+log(O/H)} &
\colhead{citation} \\
\colhead{} &
\colhead{} &
\colhead{} &
\colhead{} &
\colhead{{\AA}} &
\colhead{{\AA}} &
\colhead{} &
\colhead{} \\
\colhead{(1)} &
\colhead{(2)} &
\colhead{(3)} &
\colhead{(4)} &
\colhead{(5)} &
\colhead{(6)} &
\colhead{(7)} &
\colhead{(8)} 
}
\startdata
1  & J012534.2$+$075924.5 & EMPG & 0.010 & 1351 & 242 & \textbf{7.58} & P12 \\
2  & J080758.0$+$341439.3 & EMPG & 0.022 & 1277 & 252 & \textbf{7.69} & SA16 \\
3  & J082555.5$+$353231.9 & EMPG & 0.003 & 1441 & 238 & \textbf{7.45} & K03 \\
4  & J104457.8$+$035313.1 & EMPG & 0.013& 1462 & 276 & \textbf{7.44}  & K03 \\
5  & J141851.1$+$210239.7 & EMPG & 0.009 & 1153 & 215 & \textbf{7.50} & SA16 \\
6  & J223831.1$+$140029.8 & EMPG & 0.021 & 953 & 183 & \textbf{7.43} & P12 \vspace{2pt}  \\  \hline 
7  & J001428.8$-$004443.9 & Galaxy & 0.014 & 995 & 184 & 8.05 & P12 \\
8  & J025346.7$-$072344.0 & Galaxy & 0.005 & 787 & 138 & 7.97 & K04 \\
9  & J115804.9$+$275227.2 & Galaxy & 0.011 & 750 & 119 & 8.34 & B08 \\
10 & J125306.0$-$031258.8 & Galaxy & 0.023 & 1291\footnotemark[1] & 236 & 8.08 & K04 \\
11 & J131447.4$+$345259.7 & Galaxy & 0.003& 1700 & 293 & 8.14  & B08 \\
12 & J132347.5$-$013252.0 & Galaxy & 0.022& 1458 & 248 & 7.77  & K04 \\
13 & J143905.5$+$364821.9 & Galaxy & 0.002& 766\footnotemark[1] & 140 & 7.94 & P12 
\enddata
\tablecomments{
(1): Number.
(2): ID.
(3): Object class. Galaxies with {\metal}$<$7.69 are classified as an EMPG.
(4): Redshift.
(5), (6): Rest-frame equivalent widths of H$\alpha$ and H$\beta$ emission lines. These values are obtained from the SDSS DR13 catalog.
(7): Gas-phase metallicity obtained with the electron temperature measurement.
(8): Citation from which the metallicity values are derived---P12: \citet{Pilyugin2012}, K03: \citet{Kniazev2003}, K04: \citet{Kniazev2004}, B04: \citet{Brinchmann2008b}, SA16: \citet{SanchezAlmeida2016}.\\
{$^a$} No reliable $EW_\mathrm{0}$(H$\alpha$) measurements are given due to pixel issues on the spectrum.
Instead, we estimate $EW_\mathrm{0}$(H$\alpha$) values from $EW_\mathrm{0}$(H$\beta$) measurements and the empirical relation of $EW_\mathrm{0}$(H$\alpha$)=5.47$\times$$EW_\mathrm{0}$(H$\beta$), which is obtained from metal-poor galaxies in the literature (Figure \ref{fig:EWHa_EWHb}).\\ 
Note that we highlight values of equivalent widths and metallicities that satisfy the EMPG conditions in bold font.
}
\end{deluxetable*}

\section{SELECTION} \label{sec:selection}
In Section \ref{subsec:test}, we have confirmed that our object classifier works well with the SDSS test catalog.
Thus, we expect that our object classifier can also select EMPGs in the HSC-SSP data.
In Sections \ref{subsec:HSC_EMPG} and \ref{subsec:SDSS_EMPG}, we choose EMPG candidates from the HSC-SSP and SDSS source catalogs (Sections \ref{subsec:HSCcatalog} and \ref{subsec:SDSScatalog}) with our ML classifier (Section \ref{subsec:classifier}).
Hereafter, these candidates chosen from the HSC-SSP and SDSS source catalogs are called ``HSC-EMPG candidates''  and ``SDSS-EMPG candidates'', respectively.

\subsection{EMPG Candidates from the HSC Data} \label{subsec:HSC_EMPG}
In Section \ref{subsec:HSCcatalog}, we have created the HSC-SSP source catalog, which consists of 17,912,612 and 40,407,765 sources in the S17A and S18A source catalogs, respectively.
As noted in Section \ref{subsec:HSCcatalog}, the sources selected from S17A and S18A data at this point are partly duplicated, but the duplication will be removed in the last step of the selection.
In this section, we select EMPG candidates from the HSC-SSP source catalog in four steps described below.

In the first step, we coarsely remove sources based on blending, extendedness, and color before we apply our ML classifier.
We remove sources whose photometry is strongly affected by back/foreground objects as follows.
Fluxes of a source and a back/foreground object are measured at the central position of the source, and when a flux of the back/foreground object exceeds 50\% of the source flux, the source is removed.
We only select extended sources whose {\tt extendedness\_value} flags are 1 in all of the $griz$ bands.
The {\tt hscPipe} labels a point source and an extended source as {\tt extendedness\_value}$=$0 and 1, respectively.
The {\tt hscPipe} defines a point source as a source whose PSF magnitude ($m_\mathrm{psf}$) and {\tt cmodel} magnitude ($m_\mathrm{cmodel}$) match within $m_\mathrm{psf}$$-$$m_\mathrm{cmodel}$$<$0.0164 \citep[][]{Bosch2018}.
To save calculation time, we remove part of the sources before we apply the DL classifier.
To roughly remove sources whose colors are apparently different from EMPGs, we apply
\begin{eqnarray}
  r-i&<&-0.3 \label{eq:color_a}\\
  i-z&<&0.2. \label{eq:color_b}
\end{eqnarray}
To remove possible contamination from normal galaxies, we also apply
\begin{eqnarray}
  g-r&<&-0.3125(r-i)+0.1375. \label{eq:color_c}
\end{eqnarray}
In other words, we choose sources that satisfy all of Equations (\ref{eq:color_a})--(\ref{eq:color_c}) here. 
We show Equations (\ref{eq:color_a})--(\ref{eq:color_c}) in Figure \ref{fig:mock}.
After these selection criteria, 680 and 2,494  sources remain from the S17A and S18A data, respectively. 
The source removal in the fist step effectively reduces the calculation time in the ML classifier in the second step below. 

In the second step, we apply the ML classifier constructed in Section \ref{subsec:classifier} to the sources selected above. 
The ML classifier selects 32 and 57 sources out of the 680 (S17A) and 2,494 (S18A), respectively. 

In the third step, we remove transient objects by checking the $g, r$-band multi-epoch images.
We measure fluxes in each epoch and calculate an average and a standard deviation of these flux values.
If the standard deviation becomes larger than 25\% of the average value, we regard the source as a transient object and eliminate it from the sample.
Removing 10 and 15 sources, we obtain 22 and 42 sources after the third step for S17A and S18A data, respectively.

In the last step, we inspect a $gri$-composite image.
Here we remove apparent {\HII} regions inside a large star-forming galaxy, sources affected by a surrounding bright star, and apparently red sources.
The apparently red sources are mistakenly selected due to an issue in the {\tt cmodel} photometry.
Indeed, they show red colors ($r-i>0.0$) in the 1.0-arcsec aperture photometry, while they become blue in the {\tt cmodel} photometry.
In the inspection of multi-epoch images and $gri$-composite images, we have removed 10 and 21 sources from the S17A and S18A data, respectively.

Eventually, we thus obtain 12 and 21 HSC-EMPG candidates from the S17A and S18A catalogs, respectively.
We find that 6 out of the HSC-EMPG candidates are duplicated between the S17A and S18A catalogs.
Thus, the number of our independent HSC-EMPG candidates is 27 ($=$12$+$21$-$6).
A magnitude range of the 27 HSC-EMPG candidates is $i=19.3$--$24.3$ mag. 

Out of the 27 candidates, we find 6 candidates that are selected in S17A but not selected again in S18A.
Four out of the 6 candidates are slightly redder in S18A than in S17A and thus not selected in S18A.
The other two are removed in S18A due to flags related to a cosmic ray or a nearby bright star.
Such differences arise probably due to the different pipeline versions between S17A and S18A.
We check images and photometry of these 6 candidates individually.
Then we confirm that these 6 candidates seem to have no problem as an EMPG candidate.

\subsection{EMPG Candidates from the SDSS Data} \label{subsec:SDSS_EMPG}
In Section \ref{subsec:SDSScatalog}, we have constructed the SDSS source catalog consisting of 31,658,307 sources.
In this section, we select EMPG candidates from the SDSS source catalog similarly to the HSC source catalog in Section \ref{subsec:HSC_EMPG}.

First, we remove sources that have colors apparently different from EMPGs with equations (\ref{eq:color_a})--(\ref{eq:color_c}). 
Then we apply our ML classifier to the SDSS source catalog and our classifier has selected 107 sources. 
Checking $gri$-composite images, we eliminate apparent {\HII} regions in a spiral galaxy, sources affected by a surrounding bright star, and apparently red sources.
We also remove sources if the corresponding composite image shows an apparent problem that may be caused by an incorrect zero-point magnitude.
In the visual inspection above, 21 sources have been removed.
These steps leave us with 86 SDSS-EMPG candidates from the SDSS source catalog, whose $i$-band magnitudes range $i=14.8$--$20.9$ mag. 

Cross-matching the SDSS-EMPG candidates with the SDSS spectra data, we find that 17 out of the 86 candidates already have an SDSS spectrum.
These 17 spectra show strong nebular emission lines from galaxies at $z=0.002$--$0.026$, 15 out of which have been already reported with a metallicity measurement in the range of {\metal}$=$7.44--8.22 \citep{Kniazev2003, Kniazev2004, Izotov2007a, Engelbracht2008, Izotov2012, Shirazi2012, SanchezAlmeida2016, Izotov2016a}.
Seven out of the 15 galaxies satisfy the EMPG condition, {\metal}$<$7.69. 
All of the 6 EMPGs chosen in our classifier test (Section \ref{subsec:test}) are selected again here. 
Another object out of the 86 candidates is {\HSCEMPGa}, which is also selected as an HSC-EMPG candidate in Section \ref{subsec:HSC_EMPG}. 

\section{SPECTROSCOPY} \label{sec:spectroscopy}
We have carried out spectroscopy for the 10 EMPG candidates with 4 spectrographs: Low Dispersion Survey Spectrograph 3 (LDSS-3) and the Magellan Echellette Spectrograph \citep[MagE,][]{Marshall2008} on Magellan telescope, the Deep Imaging Multi-Object Spectrograph \citep[DEIMOS,][]{Faber2003} on Keck-II telescope, and the Faint Object Camera And Spectrograph \citep[FOCAS,][]{Kashikawa2002} on Subaru telescope. 
In this section, we explain the spectroscopy for the 10 EMPG candidates.. 
\begin{deluxetable*}{ccccccc}
\tablecolumns{6}
\tabletypesize{\scriptsize}
\tablecaption{Summary of LDSS3, MagE, DEIMOS, and FOCAS observations %
\label{table:obs}}
\tablehead{%
\colhead{ID} &
\colhead{R.A.} &
\colhead{Dec.} &
\colhead{slit width} &
\colhead{P.A.} &
\colhead{exposure} &
\colhead{seeing} \\
\colhead{} &
\colhead{(hh:mm:ss)} &
\colhead{(dd:mm:ss)} &
\colhead{(arcsec)} &
\colhead{(deg)} &
\colhead{(sec)} &
\colhead{(arcsec)} \\
\colhead{(1)} &
\colhead{(2)} &
\colhead{(3)} &
\colhead{(4)} &
\colhead{(5)} &
\colhead{(6)} &
\colhead{(7)} 
}
\startdata
\multicolumn{6}{c}{LDSS3 observation}\\
\hline
{\HSCEMPGa}  & 14:29:48.61 & $-$01:10:09.67 & $0.75$ & $+33.8$ & 3,600 &  $0.8$\\ 
\hline\hline
\multicolumn{6}{c}{MagE observation}\\
\hline
{\HSCEMPGb}  & 23:14:37.55 & $+$01:54:14.27 & $0.85$ & $+84.8$ & 3,600 & $0.9$ \\ 
{\HSCEMPGc}  & 11:42:25.19 & $-$00:38:55.64 & $0.85$ & $+103.3$ & 3,600 & $0.8$ \\ 
{\SDSSEMPGa}  & 00:02:09.94 & $+$17:15:58.65 & $1.2$& $+110.0$ & 1,800 & $1.5$ \\ 
{\SDSSEMPGb} & 16:42:38.45 & $+$22:33:09.09 & $0.85$ & $+36.0$ & 1,800 & $1.0$ \\ 
{\SDSSEMPGc} & 21:15:58.33 & $-$17:34:45.09 & $0.85$ & $+144.0$ & 1,800 & $1.1$ \\ 
{\SDSSEMPGd} & 22:53:42.41 & $+$11:16:30.62 & $1.2$ & $+76.0$ & 1,800 & $1.2$ \\ 
{\SDSSEMPGe} & 23:10:48.84 & $-$02:11:05.74 & $1.2$ & $+18.8$ & 1,800 & $1.0$ \\ 
{\SDSSEMPGf} & 23:27:43.69 & $-$02:00:55.89 & $1.2$ & $+49.6$ & 1,800 & $1.0$ \\ 
\hline\hline
\multicolumn{6}{c}{DEIMOS observation}\\
\hline 
HSC J1631$+$4426 & 16:31:14.24 & $+$44:26:04.43 & $0.80$ & $+45.9$ & 2,400 &  $0.5$\\ 
\hline\hline
\multicolumn{6}{c}{FOCAS observation}\\
\hline 
HSC J1631$+$4426 & 16:31:14.24 & $+$44:26:04.43 & $2.0$ & $+45.9$ & \mod{12,000} &  $0.6$ 
\enddata
\tablecomments{
(1) ID. (2) R.A. in J2000. (3) Dec. in J2000. (4) Slit width. (5) Position angle. (6) Exposure time. (7) Seeing.
}
\end{deluxetable*}

\begin{figure*}[!htbp]
\epsscale{0.25}
\plotone{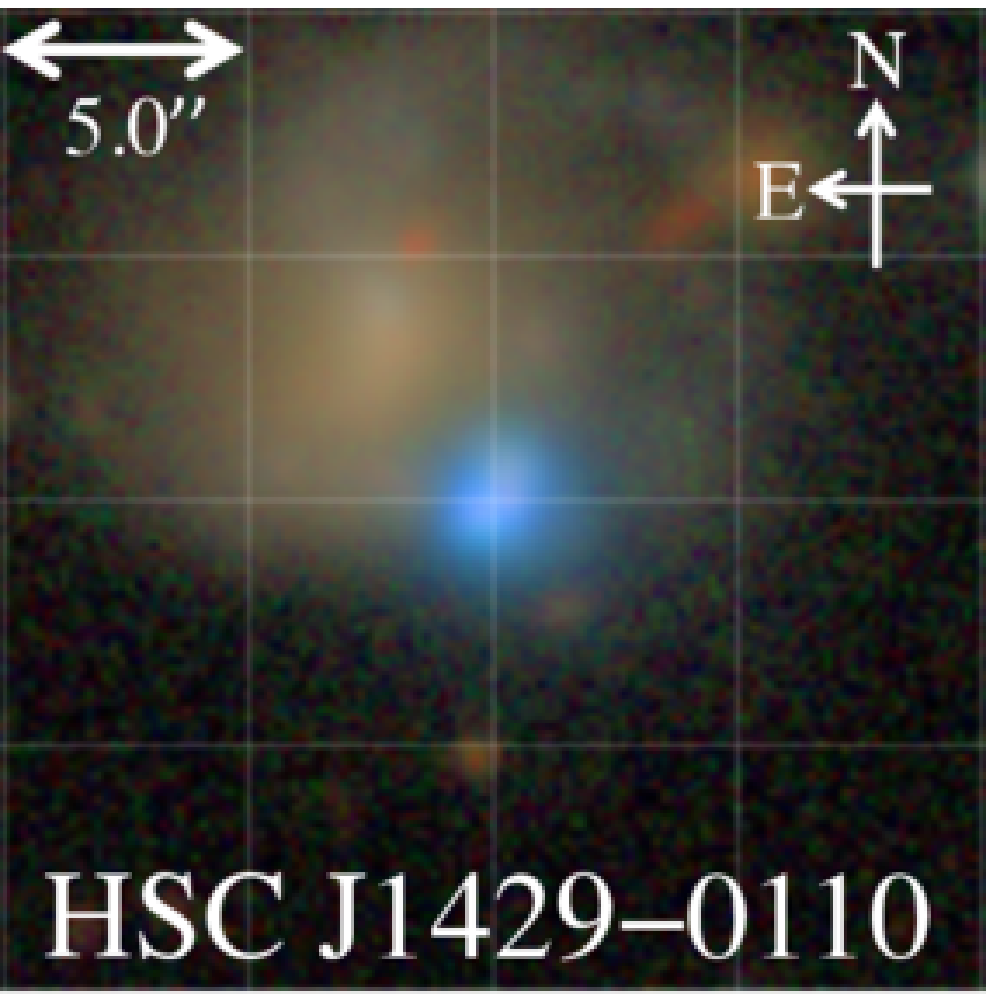}
\plotone{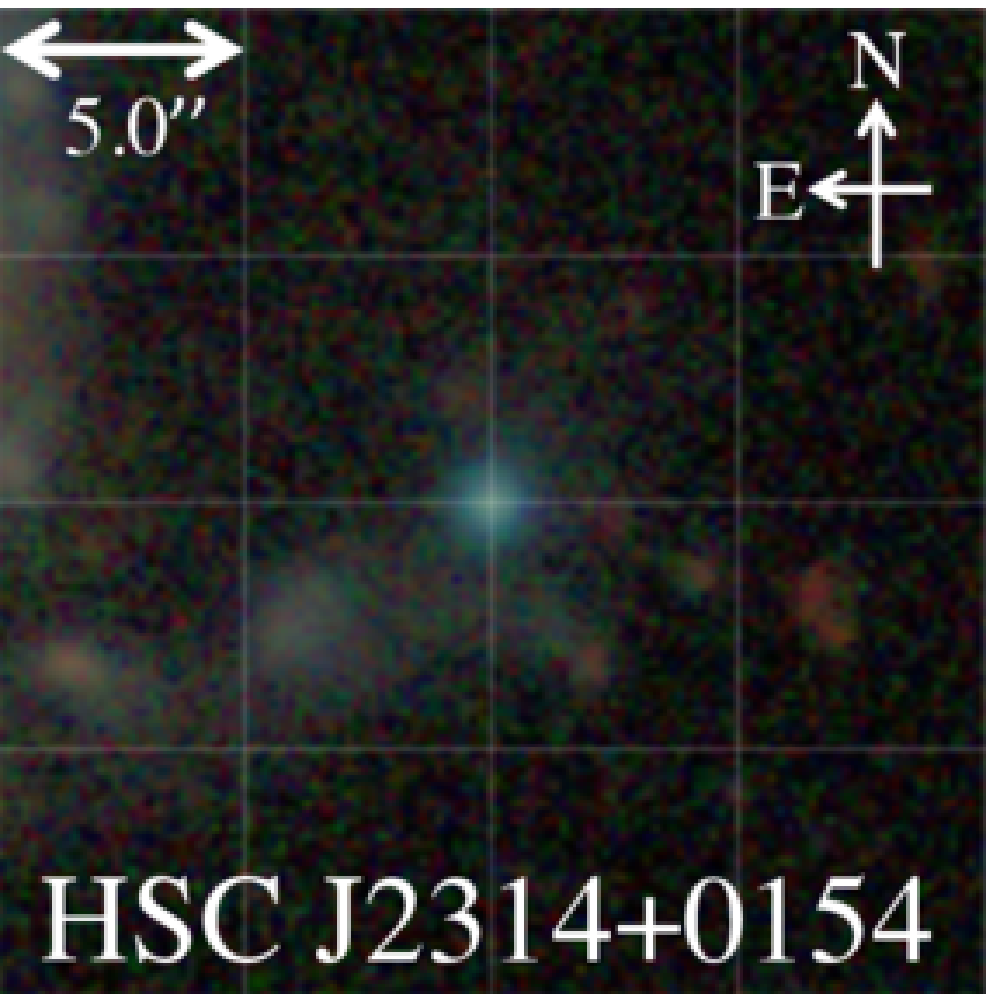}
\plotone{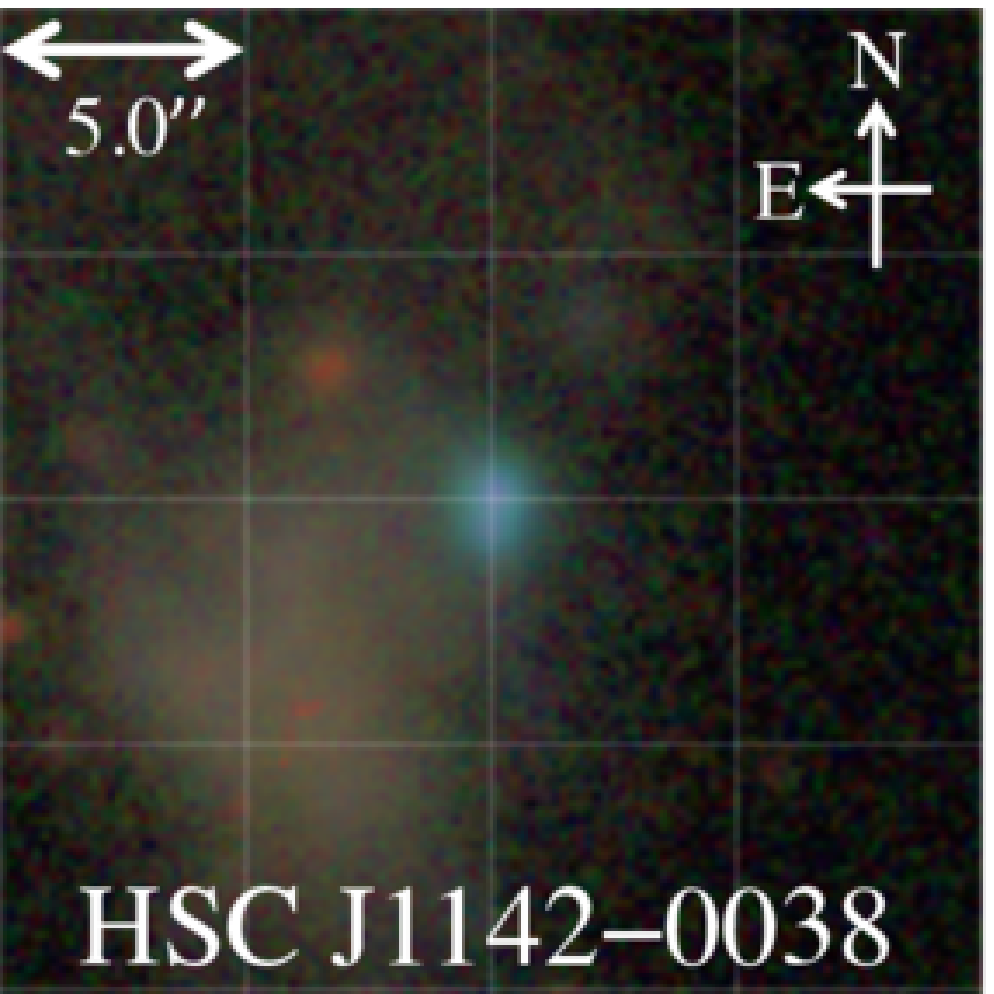}
\plotone{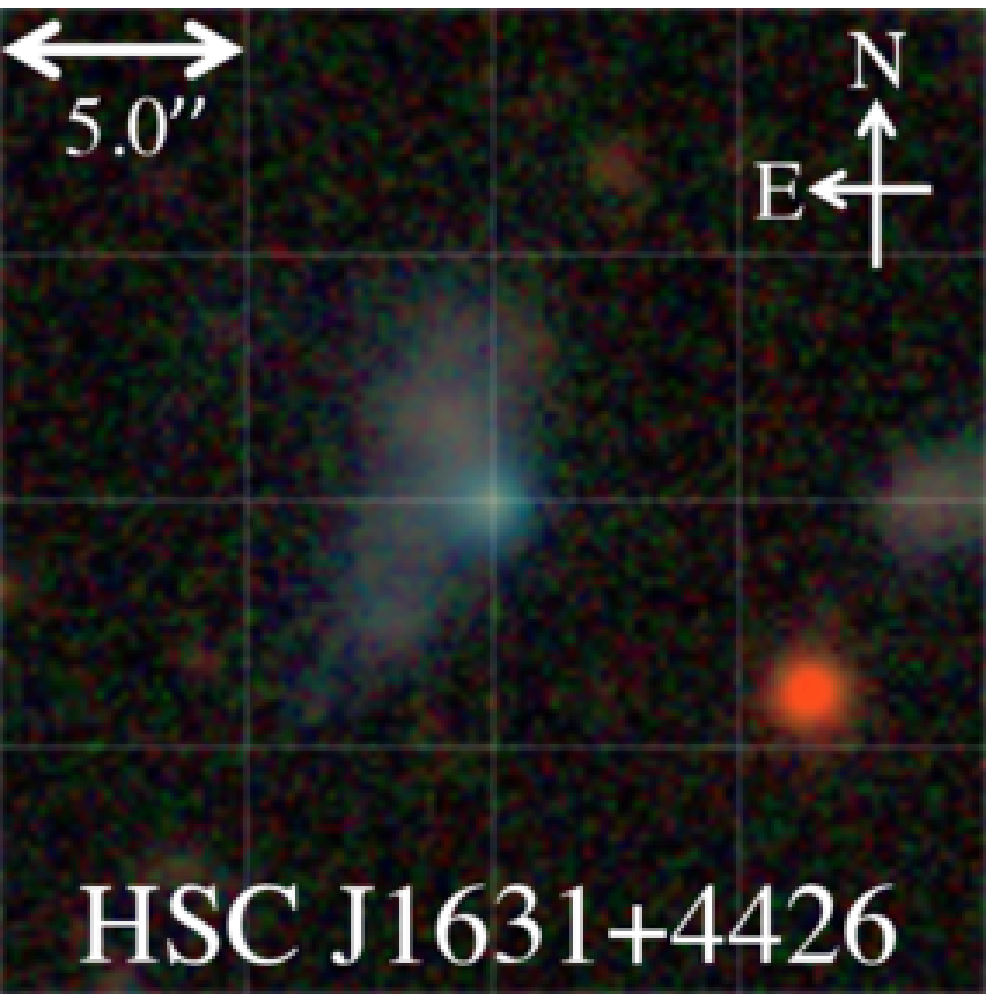}
\caption{HSC $gri$-band images of the 4 HSC-EMPG candidates, {\HSCEMPGa}, {\HSCEMPGb}, {\HSCEMPGc}, and {\HSCEMPGd}.
\label{fig:images1}}
\end{figure*}

\begin{figure*}[!htbp]
\epsscale{0.25}
\plotone{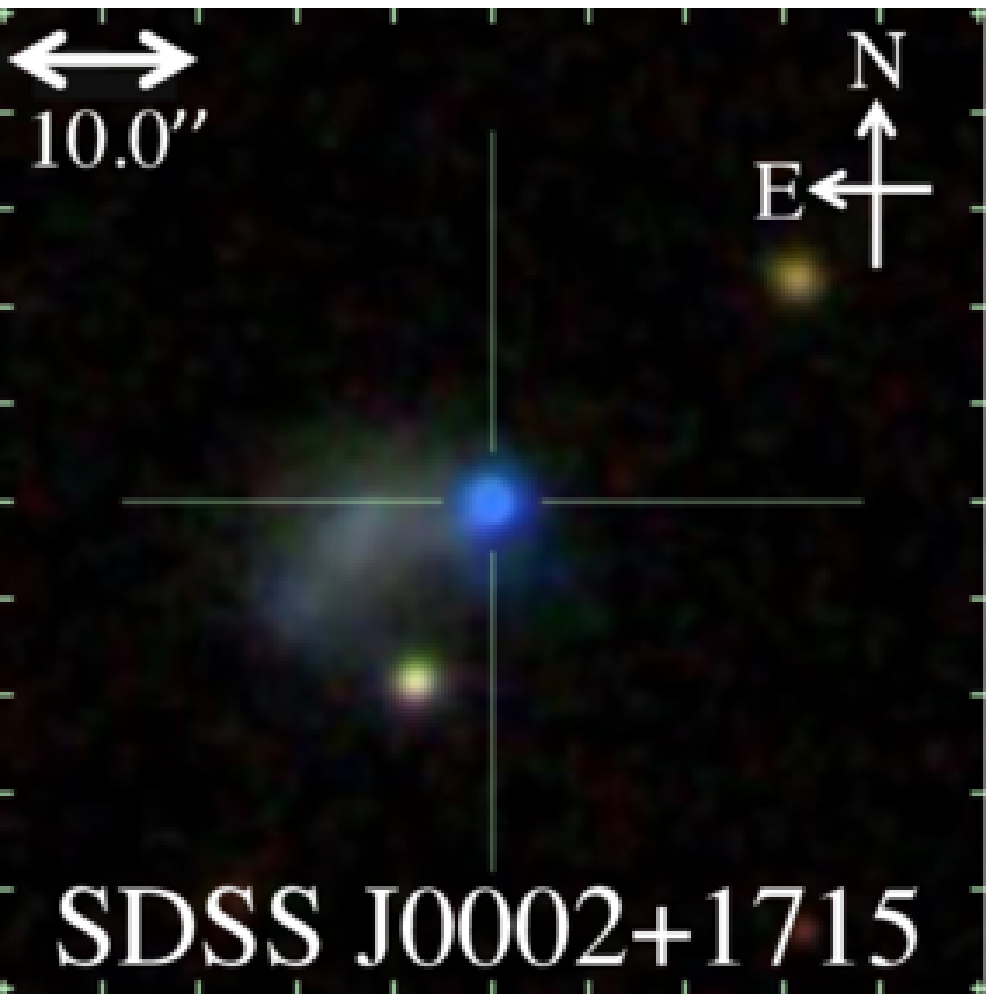}
\plotone{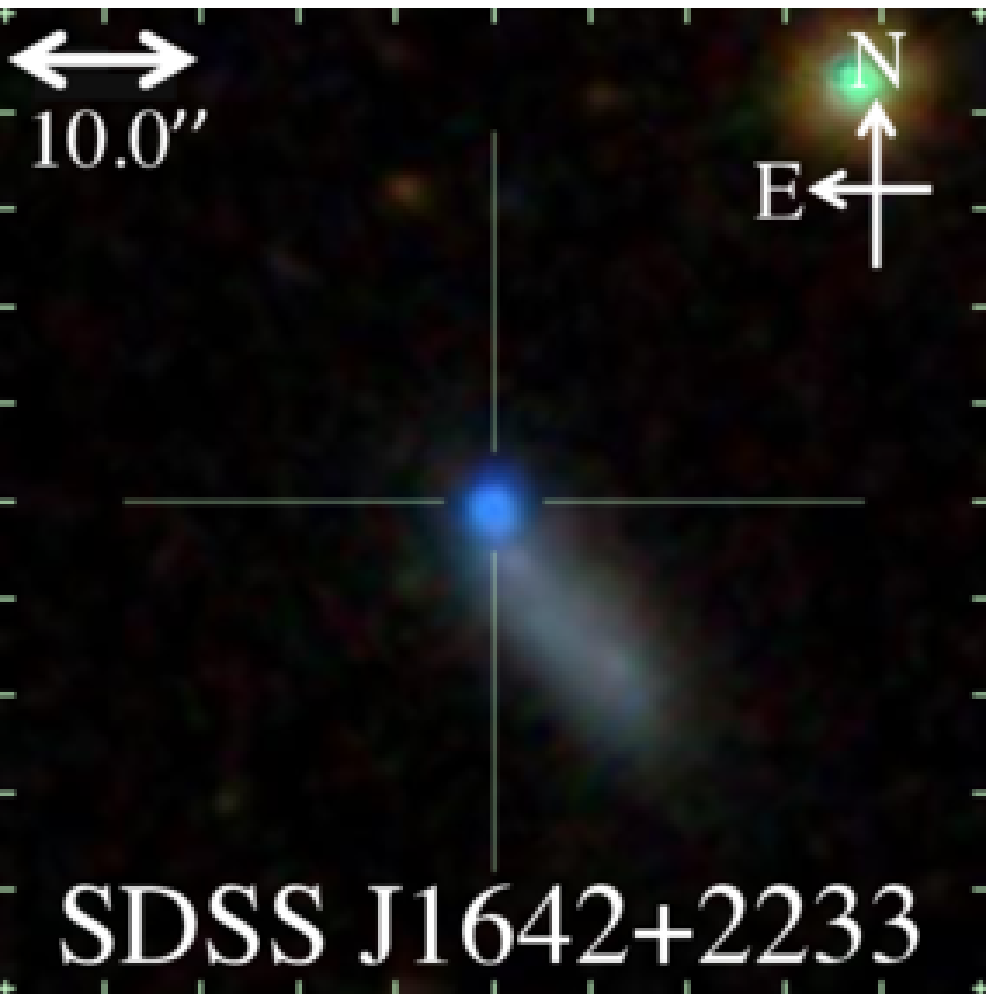}
\plotone{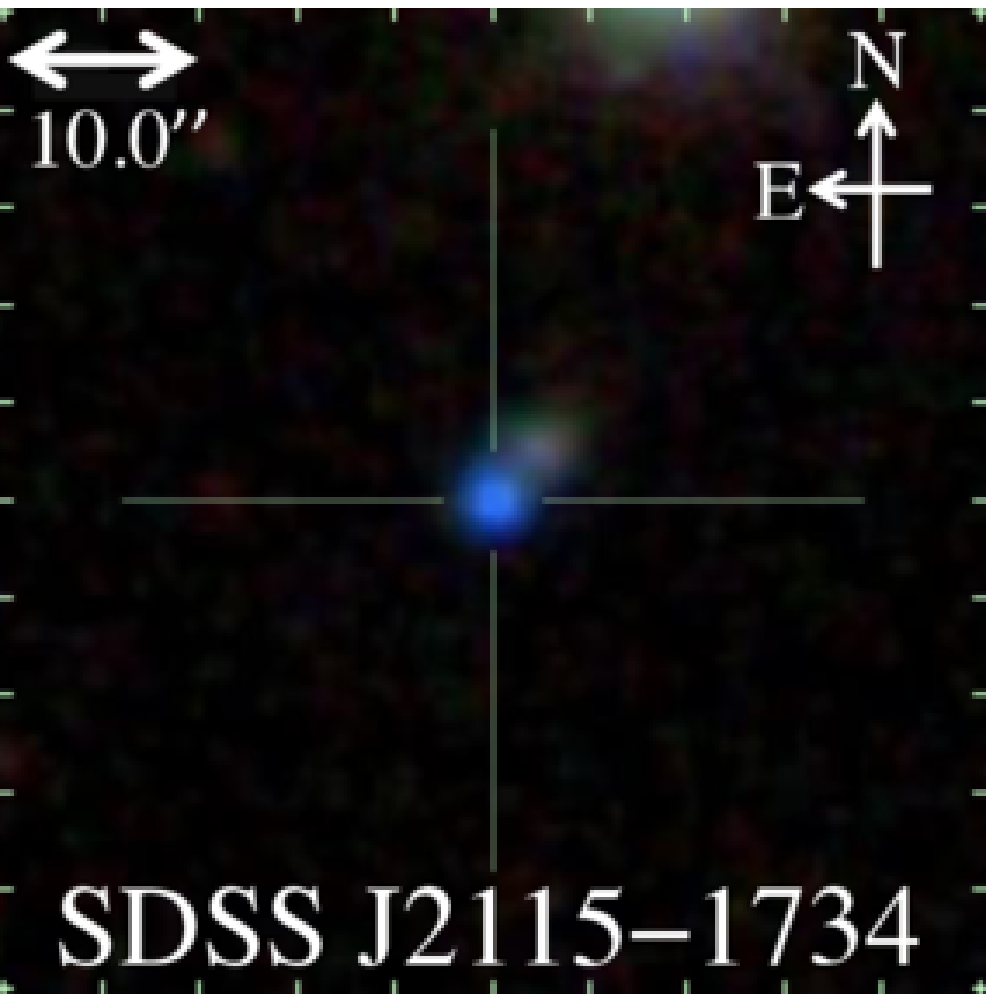}\\
\plotone{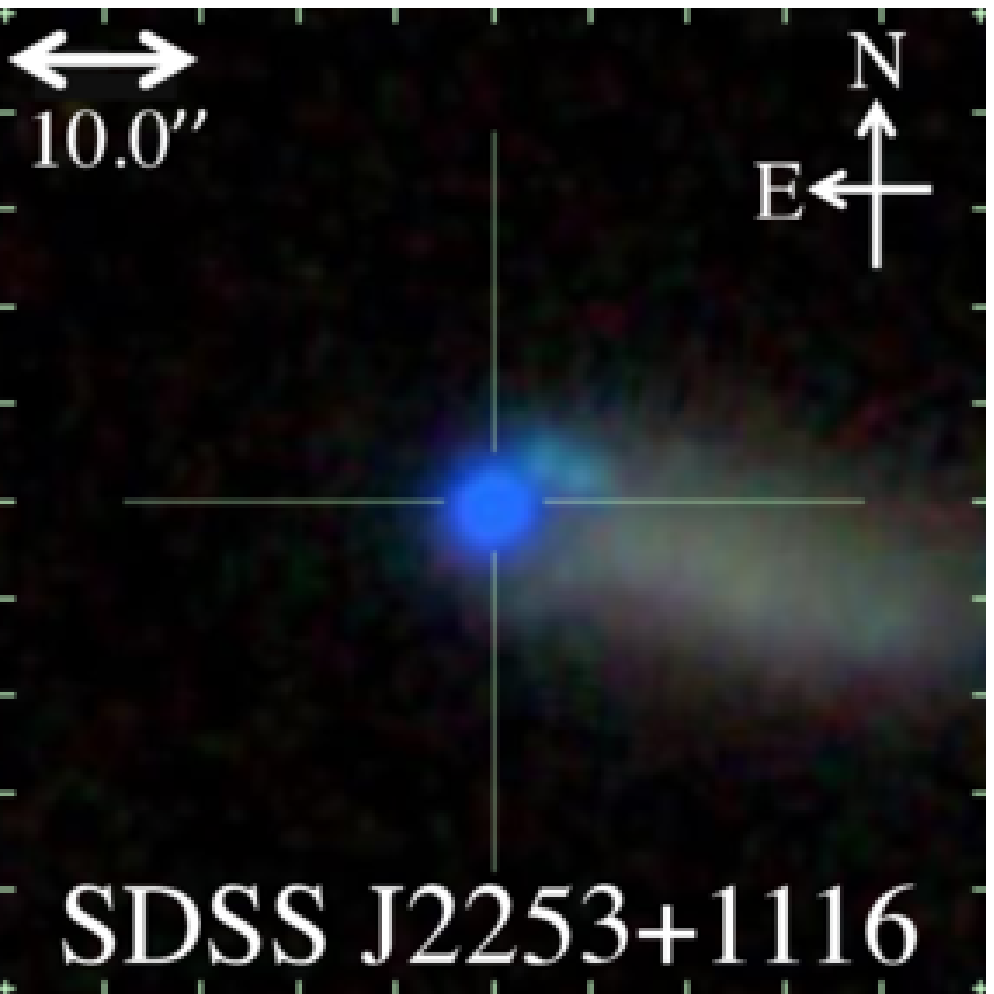}
\plotone{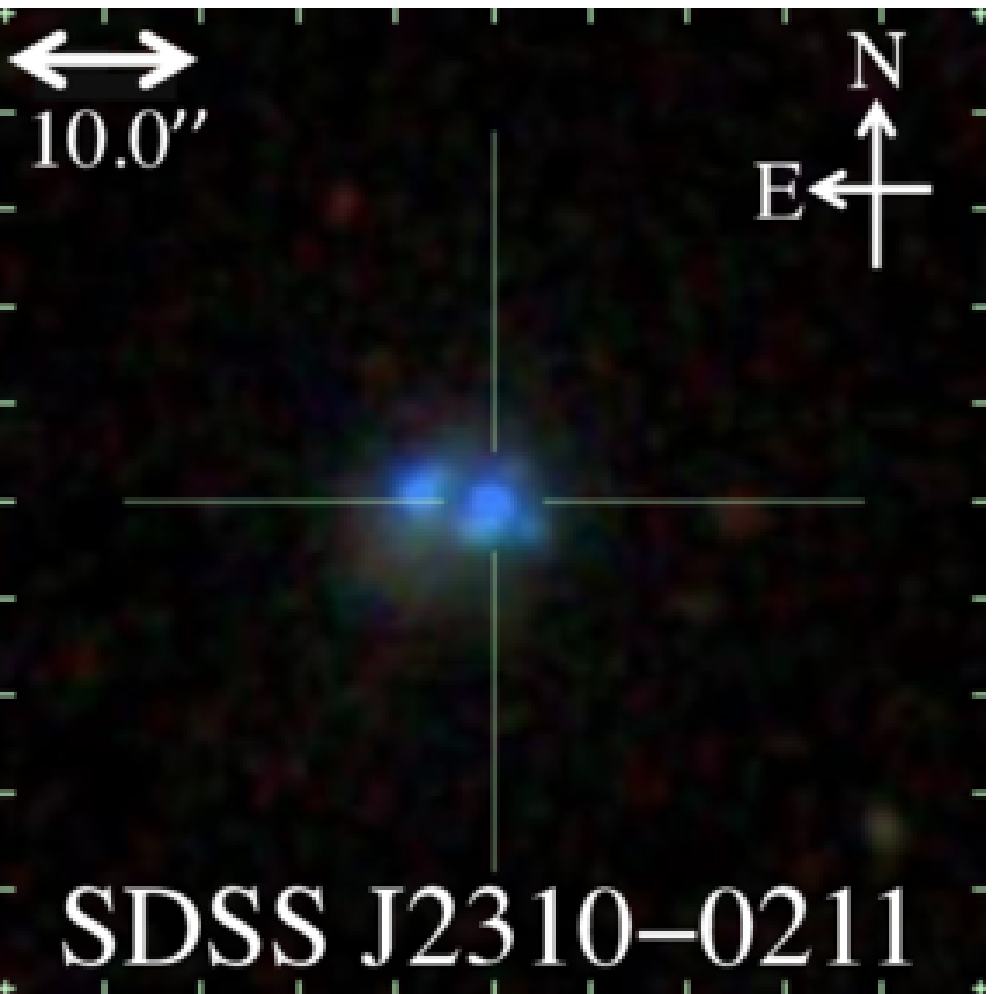}
\plotone{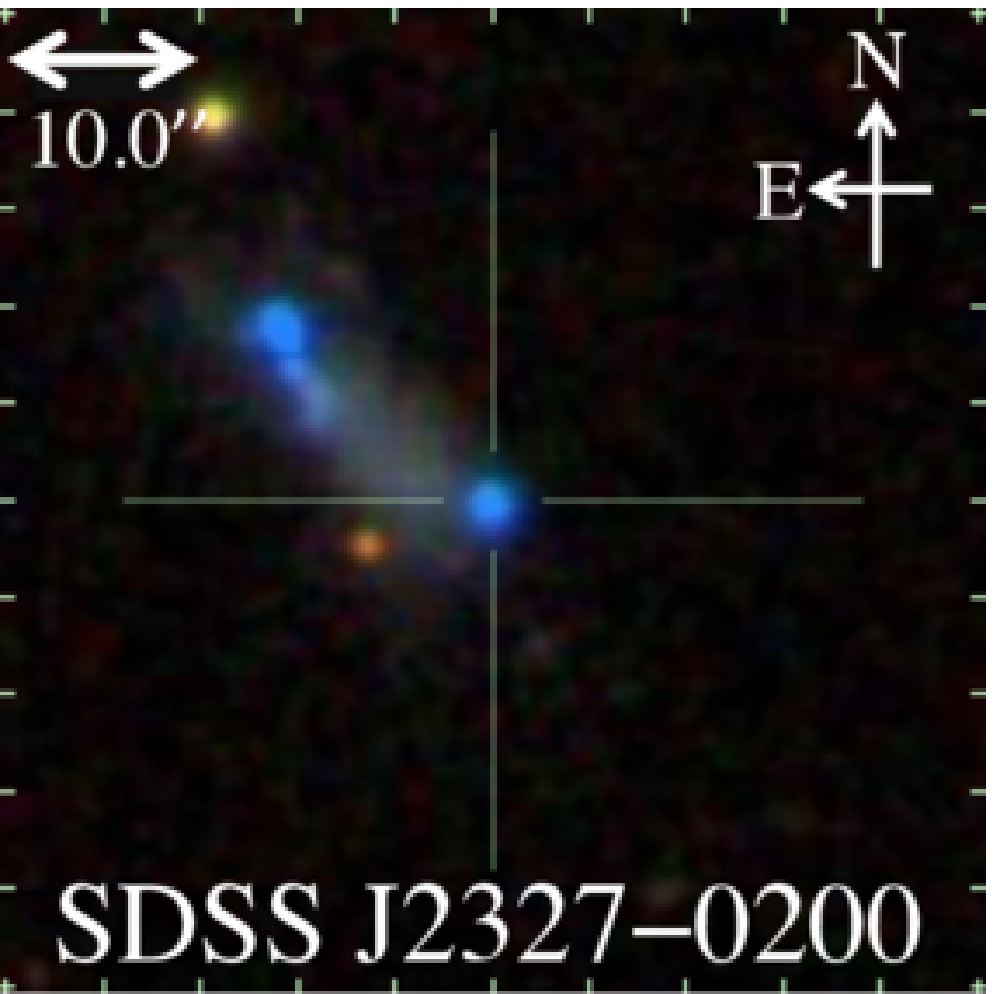}
\caption{SDSS $gri$-band images of the 6 SDSS-EMPG candidates, {\SDSSEMPGa}, {\SDSSEMPGb}, {\SDSSEMPGc}, {\SDSSEMPGd}, {\SDSSEMPGe}, and {\SDSSEMPGf}.
\label{fig:images2}}
\end{figure*}

\subsection{Magellan/LDSS-3 Spectroscopy} \label{subsec:ldss3}
We conducted spectroscopy for an HSC-EMPG candidate ({\HSCEMPGa}) on 2018 June 12 with LDSS-3 at Magellan telescope (PI: M. Rauch). 
We used the VPH-ALL grism with the $0.\!\!^{\prime\prime}75$$\times$4$^{\prime}$ long-slit, which was placed at the offset position two-arcmin away from the center of the long-slit mask so that the spectroscopy could cover the bluer side.
The exposure time was 3,600 seconds. 
The spectroscopy covered $\lambda\sim$3,700--9,500 {\AA} with the spectral resolution of $R \equiv \lambda/\Delta\lambda \sim$ 860.
The A0-type standard star CD-32\,\,9972 (RA=14:11:46.37, Dec.=$-$33:03:14.3 in J2000) was also observed. 
The sky was clear during the observation with seeing sizes of $0.\!\!^{\prime\prime}6$--$0.\!\!^{\prime\prime}9$.

\subsection{Magellan/MagE Spectroscopy} \label{subsec:mage}
We carried out spectroscopy for two HSC-EMPG candidates ({\HSCEMPGb} and {\HSCEMPGc}) and six SDSS-EMPG candidates ({\SDSSEMPGa}, {\SDSSEMPGb}, {\SDSSEMPGc}, {\SDSSEMPGd}, {\SDSSEMPGe}, and {\SDSSEMPGf}) on 2018 June 13 with MagE of Magellan telescope (PI: M. Rauch). 
We used the echellette grating with the $0.\!\!^{\prime\prime}85$$\times$10$^{\prime\prime}$ or $1.\!\!^{\prime\prime}2$$\times$10$^{\prime\prime}$ longslits.
The exposure time was 1,800 or 3,600 seconds, depending on luminosities of the candidates. 
The MagE spectroscopy covered $\lambda\sim$3,100--10,000 {\AA} with the spectral resolution of $R \equiv \lambda/\Delta\lambda \sim$ 4,000.
The A0-type standard star CD-32\,\,9972 (RA=14:11:46.37, Dec.=$-$33:03:14.3 in J2000) and the DOp-type standard star Feige\,110 (RA=23:19:58.39, Dec.=$-$05:09:55.8 in J2000) were also observed. 
The sky was clear during the observation with seeing sizes of $0.\!\!^{\prime\prime}8$--$1.\!\!^{\prime\prime}5$.

\subsection{Keck/DEIMOS Spectroscopy} \label{subsec:deimos}
We conducted spectroscopy for an HSC-EMPG candidate ({\HSCEMPGd}) as a filler target on 2018 August 10 with DEIMOS of the Keck-II telescope (PI: Y. Ono). 
We used the multi-object mode with the $0.\!\!^{\prime\prime}8$ slit width.
The exposure time was 2,400 seconds. 
We used the 600ZD grating and the BAL12 filter with a blaze wavelength at 5,500 {\AA}. 
The DEIMOS spectroscopy covered $\lambda\sim$3,800--8,000 {\AA} with the spectral resolution of $R \equiv \lambda/\Delta\lambda \sim$1,500. 
The A0-type standard star G191B2B (RA=05:05:30.6, Dec.=$+$52:49:54 in J2000) was also observed. 
The sky was clear during the observation with seeing sizes of $0.\!\!^{\prime\prime}5$.

\subsection{Subaru/FOCAS Spectroscopy} \label{subsec:focas}
We carried out deep spectroscopy for an HSC-EMPG candidate ({\HSCEMPGd}) on 2019 May 13 with FOCAS installed on the Subaru telescope (PI: T. Kojima). 
{\HSCEMPGd} was observed again with FOCAS in a longer exposure time of 10,800 seconds ($=$3 hours).
We used the long slit mode with the $2.\!\!^{\prime\prime}0$ slit width.
We used the 300R grism and the L550 filter with a blaze wavelength at 7,500 {\AA} in a 2nd order. 
The FOCAS spectroscopy covered $\lambda\sim$3,400--5,250 {\AA} with the spectral resolution of $R$$\equiv$$\lambda/\Delta\lambda$$=$400 with the $2.\!\!^{\prime\prime}0$ slit width.
The O-type subdwarf BD$+$28\,4211 (RA=21:51:11.07, Dec.=$+$28:51:51.8 in J2000) was also observed as a standard star. 
The sky condition was clear during the observation with a seeing size of $0.\!\!^{\prime\prime}6$.

The LDSS-3, MagE, DEIMOS, and FOCAS observations are summarized in Table \ref{table:obs}.

\section{REDUCTION AND CALIBRATION OF SPECTROSCOPIC DATA} \label{sec:reduction}
We explain how we reduced and calibrated the spectroscopic data of Magellan/LDSS-3, Magellan/MagE, Keck/DEIMOS, Subaru/FOCAS in Sections \ref{subsec:ldss3_rc}--\ref{subsec:focas_rc}, respectively.

\begin{figure*}[!htbp]
\epsscale{0.5}
\plotone{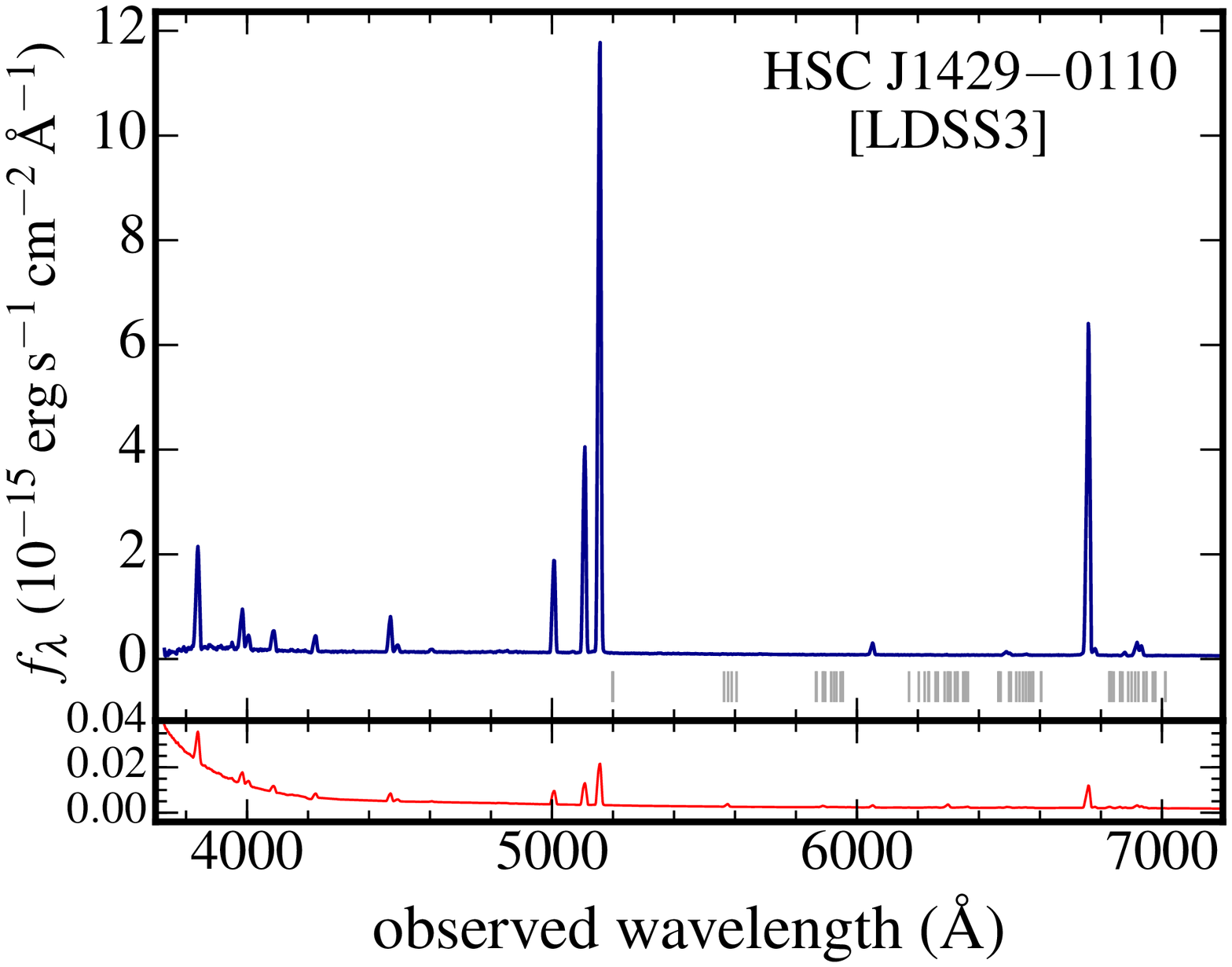}
\plotone{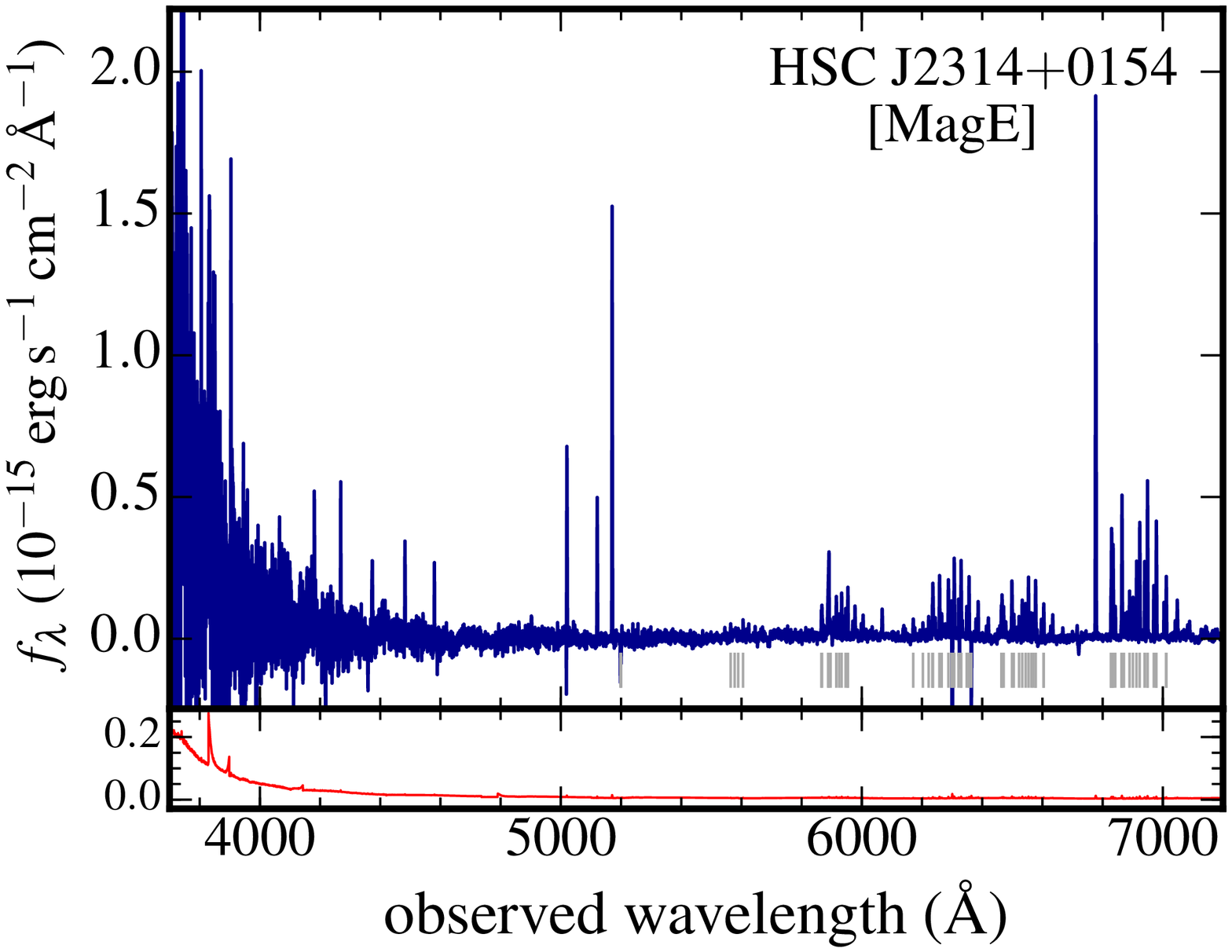}
\plotone{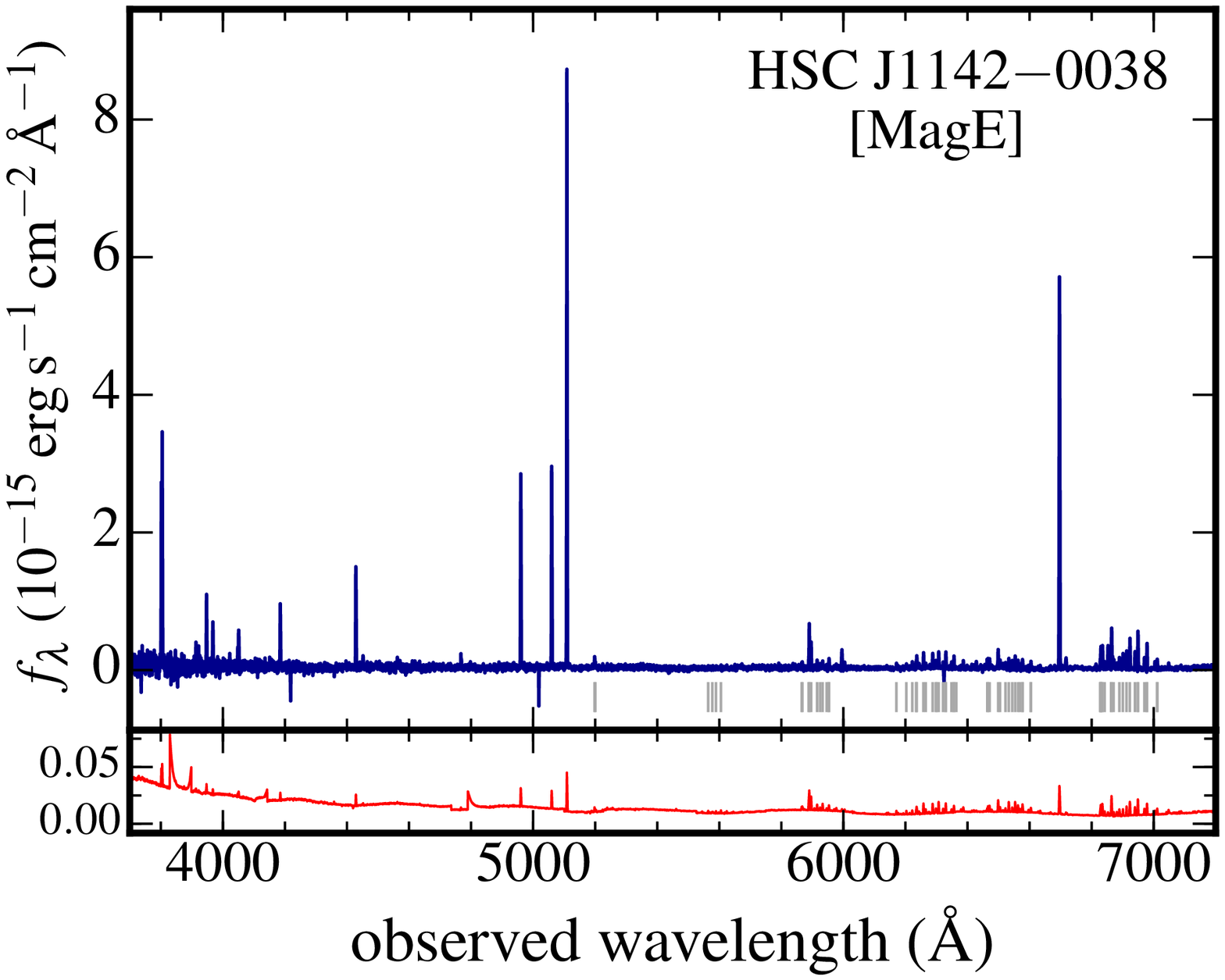}
\plotone{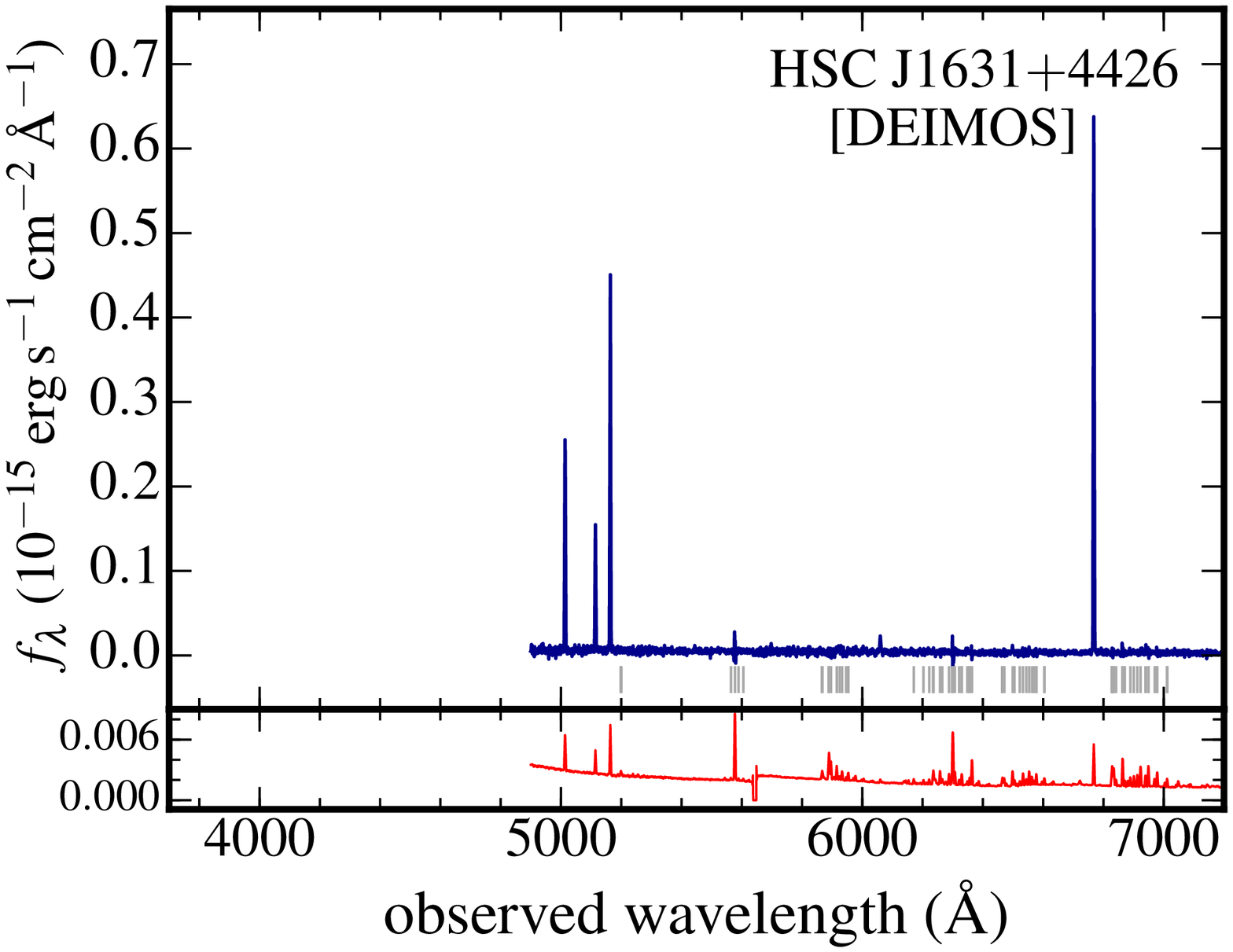}
\plotone{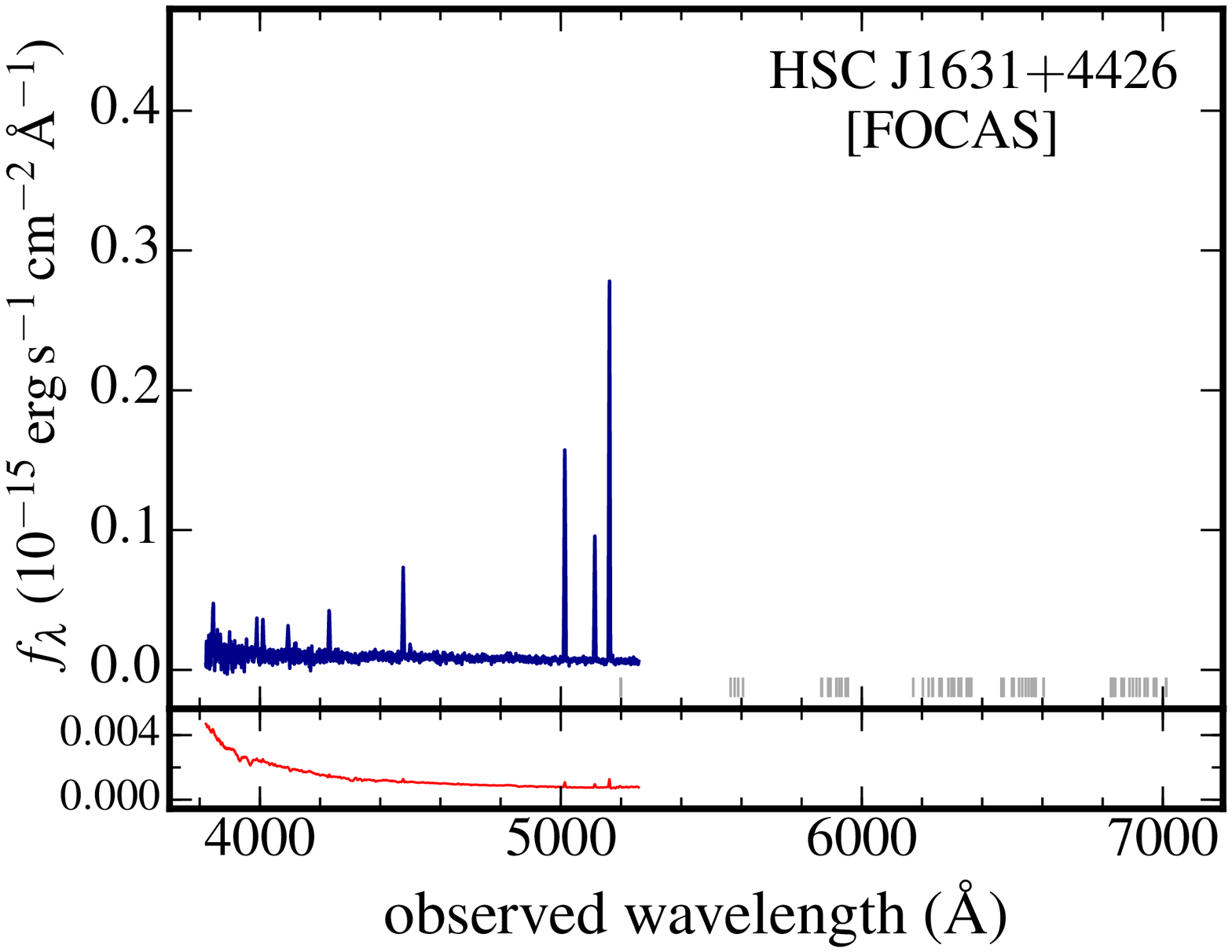}
\caption{Spectra of our 4 HSC EMPGs. 
Positions of sky emission lines are indicated by gray vertical lines at the bottom of each panel.
We mask part of strong sky lines that are not subtracted very well.
We also show noise spectra at the bottom of each panel with red lines.
Here we exhibit two spectra of the same target, {\HSCEMPGd}, for which we conduct spectroscopy both with Keck/DEIMOS (red side, $\lambda$$\gtrsim$5,000 {\AA}) and Subaru/FOCAS (blue side, $\lambda$$\lesssim$5,000 {\AA}). 
\label{fig:spec_hscempg}}
\end{figure*}

\begin{figure*}[!htbp]
\epsscale{0.5}
\plotone{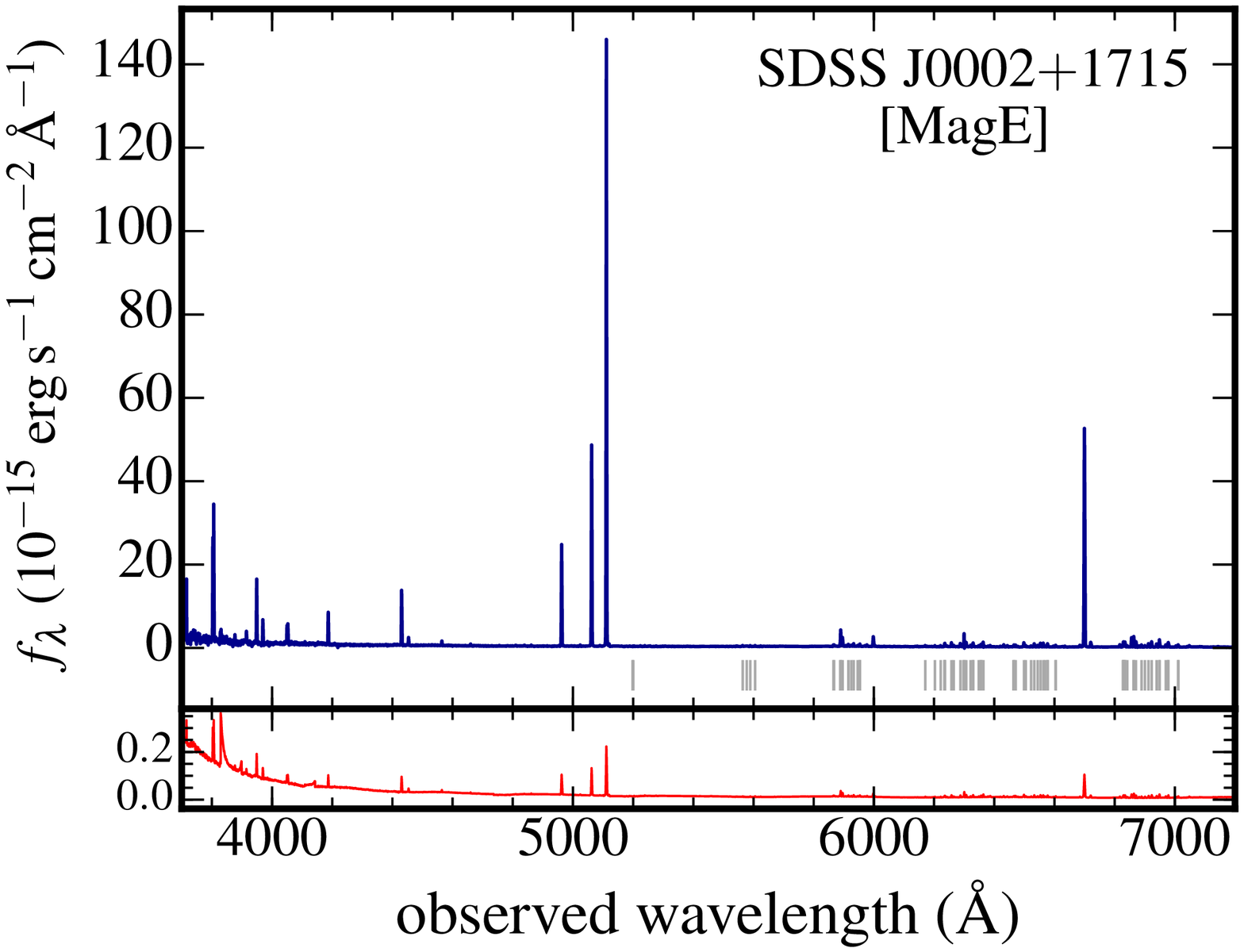}
\plotone{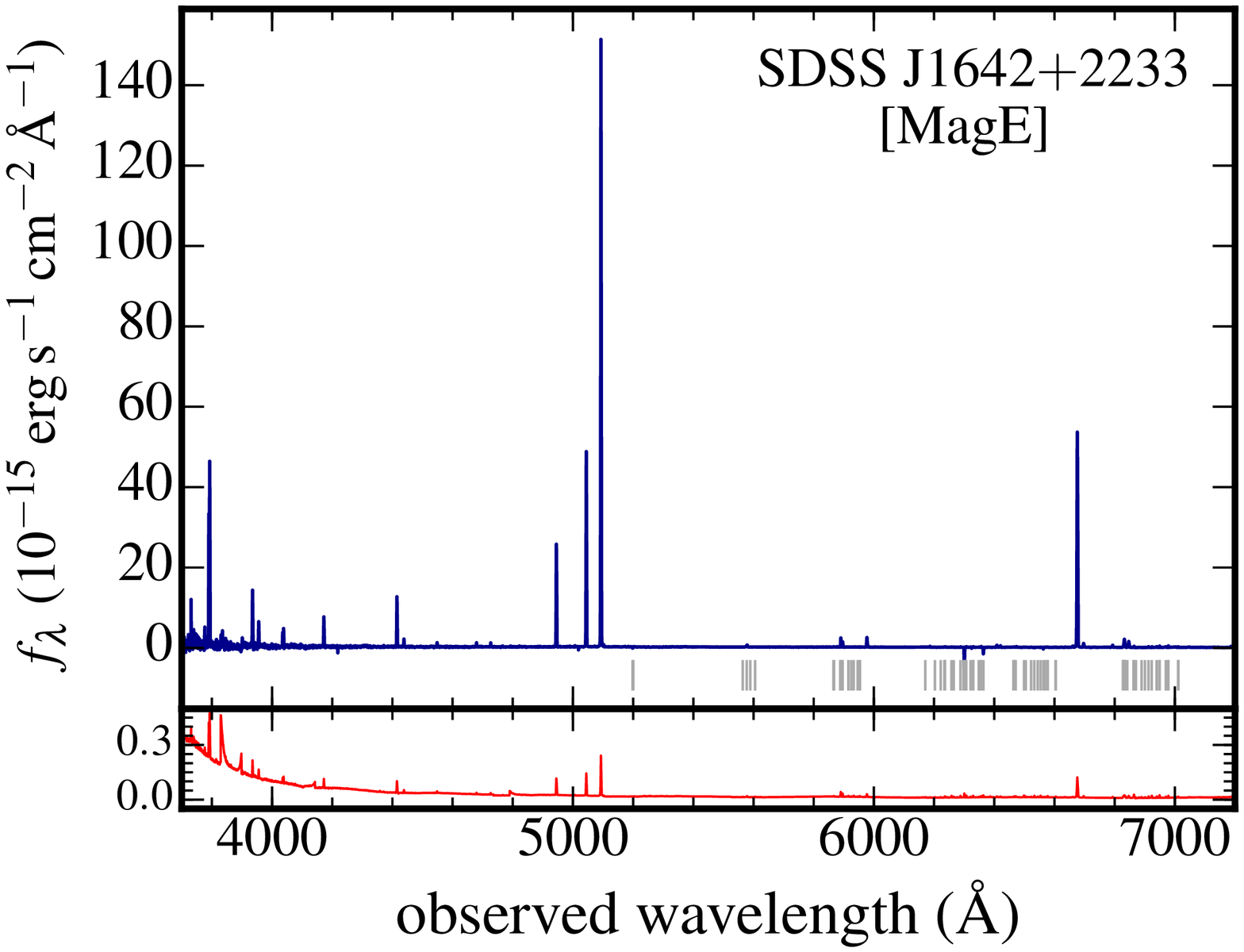}
\plotone{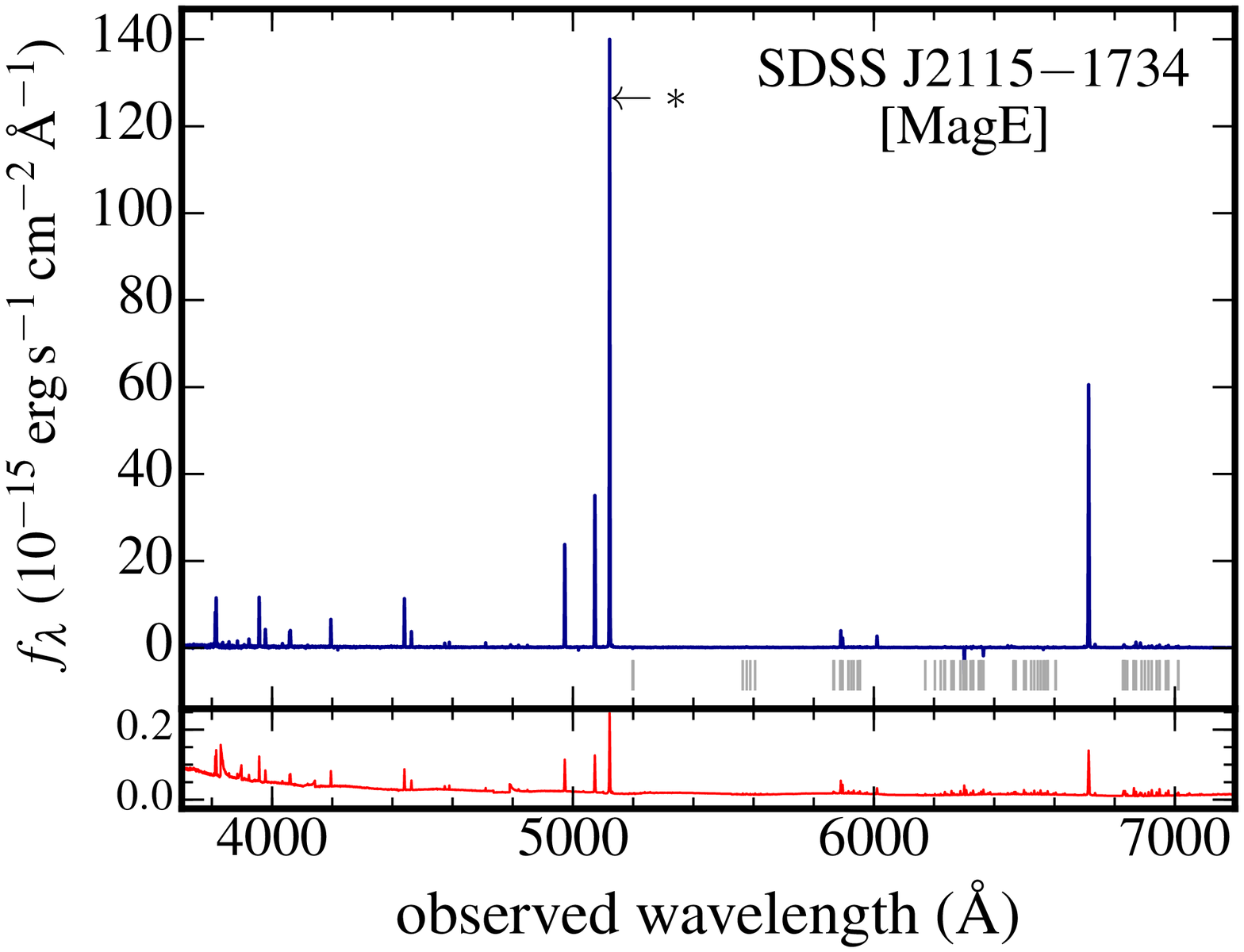}
\plotone{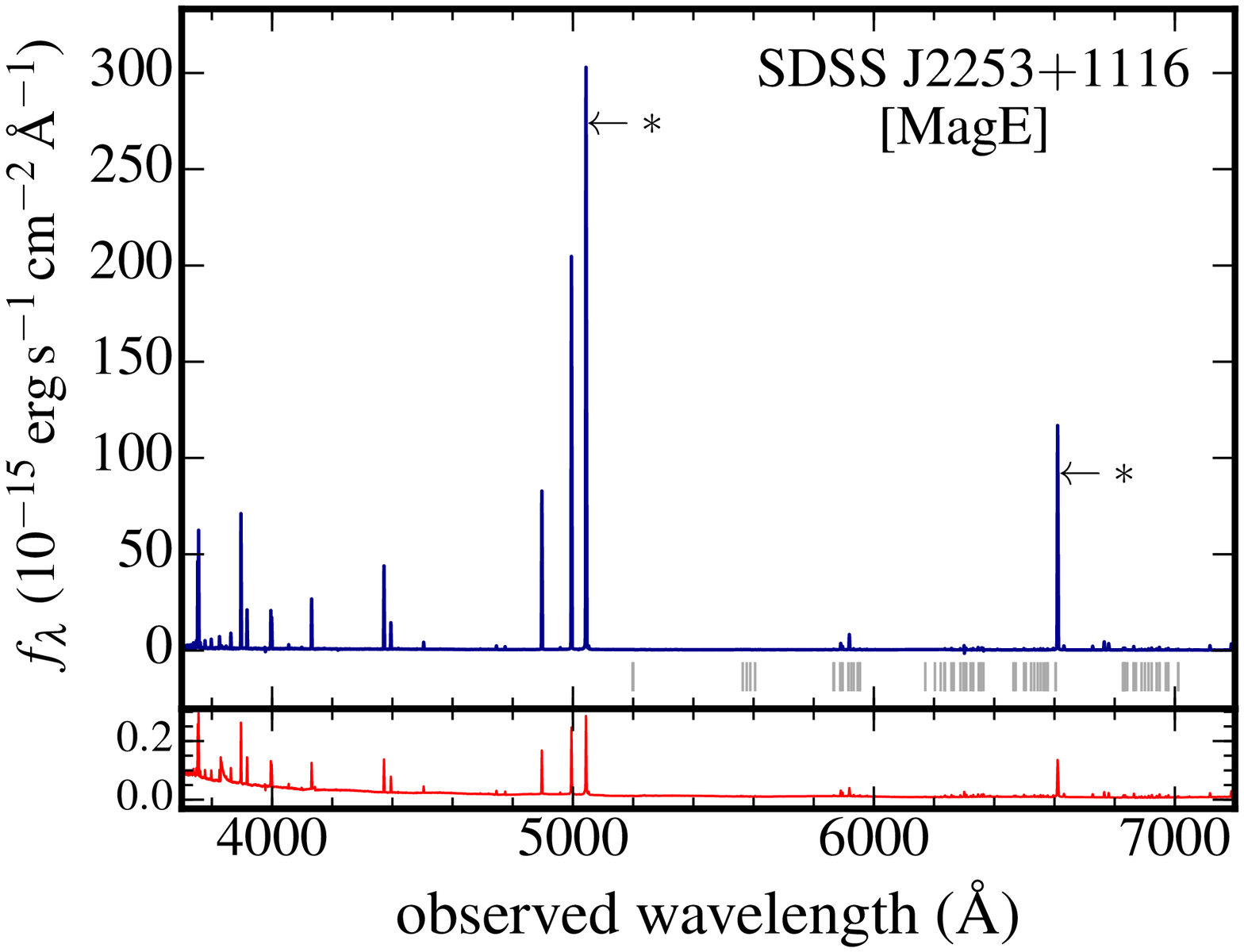}
\plotone{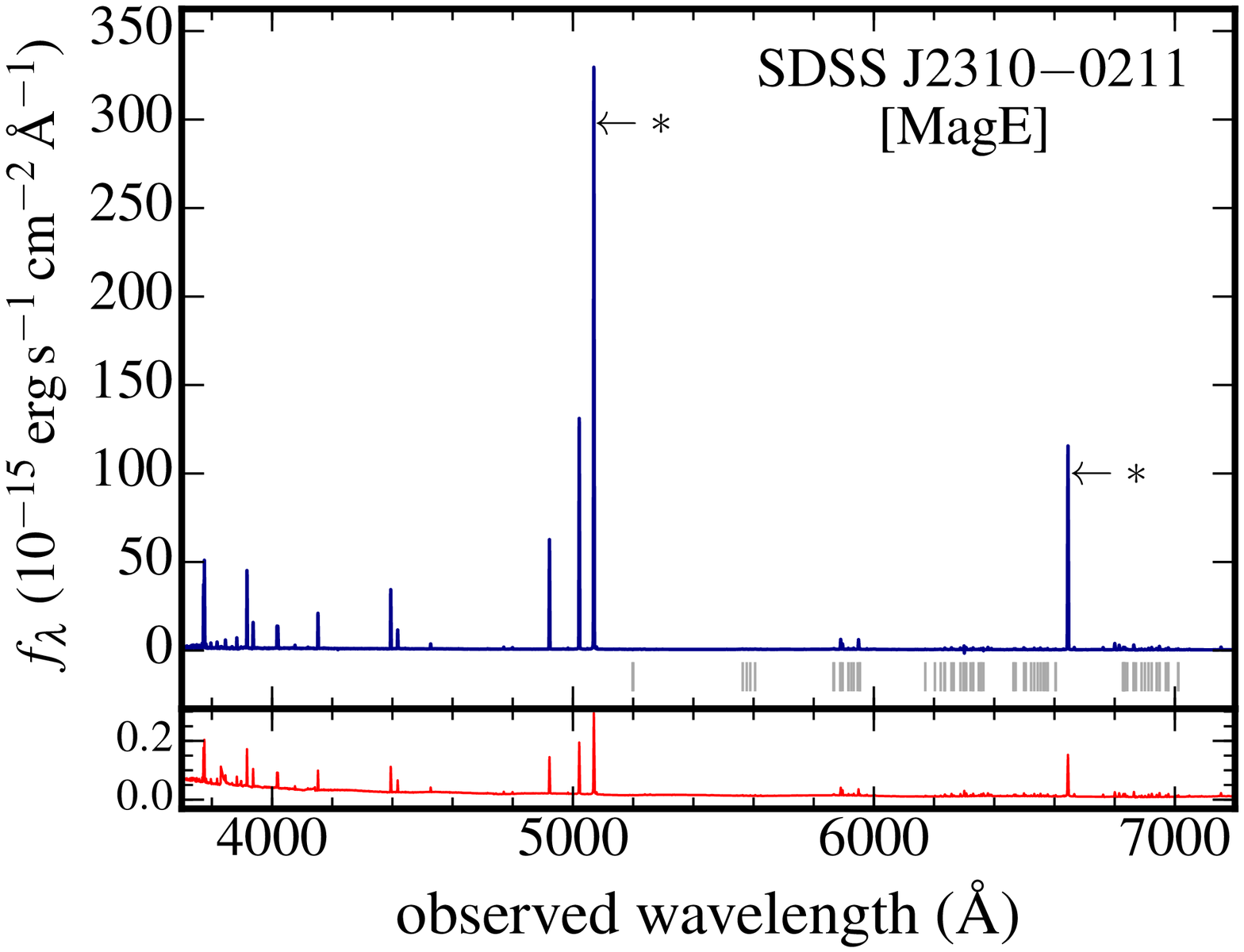}
\plotone{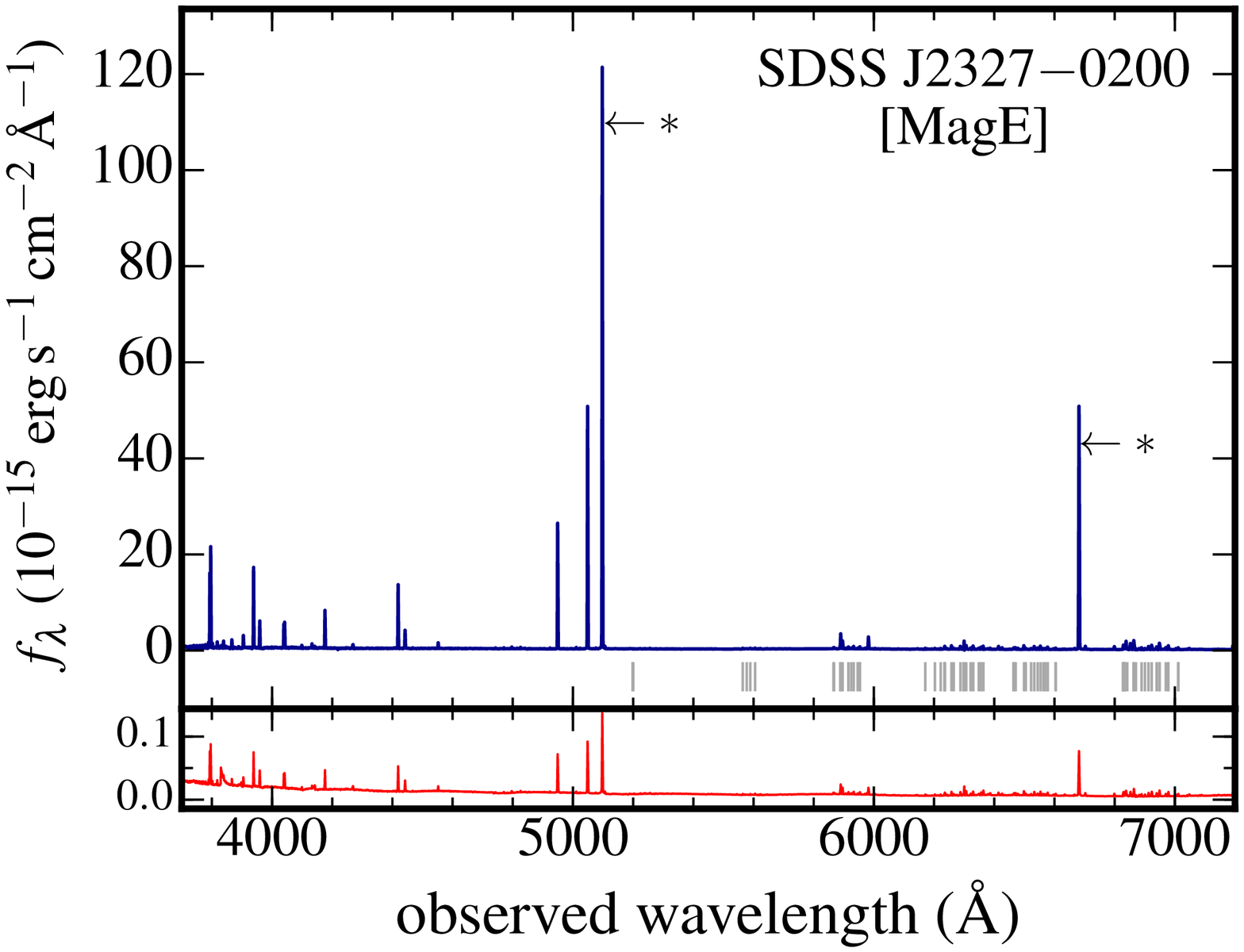}
\caption{Same as Figure \ref{fig:spec_hscempg}, but for our 6 SDSS EMPGs.
We indicate part of emission lines with asterisks that may be underestimated because of the saturation. 
The saturation depends on the strength of an emission line and its position in each spectral order of the echellette spectroscopy because an edge (a center) of each order has a low (high) sensitivity. 
\label{fig:spec_sdssempg}}
\end{figure*}

\begin{figure}[!htbp]
\epsscale{1.0}
\plotone{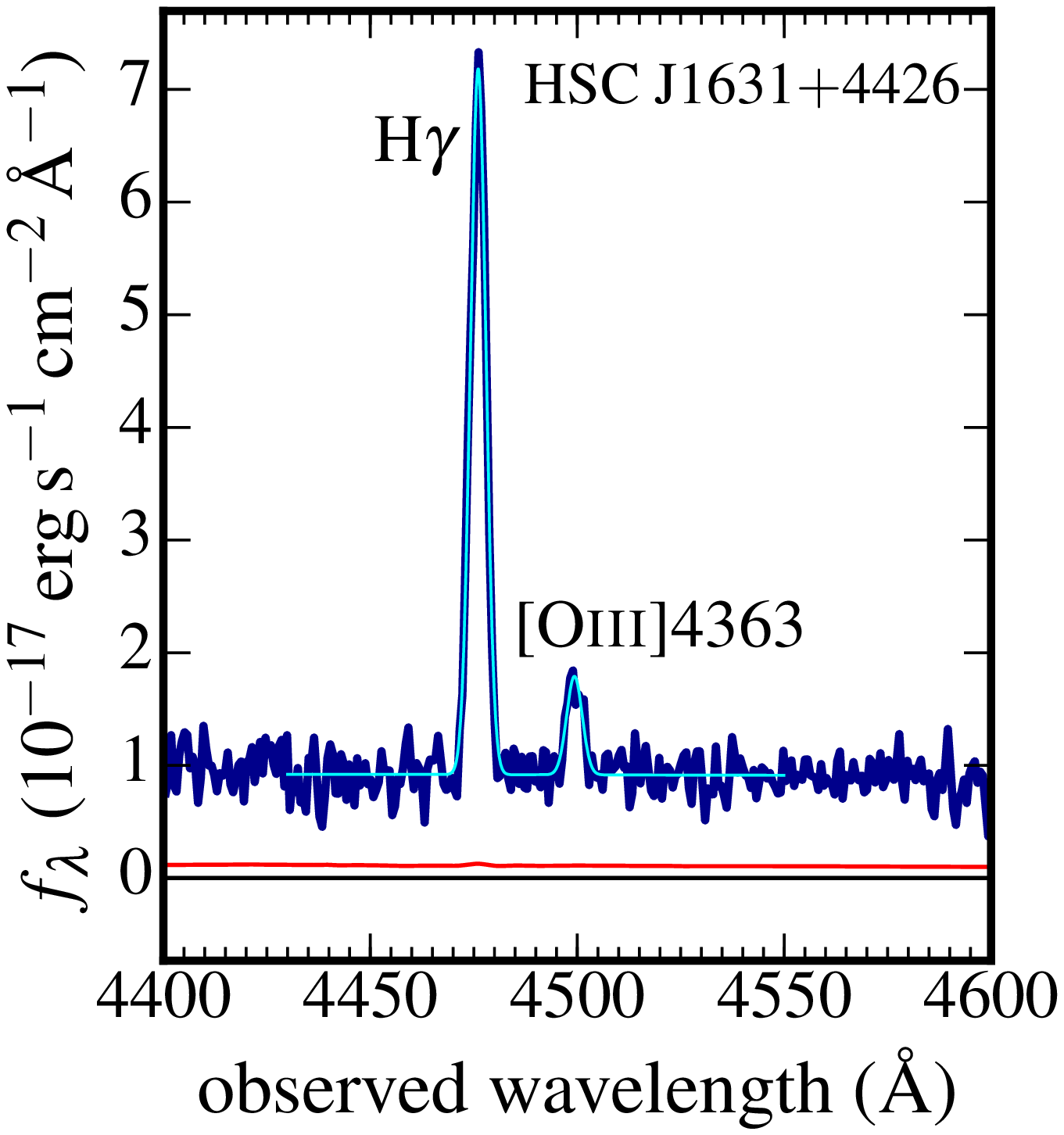}
\caption{Spectrum of {\HSCEMPGd} around an {\Hc} and an {\OIII}4363 emission lines.
\mod{The cyan curve is the best fit Gaussian functions of the {\Hc} and {\OIII}4363 emission lines.}
The red line shows noise levels at each wavelength.
\label{fig:spec_hscempg_zoom}}
\end{figure}

\subsection{LDSS-3 Data} \label{subsec:ldss3_rc}
We used the {\sc iraf} package to reduce and calibrate the data taken with LDSS-3 (Section \ref{subsec:ldss3}). 
The reduction and calibration processes include the bias subtraction, flat fielding, one-dimensional (1D) spectrum subtraction, sky subtraction, wavelength calibration, flux calibration, and atmospheric-absorption correction.
A one-dimensional spectrum was extracted from an aperture centered on the blue compact component of our EMPG candidates.
A standard star, CD-32\,\,9972 was used in the flux calibration.
The wavelengths were calibrated with the HeNeAr lamp.
Atmospheric absorption was corrected with the extinction curve at Cerro Tololo Inter-American Observatory (CTIO).
We used the CTIO extinction curve because Magellan Telescopes were located at Las Campanas Observatory, which neighbored the site of CTIO in Chili at a similar altitude.
We also calculate read-out noise and photon noise of sky$+$object emission on each CCD pixel, which are propagated to the 1D spectrum.

In our spectroscopy, a slit was not necessarily placed perpendicular to the horizon (i.e., at a parallactic angle), but instead chosen to include extended substructure in our EMPG candidates.
Thus, part of our spectra may have been affected by atmospheric refraction.
Because targets are acquired with an $R$-band camera in the LDSS-3 observation, red light falls on the center of the slit while blue light might drop out of the slit.
Thus, the atmospheric refraction can cause a wavelength-dependent slit loss.
To estimate the wavelength-dependent slit loss $SL(\lambda)$ carefully, we made a model of the atmospheric refraction.
We assumed the atmospheric refraction measured at La Silla in Chile \citep{Filippenko1982}, where the atmospheric condition was similar to Las Campanas in terms of the altitude and humidity.
The model took into consideration a parallactic angle, a slit position angle, an airmass, and a seeing size at the time of exposures.
An object size was broadened with a Gaussian convolution.
We assumed a wavelength dependence for the seeing size $\propto \lambda^{-0.2}$, where the seeing size was measured in $R$-band.
We integrated a model surface brightness $B(\lambda)$ on the slit to estimate an observed flux density  $F_{\lambda}^{obs}$ as a function of wavelength.
Then we estimated the $SL(\lambda)$ by comparing the observed flux density $F_{\lambda}^{obs}$ and total flux density $F_{\lambda}^{tot}$ predicted in the model.
\begin{equation}
SL(\lambda) = 1 - F_{\lambda}^{obs} / F_{\lambda}^{tot}
\end{equation}
Then we corrected the spectrum with SL($\lambda$) and obtained the slit-loss corrected spectrum.
The obtained SL values for {\HSCEMPGa} were SL(4,000{\AA})$=$1.74 and SL(7,000{\AA})$=$1.61, for example, giving a SL ratio of SL(4,000{\AA})/SL(7,000{\AA})$=$1.08.
The SL ratio suggests that emission line ratios were corrected up to $\sim$10$\%$ between 4,000 and 7,000{\AA}.
We estimated multiple color excesses {\EBV} from multiple pairs of Balmer lines and confirmed that these {\EBV} values were consistent between them within error bars.

\subsection{MagE Data} \label{subsec:mage_rc}
To reduce the raw data taken with MagE, we used the MagE pipeline from Carnegie Observatories Software Repository\footnote{\texttt{https://code.obs.carnegiescience.edu}}. 
The MagE pipeline has been developed on the basis of the {\tt Carpy} package \citep{Kelson2000, Kelson2003}.
The bias subtraction, flat fielding, scattered light subtraction, two-dimensional (2D) spectrum subtraction, sky subtraction, wavelength calibration, cosmic-ray removal, 1D-spectrum subtraction were conducted with the MagE pipeline.
Details of these pipeline processes are described on the web site of Carnegie Observatories Software Repository mentioned above. 
In the sky subtraction, we used a sky-line reference mask (i.e., a mask targeting a blank sky region with no object).
One-dimensional spectra were subtracted by summing pixels along the slit-length direction on a 2D spectrum.
The read-out noise and photon noise of sky$+$object emission are calculated on each CCD pixel, and propagated to the 1D spectrum.

We conducted the flux calibration with the standard star, Feige\,110, using {\sc iraf} routines.
Wavelengths were calibrated with emission lines of the ThAr lamp.
Spectra of each order were calibrated separately and combined with the weight of electron counts to generate a single 1D spectrum.
Atmospheric absorption was corrected in the same way as in Section \ref{subsec:ldss3_rc}.

In the MagE spectroscopy, we also placed a slit along a sub-structure of our EMPGs regardless of a parallactic angle. 
We also corrected the wavelength-dependent slit loss carefully in the same manner as the LDSS-3 spectroscopy described in Section \ref{subsec:ldss3_rc}.

\subsection{DEIMOS Data} \label{subsec:deimos_rc}
We used the {\sc iraf} package to reduce and calibrate the data taken with DEIMOS (Section \ref{subsec:deimos}). 
The reduction and calibration processes were the same as the LDSS-3 data explained in Section \ref{subsec:ldss3_rc}.
A standard star, G191B2B was used in the flux calibration.
Wavelengths were calibrated with the NeArKrXe lamp.
Atmospheric absorption was corrected under the assumption of the extinction curve at Mauna Kea Observatories.
It should be noted that part of flat and arc frame have been affected by stray light\footnote{As of September 2018, a cause of the stray light has not yet been identified according to a support astronomer at W. M. Keck Observatory (private communication). It is reported that the stray light pattern appears on a blue-side of CCD chips when flat and arc frames are taken with a grating tilted towards blue central wavelengths.}.
In our observation, a spectrum was largely affected in the wavelength range of $\lambda=$4,400--4,900 {\AA}.
Thus, we only used a spectrum within the wavelength range of $\lambda>$4,900 {\AA}, which was free from the stray light.
We ignore the effect of the atmospheric refraction here because we only use a red side ($\lambda>$4,900 {\AA}) of DEIMOS data, which is insensitive to the atmospheric refraction. 
We also confirm that the effect of the atmospheric refraction is negligible with the models described in Section \ref{subsec:ldss3_rc}.
We calculate noises in the same way as in \ref{subsec:ldss3_rc}.
In the DEIMOS data, we only used line flux ratios normalized to an {\Hb} flux.
Emission line fluxes were scaled with an {\Hb} flux by matching an {\Hb} flux obtained with DEIMOS to one obtained with FOCAS (see Section \ref{subsec:focas_rc}).
Note again that we have conducted spectroscopy for {\HSCEMPGd} both with DEIMOS and FOCAS.

\subsection{FOCAS Data} \label{subsec:focas_rc}
We used the {\sc iraf} package to reduce and calibrate the data taken with FOCAS (Section \ref{subsec:focas}). 
The reduction and calibration processes were the same as the LDSS-3 data explained in Section \ref{subsec:ldss3_rc}.
A standard star, BD$+$28\,4211 was used in the flux calibration.
Wavelengths were calibrated with the ThAr lamp.
Atmospheric absorption was corrected in the same way as in Section \ref{subsec:deimos_rc}.
Our FOCAS spectroscopy covered $\lambda\sim$3,800--5,250 {\AA}, which was complementary to the DEIMOS spectroscopy described in Section \ref{subsec:deimos_rc}, whose spectrum was reliable only in the range of  $\lambda>$4900 {\AA}.
We ignore the atmospheric refraction here because FOCAS is equipped with the atmospheric dispersion corrector.
We also calculate noises in the same manner as in \ref{subsec:ldss3_rc}. 
Because an {\Hb} line was overlapped in FOCAS and DEIMOS spectroscopy, we used an {\Hb} line flux to scale the emission line fluxes obtained in the DEIMOS observation (see Section \ref{subsec:deimos_rc}).\\

We show spectra of the 4 HSC-EMPG candidates and 6 SDSS-EMPG candidates obtained with LDSS-3, MagE, DEIMOS, and FOCAS spectrographs in Figures \ref{fig:spec_hscempg} and \ref{fig:spec_sdssempg}.
In the spectra of Figures \ref{fig:spec_hscempg} and \ref{fig:spec_sdssempg}, we find redshifted emission lines, confirming that these 10 candidates are real galaxies.
In Figure \ref{fig:spec_hscempg_zoom}, we also show a FOCAS spectrum of {\HSCEMPGd} around an {\OIII}4363 emission line.
The {\OIII}4363 emission line is detected significantly.

\section{ANALYSIS} \label{sec:analysis}
In this section, we explain the emission line measurement (Section \ref{subsec:line}) and the estimation of galaxy properties (Section \ref{subsec:prop}).
Here we estimate stellar masses, star-formation rates, stellar-ages, emission-line equivalent widths, electron temperatures, and metallicities of our 10 EMPG candidates confirmed in our spectroscopy.

\subsection{Emission Line Measurements} \label{subsec:line}
We measure central wavelengths and emission-line fluxes with a best-fit Gaussian profile using the {\sc iraf} routine, {\tt splot}.
In Sections \ref{subsec:ldss3_rc}--\ref{subsec:focas_rc}, we have calculated read-out noise and photon noise of sky$+$object emission on each CCD pixel and propagated them to 1D spectra.
We estimate flux errors from the 1D noise spectra with the same full width half maximum of emission lines.
As described in Section \ref{subsec:mage_rc}, we correct fluxes of the LDSS-3/MagE spectra assuming the wavelength-dependent slit-loss with the model of the atmospheric refraction.
We also include uncertainties of the slit-loss correction into the flux errors.
We measure observed equivalent widths (EWs) of emission lines with the same {\sc iraf} routine, {\tt splot} and convert them into the rest-frame equivalent widths (EW$_0$). 
Redshifts are estimated by comparing the observed central wavelengths and the rest-frame wavelengths in the air of strong emission lines.
Generally speaking, when the slit spectroscopy is conducted for a spatially-resolved object, one obtains a  spectrum only inside a slit, which may not represents an average spectrum of its whole system.
However, because our metal-poor galaxies have a size comparable to or slightly larger than the seeing size, our emission-line estimation represents an average of the whole system.
The sizes of our metal-poor galaxies will be discussed in Paper-III (Isobe et al. in prep.).

Color excesses, {\EBV} are estimated with the Balmer decrement of {\Ha}, {\Hb}, {\Hc}, {\Hd},..., and {H13} lines under the assumptions of the dust extinction curve given by \citet{Cardelli1989} and the case B recombination.
We do not use Balmer emission lines affected by a systematic error such as cosmic rays and other emission lines blending with the Balmer line.
In the case B recombination, we carefully assume electron temperatures ($T_\mathrm{e}$) so that the assumed electron temperatures become consistent with electron temperature measurements of O$^{2+}$, {\TeOIII}, which will be obtained in Section \ref{subsec:prop}.
We estimate the best {\EBV} values and their errors with the $\chi^2$ method \citep{Press2007}.
The {\EBV} estimation process is detailed as follows.
\begin{enumerate}
\item We predict Balmer emission line ratios based on the case B recombination models with an initial $T_\mathrm{e}$ guess selected from $T_\mathrm{e}$$=$10,000, 15,000, 20,000, or 25,000 K. 
We use the case B recombination models calculated with PyNeb \citep[][v1.0.14]{Luridiana2015}, a modern python tool to compute emission line emissivities based on the $n$-level atom/ion models. 
Here we fix an electron density of $n_\mathrm{e}$=100 cm$^{-3}$ for all of our 10 galaxies, which is roughly consistent with $n_\mathrm{e}$ measurements obtained in Section \ref{subsec:prop} (see also Table \ref{table:temp}).
Note that Balmer emission line ratios are insensitive to the electron density variance. 
\item We calculate $\chi^2$ from Balmer emission line ratios of our measurements and the case B models in a wide range of {\EBV}.
We find the best {\EBV} value, which gives the least $\chi^2$. 
We also obtain 16th and 84th percentiles of {\EBV} based on the $\chi^2$ calculation, and regard the 16th and 84th percentiles as {\EBV} errors.
\item We apply the dust correction to the emission line fluxes with the obtained {\EBV} value and the \citet{Cardelli1989} dust extinction curve.
\item We estimate electron temperatures {\TeOIII}, which will be described in Section \ref{subsec:prop}.
We then compare the {\TeOIII} estimates with the initial $T_\mathrm{e}$ guess to check the consistency between them.
\item Changing the initial $T_\mathrm{e}$ guess, we repeat the processes from 1) to 4) until we get a {\TeOIII} value roughly consistent with the initial $T_\mathrm{e}$ guess given in the process 1). 
\end{enumerate}
We eventually assume $T_{e}$=10,000 K ({\SDSSEMPGa} and {\SDSSEMPGb}), 15,000 K ({\HSCEMPGa}, {\HSCEMPGb}, {\SDSSEMPGd}, {\SDSSEMPGe}, and {\SDSSEMPGf}), 20,000 K ({\HSCEMPGc} and  {\SDSSEMPGc}), 25,000 K ({\HSCEMPGd}), which are roughly consistent with {\TeOIII} measurements.
In the flux estimation, we ignore the contribution of stellar atmospheric absorption around Balmer lines  because our galaxies have very large equivalent widths compared to the expected absorption equivalent width.
We summarize redshifts and dust-corrected fluxes in Tables \ref{table:mag}, \ref{table:flux}, and \ref{table:prop}.

We have confirmed consistency between observed emission-line fluxes and those estimated in the process of photometric SED fitting.
The photometric SED fitting will be detailed in Section \ref{subsec:prop}.

\begin{deluxetable*}{lcccccccccc}
\tablecolumns{11}
\tabletypesize{\scriptsize}
\tablecaption{R.A., Dec., redshifts, and photometric magnitudes of our targets %
\label{table:mag}}
\tablehead{%
\colhead{\#} &
\colhead{ID} &
\colhead{R.A.} &
\colhead{Dec.} &
\colhead{redshift} &
\colhead{$u$} &
\colhead{$g$} &
\colhead{$r$} & 
\colhead{$i$} & 
\colhead{$z$} & 
\colhead{$y$} \\
\colhead{} &
\colhead{} &
\colhead{(hh:mm:ss)} &
\colhead{(dd:mm:ss)} &
\colhead{} &
\colhead{(mag)} &
\colhead{(mag)} &
\colhead{(mag)} &
\colhead{(mag)} &
\colhead{(mag)} &
\colhead{(mag)} \\
\colhead{(1)} &
\colhead{(2)} &
\colhead{(3)} &
\colhead{(4)} &
\colhead{(5)} &
\colhead{(6)} &
\colhead{(7)} &
\colhead{(8)} &
\colhead{(9)} &
\colhead{(10)}  &
\colhead{(11)}  
}
\startdata
1  &  {\HSCEMPGa}   & 14:29:48.61 & $-$01:10:09.67 & 0.02980 & --- & 18.14 & 18.65 & 19.38 & 19.47 & 18.92  \\ 
2  &  {\HSCEMPGb}   & 23:14:37.55 & $+$01:54:14.27 & 0.03265 & --- & 21.94 & 21.95 & 22.76 & 22.57 & 22.38  \\ 
3  &  {\HSCEMPGc}   & 11:42:25.19 & $-$00:38:55.64 & 0.02035 & --- & 21.39 & 21.62 & 22.42 & 22.28 & 22.01   \\ 
4  &  {\HSCEMPGd}   & 16:31:14.24 & $+$44:26:04.43 & 0.03125 & --- & 21.84 & 21.88 & 22.52 & 22.75 & 22.39  \\ 
5  &  {\SDSSEMPGa} & 00:02:09.94 & $+$17:15:58.65 & 0.02083 & 18.48 & 17.61 & 18.05 & 18.61 & 18.57 & ---  \\  
6  &  {\SDSSEMPGb} & 16:42:38.45 & $+$22:33:09.09 & 0.01725 & 18.50 & 17.99 & 18.38 & 19.01 & 19.14 & ---  \\  
7  &  {\SDSSEMPGc} & 21:15:58.33 & $-$17:34:45.09 & 0.02296 & 19.59 & 18.49 & 19.00 & 19.67 & 19.57 & ---  \\  
8  &  {\SDSSEMPGd} & 22:53:42.41 & $+$11:16:30.62 & 0.00730 & 17.91 & 16.62 & 17.07 & 18.08 & 18.12 & ---  \\  
9  &  {\SDSSEMPGe} & 23:10:48.84 & $-$02:11:05.74 & 0.01245 & 18.12 & 17.19 & 17.46 & 17.97 & 18.02 & ---  \\  
10 & {\SDSSEMPGf}  & 23:27:43.69 & $-$02:00:55.89 & 0.01812 & 19.02 & 18.16 & 18.47 & 19.26 & 19.25 & ---    
\enddata
\tablecomments{
(1): Number.
(2): ID.
(3): R.A. in J2000.
(4): Dec. in J2000.
(5): Redshift. Typical uncertainties are $\Delta z \sim10^{-6}$.
(6)--(11): Magnitudes of $ugrizy$ broad-bands photometry. Photometry of our HSC-EMPGs is given with HSC {\tt cmodel} magnitudes, while we use SDSS {\tt model} magnitudes in the photometry of our SDSS-EMPG.
}
\end{deluxetable*}
\begin{deluxetable*}{lccccc}
\tablecolumns{5}
\tabletypesize{\scriptsize}
\tablecaption{Electron temperature and density %
\label{table:temp}}
\tablehead{%
\colhead{\#} &
\colhead{ID} &
\colhead{\TeOIII} &
\colhead{\TeOII} & 
\colhead{\neOII} \\
\colhead{} &
\colhead{} &
\colhead{(10$^{4}$\,K)} &
\colhead{(10$^{4}$\,K)} &
\colhead{(cm$^{-3}$)} \\
\colhead{(1)} &
\colhead{(2)} &
\colhead{(3)} &
\colhead{(4)} &
\colhead{(5)} 
}
\startdata
1 & {\HSCEMPGa} & $1.318^{+0.014}_{-0.013}$ & $0.909^{+0.015}_{-0.013}$ & ---  \\
2 & {\HSCEMPGb} & --- & ---  \\
3 & {\HSCEMPGc} & $1.492^{+0.044}_{-0.038}$ & ---$^\mathrm{a}$ & $109^{+20}_{-17}$\\
4 & {\HSCEMPGd} & $2.557^{+0.119}_{-0.110}$ & ---$^\mathrm{a}$ & ---\\
5 & {\SDSSEMPGa} & $1.179^{+0.006}_{-0.005}$ & ---$^\mathrm{a}$ & ---\\
6 & {\SDSSEMPGb} & $1.210\pm0.007$ & $0.879^{+0.007}_{-0.008}$ & $77^{+9}_{-6}$  \\
7 & {\SDSSEMPGc} & $1.807^{+0.008}_{-0.007}$ & $1.210^{+0.023}_{-0.016}$ & $29\pm10$  \\
8 & {\SDSSEMPGd} & $1.480^{+0.001}_{-0.003}$ & $1.333\pm0.008$ & $70^{+5}_{-3}$  \\
9 & {\SDSSEMPGe} & $1.632\pm0.003$ & $1.151\pm0.007$ & $109^{+3}_{-5}$  \\
10 & {\SDSSEMPGf} & $1.566^{+0.003}_{-0.002}$ & $1.260^{+0.010}_{-0.012}$ & $128^{+4}_{-6}$  
\enddata
\tablecomments{
(1): Number.
(2): ID.
(3), (4): Electron temperatures of O$^{2+}$ and O$^{+}$.
(5): Electron density of O$^{+}$.
{$^a$} The {\TeOII} values are obtained with the empirical relation of \citet{Campbell1986} and \citet{Garnett1992} because we cannot estimate {\TeOII} directly due to the non-detection of {\OII}7320,7330 emission lines.
}
\end{deluxetable*}

\begin{deluxetable*}{cccccccc}
\tablecolumns{11}
\tabletypesize{\scriptsize}
\tablecaption{Flux measurements %
\label{table:flux}}
\tablehead{%
\colhead{\#} &
\colhead{ID} &
\colhead{{\OII}3727} &
\colhead{{\OII}3729} &
\colhead{{\OII}$_\mathrm{tot}$} &
\colhead{H13} &
\colhead{H12} &
\colhead{H11} \\
\colhead{(1)} &
\colhead{(2)} &
\colhead{(3)} &
\colhead{(4)} &
\colhead{(5)} &
\colhead{(6)} &
\colhead{(7)} &
\colhead{(8)} 
}
\startdata
1 & {\HSCEMPGa} & --- & --- & 166.62$\pm$1.99 & --- & --- & --- \\
2 & {\HSCEMPGb} & $<$14.21 & $<$13.50 & $<$19.60 & --- & --- & --- \\
3 & {\HSCEMPGc} & 68.06$\pm$0.81 & 91.82$\pm$0.84 & 159.88$\pm$1.16 & 3.26$\pm$0.68 & 3.63$\pm$0.67 & 6.65$\pm$0.79 \\
4 & {\HSCEMPGd} & --- & --- & 50.12$\pm$2.66 & --- & --- & --- \\
5 & {\SDSSEMPGa} & 74.28$\pm$0.50 & 103.08$\pm$0.51 & 177.36$\pm$0.71 & --- & --- & --- \\
6 & {\SDSSEMPGb} & 103.35$\pm$0.72 & 149.41$\pm$0.76 & 252.75$\pm$1.05 & --- & --- & --- \\
7 & {\SDSSEMPGc} & 34.05$\pm$0.27 & 46.85$\pm$0.30 & 80.91$\pm$0.41 & 2.80$\pm$0.20 & 3.42$\pm$0.32 & 4.23$\pm$0.20 \\
8 & {\SDSSEMPGd} & 39.44$\pm$0.11 & 55.06$\pm$0.12 & 94.50$\pm$0.16 & 2.36$\pm$0.06 & 3.40$\pm$0.05 & 4.08$\pm$0.05 \\
9 & {\SDSSEMPGe} & 43.70$\pm$0.10 & 59.12$\pm$0.12 & 102.82$\pm$0.16 & 2.38$\pm$0.06 & 2.75$\pm$0.06 & 3.78$\pm$0.06 \\
10 & {\SDSSEMPGf} & 45.07$\pm$0.11 & 59.96$\pm$0.12 & 105.03$\pm$0.16 & 2.48$\pm$0.06 & 3.36$\pm$0.06 & 3.99$\pm$0.08 \\
\hline\hline 
\# & H10 & H9 & H7 & H{$\delta$} & H{$\gamma$} & {\OIII}4363 & H{$\beta$} \\ 
 (1) & (9) & (10) & (11) & (12) & (13) & (14) & (15) \\ \hline 
 1 & --- & 5.45$\pm$0.68 & 32.48$\pm$0.48 $^\mathrm{a}$& 21.64$\pm$0.33 & 45.55$\pm$0.28 & 9.06$\pm$0.23 & 100.00$\pm$0.24 \\
2 & --- & --- & --- & 26.15$\pm$3.11 & 46.56$\pm$1.67 & --- & 100.00$\pm$1.00 \\
3 & 4.39$\pm$0.66 & 6.13$\pm$0.63 & 17.16$\pm$0.54 & 27.36$\pm$0.51 & 48.20$\pm$0.43 & 5.96$\pm$0.39 & 100.00$\pm$0.52 \\
4 & --- & 4.68$\pm$1.17 & 20.03$\pm$0.87 $^\mathrm{a}$& 27.53$\pm$0.65 & 46.88$\pm$0.50 & 8.18$\pm$0.48 & 100.00$\pm$0.37 \\
5 & 5.99$\pm$0.30 & 8.42$\pm$0.25 & 16.27$\pm$0.19 & 26.04$\pm$0.17 & 46.92$\pm$0.14 & 6.38$\pm$0.09 & 100.00$\pm$0.16 \\
6 & --- & 6.74$\pm$0.33 & 15.27$\pm$0.24 & 26.44$\pm$0.20 & 44.90$\pm$0.16 & 6.59$\pm$0.10 & 100.00$\pm$0.17 \\
7 & 4.40$\pm$0.19 & 5.88$\pm$0.18 & 15.54$\pm$0.16 & 26.12$\pm$0.16 & 46.99$\pm$0.16 & 13.94$\pm$0.11 & 100.00$\pm$0.19 \\
8 & 5.59$\pm$0.05 & 7.59$\pm$0.05 & 16.12$\pm$0.06 & 25.67$\pm$0.05 & 46.52$\pm$0.06 & 14.13$\pm$0.04 & 100.00$\pm$0.07 \\
9 & 5.72$\pm$0.07 & 7.49$\pm$0.06 & 16.38$\pm$0.06 & 26.66$\pm$0.06 & 48.57$\pm$0.07 & 14.85$\pm$0.04 & 100.00$\pm$0.09 \\
10 & 5.29$\pm$0.06 & 7.19$\pm$0.06 & 17.45$\pm$0.07 & 25.93$\pm$0.07 & 47.39$\pm$0.08 & 12.81$\pm$0.05 & 100.00$\pm$0.11 \\
\hline\hline 

\# & {\OIII}4959 & {\OIII}5007 & H{$\alpha$} & {\NII}6584 & {\OII}7320 & {\OII}7330 \\ 
 (1) & (16) & (17) & (18) & (19) & (20) & (21) \\ \hline 
 1 & 210.62$\pm$0.29 & 626.89$\pm$0.46 & 246.66$\pm$0.20 & 5.08$\pm$0.18 & 1.45$\pm$0.07 & 0.92$\pm$0.07 \\ 
2 & 69.57$\pm$0.72 & 207.48$\pm$0.88 & 278.45$\pm$0.66 & 2.10$\pm$0.29 & --- & --- \\ 
3 & 102.76$\pm$0.47 & 308.14$\pm$0.65 & 272.09$\pm$0.57 & 8.64$\pm$0.26 & --- & --- \\ 
4 & 55.76$\pm$0.34 & 170.92$\pm$0.38 & 229.46$\pm$1.00 & $<$0.48 & --- & --- \\ 
5 & 196.65$\pm$0.19 & 593.18$\pm$0.32 & 280.48$\pm$0.17 & 5.68$\pm$0.04 & --- & --- \\ 
6 & 183.24$\pm$0.21 & 571.59$\pm$0.34 & 276.34$\pm$0.20 & 5.28$\pm$0.04 & 1.84$\pm$0.04 & 1.47$\pm$0.04 \\ 
7 & 165.47$\pm$0.22 & --- & 278.90$\pm$0.22 & 3.11$\pm$0.04 & 1.09$\pm$0.04 & 0.83$\pm$0.04 \\ 
8 & 250.60$\pm$0.11 & --- & --- & 3.29$\pm$0.01 & 1.55$\pm$0.01 & 1.03$\pm$0.01 \\ 
9 & 214.32$\pm$0.12 & --- & --- & 2.76$\pm$0.02 & 1.26$\pm$0.02 & 0.98$\pm$0.02 \\ 
10 & 200.95$\pm$0.14 & --- & --- & 3.21$\pm$0.02 & 1.69$\pm$0.02 & --- 
\enddata
\tablecomments{
(1): Number.
(2): ID.
(3)--(21): Dust-corrected emission-line fluxes normalized to an {\Hb} line flux in the unit of {\ergscm}. 
Upper limits are given with a 1$\sigma$ level. 
Lines suffering from saturation or affected by sky emission lines are also shown as no data here.
{\OII}$_\mathrm{tot}$ represents a sum of {\OII}3727 and {\OII}3729 fluxes. 
If the spectral resolution is not high enough to resolve {\OII}3727 and {\OII}3729 lines, we only show {\OII}$_\mathrm{tot}$ fluxes.
\tablenotetext{a}{A sum of {\NeIII}3867 and H7 fluxes because they are blended due to the low spectral resolution.}
}
\end{deluxetable*}
\begin{turnpage}
\begin{deluxetable*}{lcccccccccccccccc}
\tablecolumns{15}
\tabletypesize{\scriptsize}
\tablecaption{Parameters of our metal-poor galaxies %
\label{table:prop}}
\tablehead{%
\colhead{\#} &
\colhead{ID} &
\colhead{EMPG?} &
\colhead{$F$(H$\beta$)} &
\colhead{{\EW}(H$\beta$)} &
\multicolumn{2}{c}{12+log(O/H)}&
\colhead{$\log$(M$_{\star}$)} &
\colhead{$\log$(SFR)} &
\colhead{{\EBV}} &
\colhead{age} &
\colhead{$\sigma$} &
\colhead{D$_\mathrm{near}$}  \\
\colhead{} &
\colhead{} &
\colhead{} &
\colhead{(\ergscm)} &
\colhead{({\AA})} &
\colhead{direct} &
\colhead{empirical} &
\colhead{(M$_{\odot}$)} &
\colhead{({M$_{\odot}$\,yr$^{-1}$})} &
\colhead{(mag)} &
\colhead{(Myr)} &
\colhead{(km\,s$^{-1}$)} & 
\colhead{(Mpc)} \\
\colhead{(1)} &
\colhead{(2)} &
\colhead{(3)} &
\colhead{(4)} &
\colhead{(5)} &
\colhead{(6)} &
\colhead{(7)} &
\colhead{(8)} &
\colhead{(9)} &
\colhead{(10)} & 
\colhead{(11)} & 
\colhead{(12)} & 
\colhead{(13)} 
}
\startdata
1  &  {\HSCEMPGa}   & {\it no} & $67.1 \pm 2.2$ & $172.6^{+0.7}_{-0.6}$ & $8.27\pm0.02$ & --- & $6.55^{+0.13}_{-0.09}$ & $0.426\pm0.013$ & $0.35\pm0.02$ & 3.4 & --- & 1.56 \\ 
2  &  {\HSCEMPGb}  & {\bf \it yes} & $2.61 \pm 0.10$ & $213.3^{+23.4}_{-17.6}$ & --- & $7.225^{+0.027}_{-0.024}$ & $5.17\pm0.01$ & $-0.851\pm0.013$ & $0.28\pm0.03$ & 4.1 & $21.6^{+0.8}_{-7.7}$ & 2.25 \\ 
3  &  {\HSCEMPGc}   & {\it no} & $4.27 \pm 0.15$ & $111.9^{+1.4}_{-1.3}$ & $7.72\pm0.03$ & --- & $4.95^{+0.04}_{-0.01}$ & $-1.066\pm0.013$ & $0.00^{+0.02}_{-0.00}$ & 3.7 & $21.9^{+0.8}_{-7.6}$ & 2.28 \\ 
4  &  {\HSCEMPGd}  & {\it yes} & $1.31 \pm 0.04$  & $123.5^{+3.5}_{-2.8}$ & $6.90\pm0.03$ & $7.175^{+0.006}_{-0.005}$ & $5.89^{+0.10}_{-0.09}$ & $-1.276\pm0.013$ & $0.19\pm0.03$ &  50 & --- & 1.54 \\ 
5  &  {\SDSSEMPGa}  & {\it no} & $45.7 \pm 1.4$ & $103.9\pm0.2$ & $8.22\pm0.01$ & --- & $7.06\pm0.03$ & $-0.002\pm0.013$ & $0.00^{+0.01}_{-0.00}$ & 31 & $27.7^{+0.9}_{-5.9}$ & 3.05 \\  
6  &  {\SDSSEMPGb}  & {\it no} & $46.3 \pm 1.5$ & $153.7^{+0.5}_{-0.4}$ & $8.45\pm0.01$ & --- & $6.06^{+0.03}_{-0.13}$ & $-0.169\pm0.013$ & $0.02\pm0.02$ & 25 & $29.8^{+1.0}_{-5.6}$ & 0.49 \\  
7  &  {\SDSSEMPGc}  & {\it yes} & $69.9 \pm 2.1$ & $214.0^{+0.9}_{-0.8}$ & $7.68\pm0.01$ & --- & $6.56\pm0.02$ & $0.266\pm0.013$ & $0.17\pm0.04$ & 21 &  $28.6^{+0.9}_{-5.8}$ & 17.69 \\  
8  &  {\SDSSEMPGd}  & {\it no} & $139 \pm 4.28$ & $264.7\pm0.3$ & $7.973\pm0.002$ & --- & $5.78\pm0.01$ & $-0.541\pm0.013$ & $0.00^{+0.01}_{-0.00}$ & 4.1 & $18.9^{+0.7}_{-9.4}$ & 14.33 \\  
9  &  {\SDSSEMPGe}  & {\it no} & $99.3 \pm 3.1$ & $127.6\pm0.2$ & $7.890^{+0.003}_{-0.004}$ & --- & $6.99\pm0.03$ & $-0.155\pm0.013$ & $0.01^{+0.02}_{-0.01}$ & 51 & $12.6^{+0.5}_{-12.6}$ & 2.15 \\  
10 & {\SDSSEMPGf}  & {\it no} & $40.7 \pm 1.2$ & $111.0\pm0.2$ & $7.866^{+0.004}_{-0.005}$ & --- & $6.51^{+0.02}_{-0.03}$ & $-0.180\pm0.013$ & $0.00^{+0.02}_{-0.00}$ & 22 & $12.0^{+0.5}_{-12.0}$ & 4.00   
\enddata
\tablecomments{
(1): Number.
(2): ID.
(3): Whether or not an object satisfies the EMPG definition, {\metal}$<$7.69. If yes (no), we write {\it yes} ({\it no}) in the column.
(4): {\Hb} emission line flux normalized in the unit of 10$^{-15}$\,{\ergscm}, which is corrected for the slit loss and the dust extinction.
These errors include 3\% systematic uncertainties caused by the absolute flux calibration \citep[e.g.,][]{Oke1990}.
(5): Rest-frame equivalent width of an {\Hb} emission line.
(6)--(7): Gas-phase metallicity obtained with the direct method and the empirical relation of \citet{Izotov2019b}.  
(8): Stellar mass.
(9): Star-formation rate. 
(10): Color excess.
(11): Maximum stellar age.
(12): Velocity dispersion obtained from an {\Hb} emission line. 
An instrumental velocity dispersion is already removed.
Note that the emission line broadening from a galaxy rotation is not eliminated.
(13): Distance to the nearest neighborhood selected in the SDSS DR13 spectroscopic catalog.
}
\end{deluxetable*}
\end{turnpage}

\subsection{Galaxy Properties} \label{subsec:prop}
We estimate stellar masses, maximum stellar ages, star-formation rates (SFRs), electron densities ($n_{\mathrm e}$), electron temperatures ($T_{\mathrm e}$) and gas-phase metallicities (O/H) of our EMPG candidates.
The SFR, $n_{\mathrm e}$, $T_{\mathrm e}$, and O/H estimates and their errors are obtained with the Monte Carlo simulation.
For each of the 10 galaxies, we generate 1,000 sets of emission line fluxes based on the emission line flux measurements and errors obtained in Section \ref{subsec:line}.
Here we assume that flux measurement errors approximately follow the Gaussian profile.
The 16th and 84th percentiles in the SFR, $n_{\mathrm e}$, $T_{\mathrm e}$, and O/H distributions are regarded as lower and upper errors of SFR, $n_{\mathrm e}$, $T_{\mathrm e}$, and O/H. 

We estimate stellar masses and stellar ages of our EMPG candidates with the SED interpretation code \textsc{beagle} \citep{Chevallard2016}.
The constant star-formation history is assumed in the model. 
We run the \textsc{beagle} code with 4 free parameters of stellar mass, maximum stellar age, ionization parameter, and metallicity, while we fix a redshift determined in our spectroscopy.
In the \textsc{beagle} models, both stellar and nebular emission is calculated.
Note that, if the nebular emission is not included in the SED models, the stellar continuum (i.e., stellar mass) can be overestimated.
We confirm that emission line fluxes estimated by the \textsc{beagle} codes are consistent with the observed fluxes in our spectroscopy, suggesting that our stellar mass estimates are not overestimated for the reason above.
We assume dust free conditions to reduce calculation time, but we note that when dust extinction is added as a free parameter in a rough parameter estimation, their values are approximately zero.
Finally, we obtain estimates of stellar mass and maximum stellar age in the range of {\mass}=4.95--7.06 and $t_\mathrm{age,max}$=3.4--51 Myr.

SFRs are estimated with the dust-corrected {\Ha} fluxes under the assumption of the star-formation history of \citet{Kennicutt1998}.
Here we assume that the {\Ha} emission line is dominantly contributed by the photoionization caused by ionizing photon radiation from massive stars. 
If the {\Ha} line is saturated, we use afn {\Hb} line instead.
The estimated SFRs of our EMPGs range {\sfr}=($-1.28$)--0.43.

We estimate electron densities and electron temperatures with a PyNeb package, \texttt{getTemDen}.
We use {\OII}3727/3729 flux ratios to estimate the electron densities of O$^{+}$ ions, $n_{\mathrm e}$({\sc{Oii}}), obtaining $n_{\mathrm e}$({\sc{Oii}})$=$29--128 cm$^{-3}$ for our 7 galaxies.
We use line ratios of {\OIII}4363/5007 and {\OII}(3727+3729)/(7320+7330) to estimate the electron temperatures coupled with O$^{++}$ ions, $T_{\mathrm e}$({\sc{Oiii}}) and O$^{+}$ ions, $T_{\mathrm e}$({\sc{Oii}}), respectively.
In the electron temperature estimation, we assume $n_\mathrm{e}=100$ cm$^{-3}$, which is consistent with our $n_\mathrm{e}$ measurements.
To estimate $T_{\mathrm e}$({\sc{Oiii}}) of {\HSCEMPGd}, we use a PyNeb package, \texttt{getEmissivity} instead of \texttt{getTemDen} because {\HSCEMPGd} shows a high electron temperature.
Both the \texttt{getTemDen} and \texttt{getEmissivity} packages provide the same relation between $T_{\mathrm e}$({\sc{Oiii}}) and {\OIII}4363/5007.
\mod{The \texttt{getTemDen} package can be used only below $\sim$25,120 K because the {\sc{Oiii}} collision strengths used in \texttt{getTemDen} \citep{Storey2014} have been calculated in the range of $\sim$100--25,000 K, while the \texttt{getEmissivity} package is effective up to 200,000 K \citep[][]{Aggarwal1999, Luridiana2015}.}
If an {\OIII}5007 line is saturated, we estimate an {\OIII}5007 flux with {\OIII}5007$=$2.98$\times${\OIII}4959, which is strictly determined by the Einstein {\it A} coefficient.
If either of {\OII}7320 or {\OII}7330 line is detected, we estimate a total flux of {\OII}(7320+7330) with a relation of {\OII}7330$=$0.56$\times${\OII}7320.
Using PyNeb, we have confirmed that the {\OII} relation above holds with very little dependence on $T_\mathrm{e}$ and $n_\mathrm{e}$.
If none of {\OII}7320 and {\OII}7330 is detected, we estimate {\TeOII} from an empirical relation of  {\TeOII}=0.7$\times${\TeOIII}$+$3000 \citep{Campbell1986, Garnett1992}.
The estimates of electron densities and electron temperatures are summarized in Table \ref{table:temp}.

We also estimate $T_{\mathrm e}$-based metallicities with {\OIII}5007/H$\beta$ and {\OII}3727,3729/H$\beta$ flux ratios and electron temperatures of $T_{\mathrm e}$({\sc{Oiii}}) and $T_{\mathrm e}$({\sc{Oii}}), using a PyNeb package, \texttt{getIonAbundance}.
Here we again assume $n_\mathrm{e}=100$ cm$^{-3}$, which is consistent with our $n_\mathrm{e}$ measurements.
We obtain $T_{\mathrm e}$-based metallicities ranging from 12+log(O/H)$=$6.90 to 8.45. 
\mod{Note that metallicity estimates are sensitive to the electron temperature. 
For example, when the electron temperature increases by 1,000 K, the metallicity decreases by 0.03 dex at 12+log(O/H)$\sim$6.90.}
Because an {\OIII}4363 emission line is not detected in the spectrum of {\HSCEMPGb}, we do not estimate a $T_{\mathrm e}$-based metallicity of {\HSCEMPGb}.
For comparison, we also estimate metallicities of {\HSCEMPGb} and {\HSCEMPGd} with the empirical relation obtained from metal-poor galaxies by \citet{Skillman1989c}. 
This empirical relation is calibrated with emission line indices of $R_{23}$ ($\equiv$({\OII}3727+{\OIII}4959,5007)/{\Hb}).
\citet{Izotov2019b} have confirmed that the \citet{Skillman1989c} calibration well reproduces metallicities of the famous EMPGs, J0811$+$4730 \citep[][]{Izotov2018a}, SBS0335$-$052 \citep[e.g.,][]{Izotov2009}, LittleCub \citep[][]{Hsyu2017}, and DDO68 \citep[][]{Pustilnik2005, Annibali2019}.
This empirical relation is applicable in the low metallicity range of {\metal}$\lesssim$7.3, which corresponds to $\log$($R_{23}$)$\lesssim$0.5.
Because our galaxies except for {\HSCEMPGb} and {\HSCEMPGd} do not satisfy $\log$($R_{23}$)$\lesssim$0.5, we do not estimate metallicities of the other 8 galaxies with the empirical relation.
The estimates of stellar masses, ages, star-formation rates, and gas-phase metallicities are summarized in Table \ref{table:prop}.

\mod{One may be interested in the metallicity of {\HSCEMPGd}, which gives the lowest metallicity in our sample.
In Figure \ref{fig:spec_hscempg_zoom}, we show a FOCAS spectrum of {\HSCEMPGd} around an {\OIII}4363 emission line.
The {\OIII}4363 is a key emission line in the metallicity estimation.
As shown in Figure \ref{fig:spec_hscempg_zoom}, we significantly detect the {\OIII}4363 emission line, whose  signal-to-noise ratio (S/N) is 17.0.
To double check the reliability of the {\OIII}4363 detection, we also estimate the {\OIII}4363 flux by integrating a continuum-subtracted spectrum around the {\OIII}4363 emission line, and compare it with the {\OIII}4363 flux obtained by the Gaussian fitting (Section \ref{subsec:line} and Table \ref{table:flux}). 
We find that a ratio of fluxes obtained by the Gaussian fitting and the integration is 1.014. 
The 1.4\% uncertainty between the Gaussian fitting and the integration is smaller than the {\OIII}4363 flux error  (1/17.0$\times$100$=$5.9\%) shown in Table \ref{table:flux}.
We have confirmed above that the {\OIII}4363 flux has been reliably calculated in the two independent method.}

\section{RESULTS and DISCUSSIONS} \label{sec:results}
In Section \ref{subsec:identification}, we describe the results of the object class identification for our HSC-EMPG and SDSS-EMPG candidates and show the distribution of {\EW}({\Hb}) and metallicity to characterize our sample.
We also investigate the cosmic number density of our metal poor galaxies (Section \ref{subsec:density}) and their environment (Section \ref{subsec:environment}).
We show the stellar mass and SFR ($M_{\star}$-$SFR$) and the stellar-mass and metallicity ($M_{\star}$-$Z$) relations of our EMPG candidates in Sections \ref{subsec:msfr} and \ref{subsec:mz}.
In Section \ref{subsec:bpt}, we discuss the possibility of the AGN/shock contribution on the diagram of {\NII}/{\Ha} and {\OIII}/{\Hb} emission line ratios, so-called the Baldwin-Phillips-Terlevich diagram \citep[BPT diagram,][]{Baldwin1981}. 
The velocity dispersions of our sample is presented and discussed in Section \ref{subsec:velocity}.

\subsection{Object Class Identification} \label{subsec:identification}
As described in Section \ref{sec:spectroscopy}, we conducted spectroscopy for 4 out of 27 HSC-EMPG candidates and 6 out of 86 SDSS-EMPG candidates.
We find that all of the observed 10 EMPG candidates are confirmed as real galaxies with strong emission lines.
We show spectra of the 4 HSC-EMPG candidates and 6 SDSS-EMPG candidates that exhibit strong emission lines in Figures \ref{fig:spec_hscempg} and \ref{fig:spec_sdssempg}.
Two spectra are shown for {\HSCEMPGd} because we have conducted spectroscopy both with Keck/DEIMOS and Subaru/FOCAS for this object.

\begin{figure}
\epsscale{1.2}
\plotone{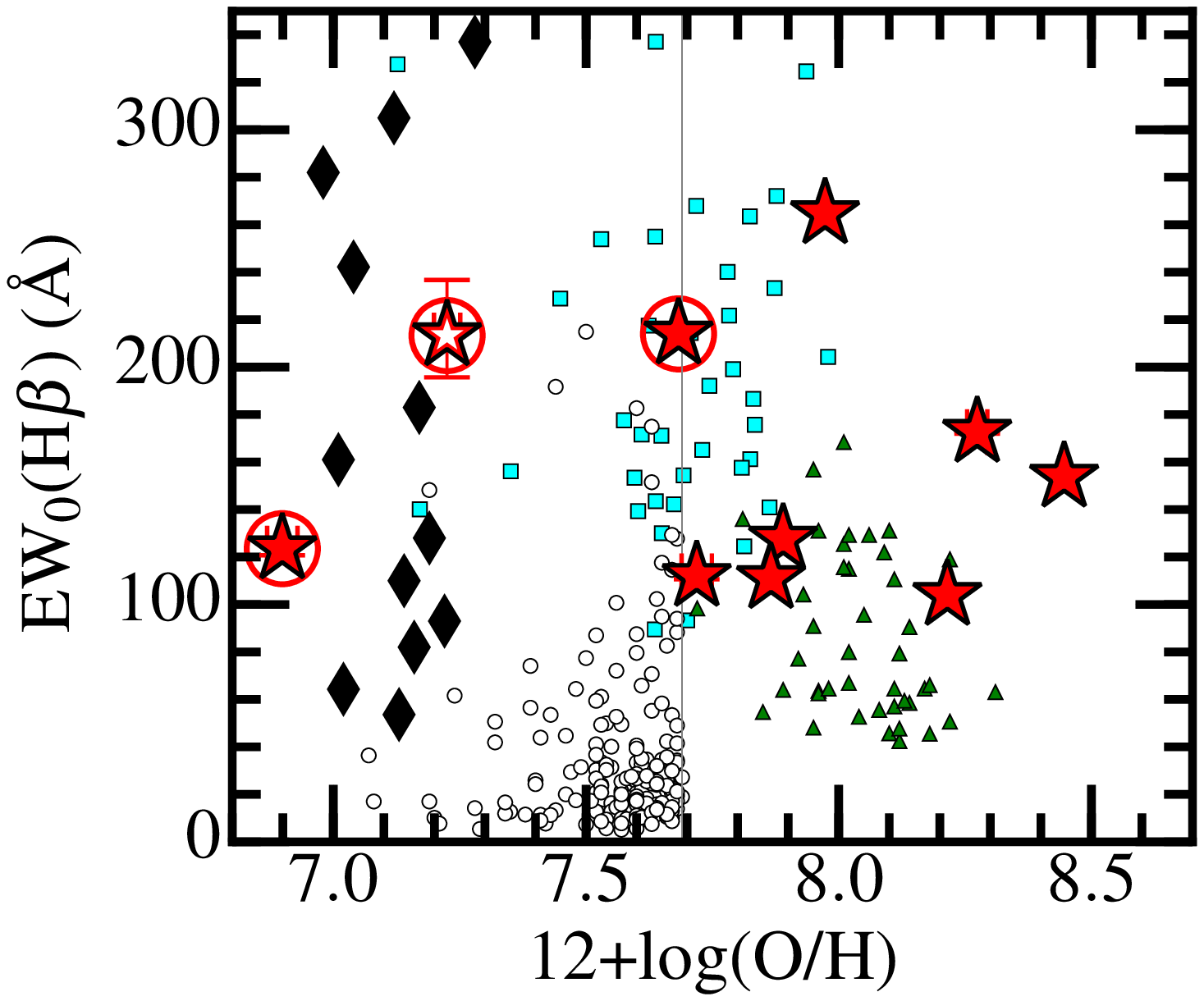}
\caption{{\EW}({\Hb}) as a function of metallicity of our metal-poor galaxies from HSC-EMPG and SDSS-EMPG  source catalogs (red stars).
The solid line indicates the EMPG criterion given in this paper,  {\metal}$<$7.69.
Galaxies that satisfy the EMPG condition in our metal poor galaxy sample are marked with a large circle.
The open star indicates a galaxy whose metallicity is obtained with the empirical relation of \citet{Izotov2019b}, not with the direct method because of the non-detection of {\OIII}4363 and {\OII}7320,7330 emission lines (Section \ref{subsec:prop}). 
We also present GPs \citep[][green triangle]{Yang2017a}, BBs \citep[][cyan square]{Yang2017b}, and metal-poor galaxies \citep[][open circle]{SanchezAlmeida2016} from the literature for comparison.
With diamonds, we show representative metal-poor galaxies (or clumps of them), J0811$+$4730 \citep[][]{Izotov2018a}, SBS0335$-$052 \citep[e.g.,][]{Izotov2009}, AGC198691 \citep[][]{Hirschauer2016}, J1234$+$3901 \citep[][]{Izotov2019a}, LittleCub \citep[][]{Hsyu2017}, DDO68 \citep[][]{Pustilnik2005, Annibali2019}, IZw18 \citep[e.g.,][]{Izotov1998a, Thuan2005a}, and LeoP \citep[][]{Skillman2013} of {\metal}$\sim$7.0--7.2.
\label{fig:ew_metal}}
\end{figure}

Figure \ref{fig:ew_metal} shows the distribution of metallicity and {\EW}({\Hb}) of our EMPG candidates (red stars).
We find that our sample covers a wide range of metallicities, {\metal}=6.9--8.5 (i.e., 0.02--0.6 Z$_{\odot}$) and that 3 out of our 10 candidates satisfy the EMPG criterion of {\metal}$<$7.69, while the other 7 candidates do not satisfy the criterion.
Remarkably, {\HSCEMPGd} has a metallicity of {\metal}=6.90$\pm0.03$ (i.e., 0.016 Z$_{\odot}$), which is one of the lowest metallicities reported ever.
We also find that 2 out of the 3 EMPGs, {\HSCEMPGb} and {\HSCEMPGd}, are selected from the HSC data and have $i$-band magnitudes of 22.8 and 22.5 mag.
We argue that these 2 faint EMPGs are selected thanks to the deep HSC data, which suggests that the deep HSC data are advantageous to select very faint EMPGs.
It should be also noted that the other 7 galaxies out of the EMPG definition still show a low metallicity ($Z/Z_{\odot}\sim0.1$--$0.3$).
We expect to find more EMPGs from the HSC-EMPG catalog in our future spectroscopy. 
This is because the pilot spectroscopy has targeted relatively bright HSC sources ($\sim$22 mag) and the our future spectroscopy will target fainter HSC sources down to $\sim$24.3 mag, which are expected to have lower metallicity.

\begin{figure}
\epsscale{1.2}
\plotone{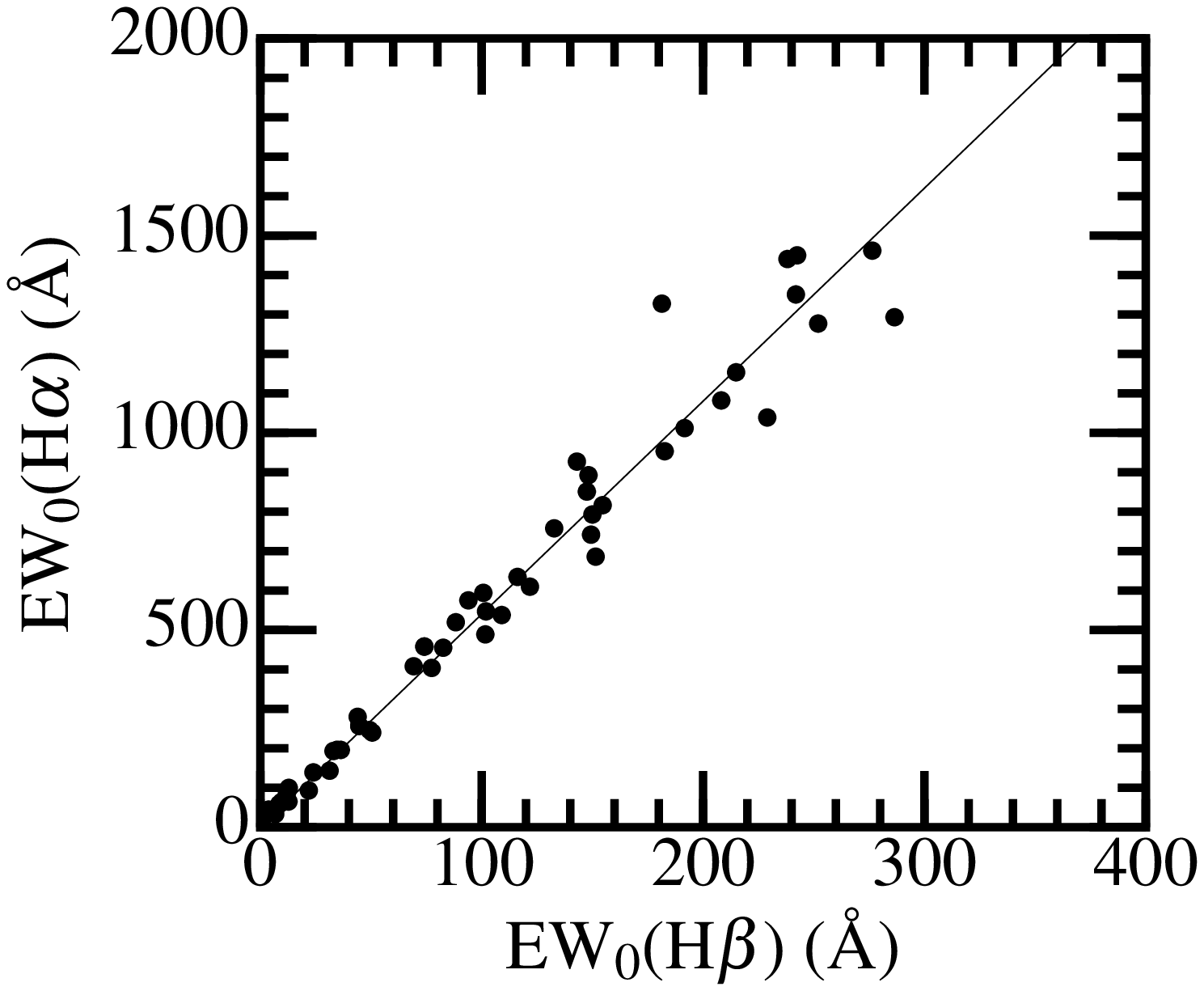}
\caption{Relation between rest-frame equivalent widths of H$\alpha$ and H$\beta$ for metal-poor galaxies in the literature \citep{Kunth2000, Kniazev2003, Guseva2007, Izotov2007, Izotov2009, Pustilnik2010, Izotov2012, Pilyugin2012, SanchezAlmeida2016, Guseva2017}.
The best least-square fit is shown with a solid line, which is {\EW}({\Ha})=5.47$\times${\EW}({\Hb}).
Because metal-poor galaxies are less dusty, a flux ratio of $F$({\Ha})/$F$({\Hb}) becomes almost constant ($\sim$2.7--3.0, determined by the case B recombination) in the most case.
In addition, a ratio of continuum level at {\Ha} and {\Hb}, $f_{\lambda, 0}$(6563 {\AA})/$f_{\lambda, 0}$(4861 {\AA}) always becomes $\sim$0.5 because the continuum slope differs little among metal-poor galaxies at wavelength of $\lambda$$>$4000 {\AA}.
Thus, the tight relation between {\EW}({\Ha}) and {\EW}({\Hb}) is only applicable to metal-poor galaxies.
\label{fig:EWHa_EWHb}}
\end{figure}

In Figure \ref{fig:ew_metal}, we also show GPs \citep[][green triangle]{Yang2017a}, BBs \citep[][cyan square]{Yang2017b}, and local metal-poor galaxies \citep[][SA16 hereafter, open circle]{SanchezAlmeida2016} for comparison.
We also compare them with the representative metal-poor galaxies in the range of {\metal}$\sim$7.0--7.2, J0811$+$4730 \citep[][]{Izotov2018a}, SBS0335$-$052 \citep[e.g.,][]{Izotov2009}, AGC198691 \citep[][]{Hirschauer2016}, J1234$+$3901 \citep[][]{Izotov2019a}, LittleCub \citep[][]{Hsyu2017}, DDO68 \citep[][]{Pustilnik2005, Annibali2019}, IZw18 \citep[e.g.,][]{Izotov1998a, Thuan2005a}, and LeoP \citep[][]{Skillman2013} with diamonds.
Although {\EW}({\Ha}) has been used to select high-EW EMPGs in the models (Section \ref{subsec:model}), we compare {\EW}({\Hb}) here because some of {\Ha} emission lines are saturated in our observation.
The EW condition used in the model, {\EW}({\Ha})$>$1,000 {\AA}, corresponds to {\EW}({\Hb})$>$200 {\AA} under the assumption of the tight correlation between {\EW}({\Ha}) and {\EW}({\Hb}) as demonstrated in Figure \ref{fig:EWHa_EWHb}.
We find that our metal-poor galaxy sample covers a high {\EW}({\Hb}) range of $\sim$100--300 {\AA}.
Most of BBs and the representative metal-poor galaxies also show high equivalent widths of $\sim$100--300 {\AA}.
These high {\EW}({\Hb}) values ($\sim$100--300 {\AA}) are in contrast to the metal-poor galaxy sample of SA16, in which most galaxies show {\EW}({\Hb})$\lesssim$100 {\AA}.
As suggested in Figure \ref{fig:age}, galaxies that consist of younger stellar population have higher equivalent widths of Balmer emission lines.
Thus, the high {\EW}({\Hb}) values may suggest that our metal-poor galaxies, BBs, and the representative metal-poor galaxies possess younger stellar population than the metal-poor galaxies of SA16.

\subsection{Number Density} \label{subsec:density}
We roughly estimate the cosmic number densities of metal-poor galaxies that we have selected from the HSC- and SDSS-photometry catalogs (i.e., HSC- and SDSS-EMPG candidates).
Here, we assume that all of the HSC- and SDSS-EMPG candidates are real metal-poor galaxies because we have spectroscopically confirmed that all of the 10 HSC- and SDSS-EMPG candidates are real metal-poor galaxies (Section \ref{subsec:identification}).
Note that we do not estimate the cosmic number densities of EMPGs here, but our sample galaxies to see if our sample galaxies are rare or not.
The HSC and SDSS broad-band filters select EMPGs at $z<0.035$ and $z<0.030$, respectively, which correspond to 149 and 128 Mpc in cosmological physical distance.
The redshift-range difference ($z<0.035$ and $z<0.030$) is caused by the different response curves of the HSC and SDSS broad-band filters.
Because we have selected 27 (86) EMPG candidates from the HSC (SDSS) data, whose effective observation area is 509 (14,555) deg$^2$, within $z<0.035$ ($z<0.030$), we obtain the number density, 1.5$\times$$10^{-4}$ (2.8$\times$$10^{-5}$) {Mpc}$^{-3}$, from the HSC (SDSS) data.
As suggested by previous surveys \citep{Cardamone2009, Yang2017b}, we confirm again that the metal-poor galaxies with strong emission lines are rare in the local universe. 
We also find that the number density of metal-poor galaxies is $\times$10 times higher in the HSC data than in the SDSS data.
This difference is explained by the facts that fainter galaxies are more abundant and that our HSC metal-poor galaxies (median: $i\sim22.5$ mag) are $\sim30$ times fainter than our SDSS metal-poor galaxies (median: $i\sim18.8$ mag).
The number density estimation may depend on the selection criteria and the completeness and purity of our EMPG candidate samples.

\subsection{Environment} \label{subsec:environment}
\begin{figure}
\epsscale{1.2}
\plotone{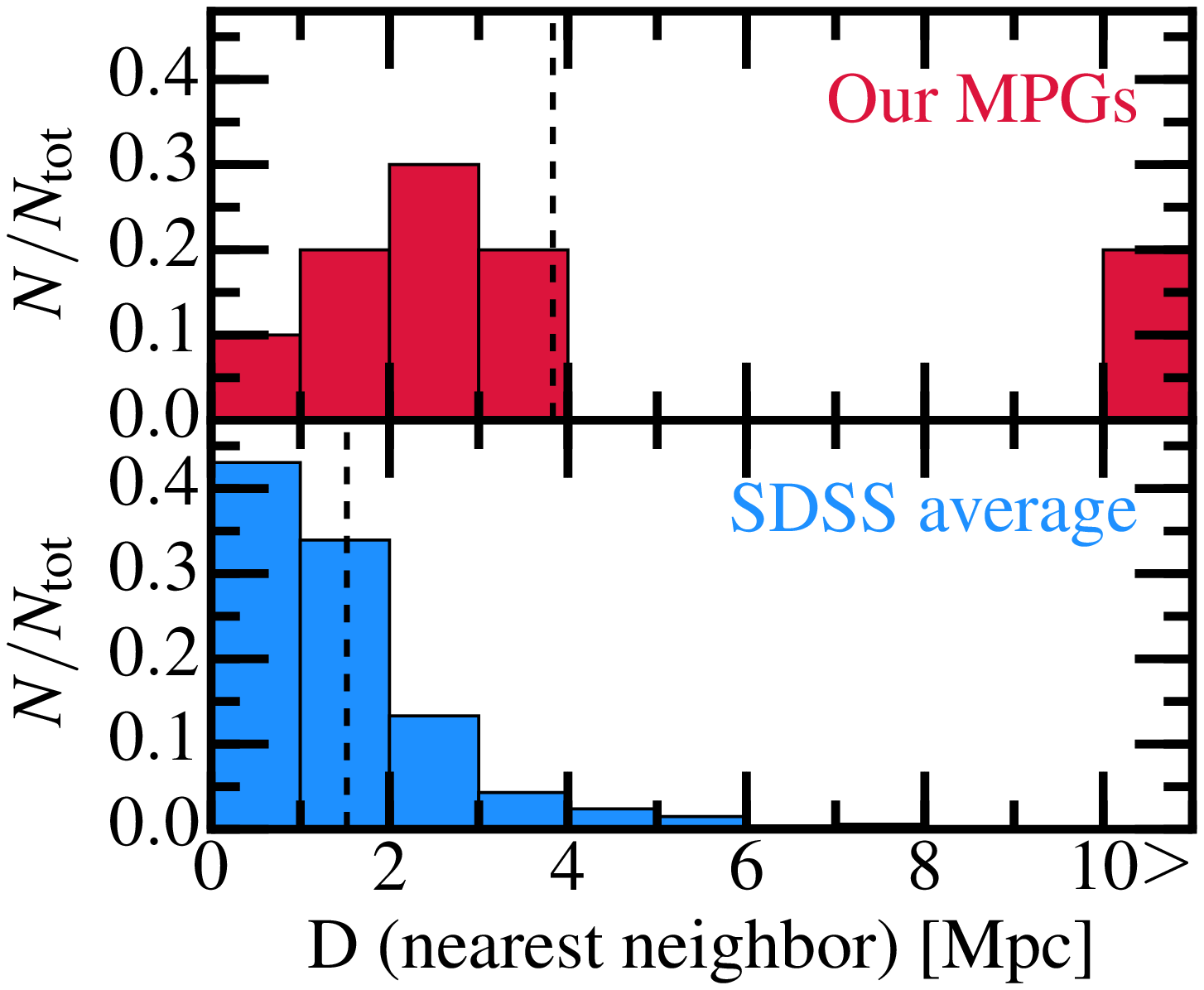}
\caption{Normalized histogram of the nearest neighborhood distances of our ten metal-poor galaxies (top panel) and local, typical SFGs obtained from SDSS (bottom panel).
The number of galaxies in each bin ($N$) is normalized by the total number of galaxies ($N_\mathrm{tot}$).
The dashed lines indicate average values of the nearest neighborhood distances of our metal-poor galaxies (3.83 Mpc) and typical SFGs (1.52 Mpc). 
The bin between 10 and 11 represents the number of galaxies whose nearest neighborhood distance is beyond 10 Mpc.
\label{fig:distance}}
\end{figure}

To characterize the environment of our metal-poor galaxies, we compare nearest neighborhood distances (D$_\mathrm{near}$) of our 10 spectroscopically-confirmed galaxies and local, typical SFGs.
One-thousand local, typical SFGs are randomly chosen from the SDSS DR13 spectroscopic catalog in the range of $z=0.03$--$0.05$. 
We calculate distances from our 10 spectroscopically-confirmed galaxies and local, typical SFGs to surrounding galaxies selected from the SDSS DR13 spectroscopic catalog.
Then we identify their nearest neighbor.
The D$_\mathrm{near}$ values of our metal-poor galaxies range from 0.49 to 17.69 Mpc, which are summarized in Table \ref{table:prop}.
We also estimate D$_\mathrm{near}$ of  typical SFGs randomly selected from the SDSS DR13 spectroscopic catalog 
Figure \ref{fig:distance} compares the D$_\mathrm{near}$ distributions of our metal-poor galaxies as well as local, typical SFGs.
The average D$_\mathrm{near}$ value of our metal-poor galaxies is 3.83 Mpc, which is about 2.5 times larger than that of local, typical SFGs (1.52 Mpc).
We also find that 9 out of our 10 metal-poor galaxies have D$_\mathrm{near}$ values larger than the average of local, typical SFGs (i.e., D$_\mathrm{near}$$>$1.52 Mpc).
Statistically, a Kolmogorov-Smirnov test rejects the null hypothesis (i.e., the distributions of the two samples are the same) with a $p$-value of $1.9\times10^{-3}$, suggesting that these distributions are significantly different. 
Thus, we conclude that our metal-poor galaxies exist in the relatively isolated environment compared to the local, typical SFGs.
According to \citet{Yang2017b}, their BB galaxy sample also shows significantly larger distances to their nearest neighborhood.
\citet{Filho2015} also report that most of metal poor galaxies are found in low-density environments.
These observational results suggest that the metal-poor galaxies have started an intensive star-formation in an isolated environment.

The formation mechanism of metal-poor strong-line galaxies in the local universe is a question. 
One possible explanation is that the cosmic UV background had prevented the star formation in metal-poor intergalactic gas until recently in low-density regions, but the star formation was suddenly triggered by the collapse or collision of the metal-poor gas.
\citet{SanchezAlmeida2013, SanchezAlmeida2015} has investigated tadpole galaxies, which is one of the typical metal-poor galaxy populations, and found that a blue head of tadpole galaxies has significantly lower metallicity than the rest of the galaxy body by factors of 3--10. 
The Northern Extended Millimeter Array (NOEMA) mm-wave interferometer has revealed that a tadpole galaxy possesses molecular gas at its head \citep{Elmegreen2018}.
\citet{Filho2013} demonstrate that metal-poor galaxies are surrounded by asymmetric {\HI} gas, which can be shaped by the accretion of metal-poor gas.
However, \citet{Filho2013} and \citet{SanchezAlmeida2016} have reported the various morphology of metal-poor galaxies, which suggests that the star-formation mechanism is different among metal-poor galaxies (i.e., multiple mechanisms exist). 
The formation mechanism of low-mass, metal-poor galaxies in the field environment is still under debate.
Statistical studies are necessary with larger samples.

\begin{figure}
\epsscale{0.9}
\plotone{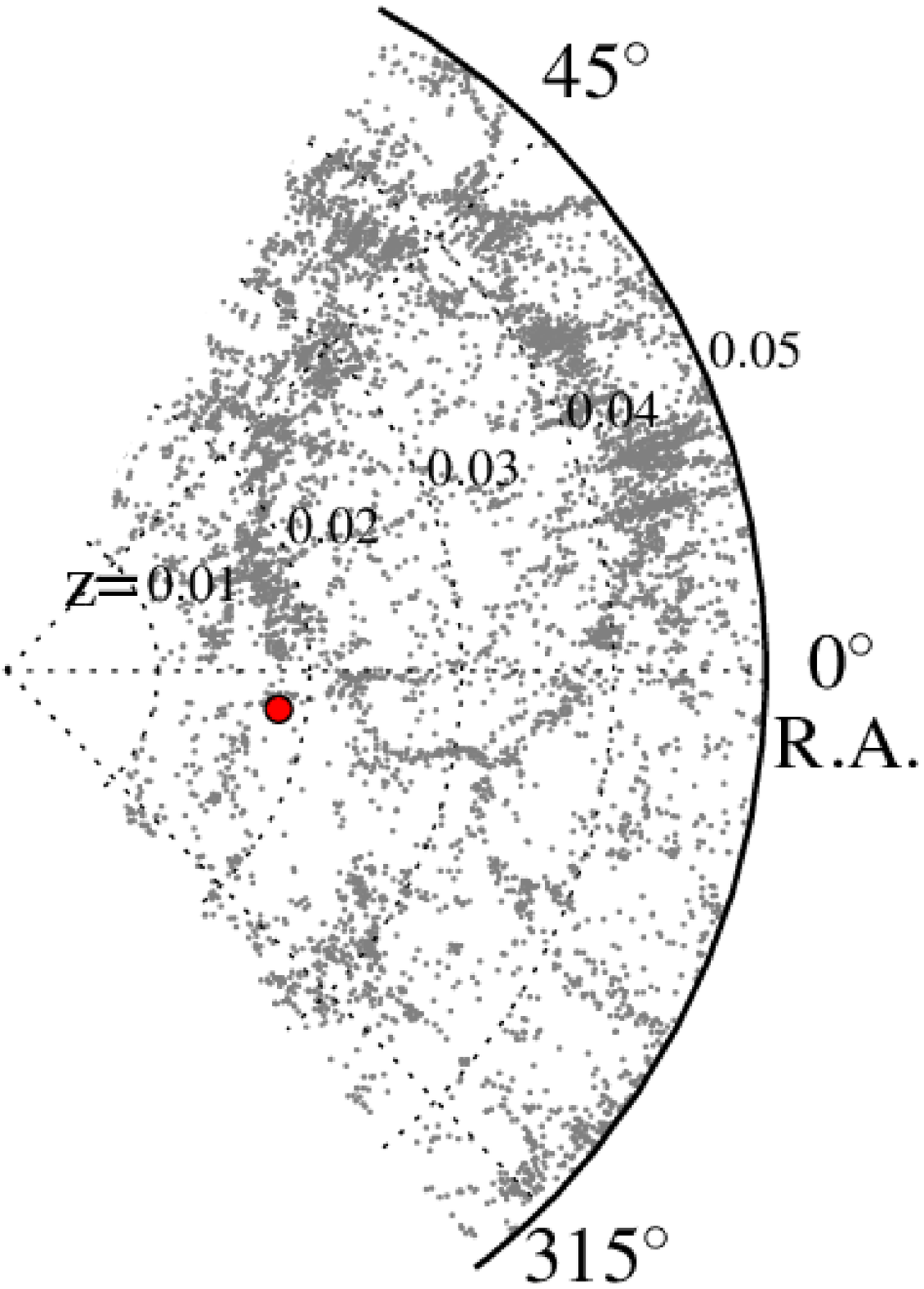}
\caption{Illustration of the large-scale structure slice around {\SDSSEMPGf} (R.A.$=$23:27:43.69 , Dec.$=$$-$02:00:55.89; red circle) projected onto an R.A.-redshift plane.
Gray dots represent galaxies selected from the SDSS DR13 spectroscopic catalog.
Here we only show galaxies falling between $\pm$5.0 degree away from {\SDSSEMPGf} in declination. 
\label{fig:env}}
\end{figure}

\subsection{M$\star$-SFR Relation} \label{subsec:msfr}
\begin{figure}
\epsscale{1.2}
\plotone{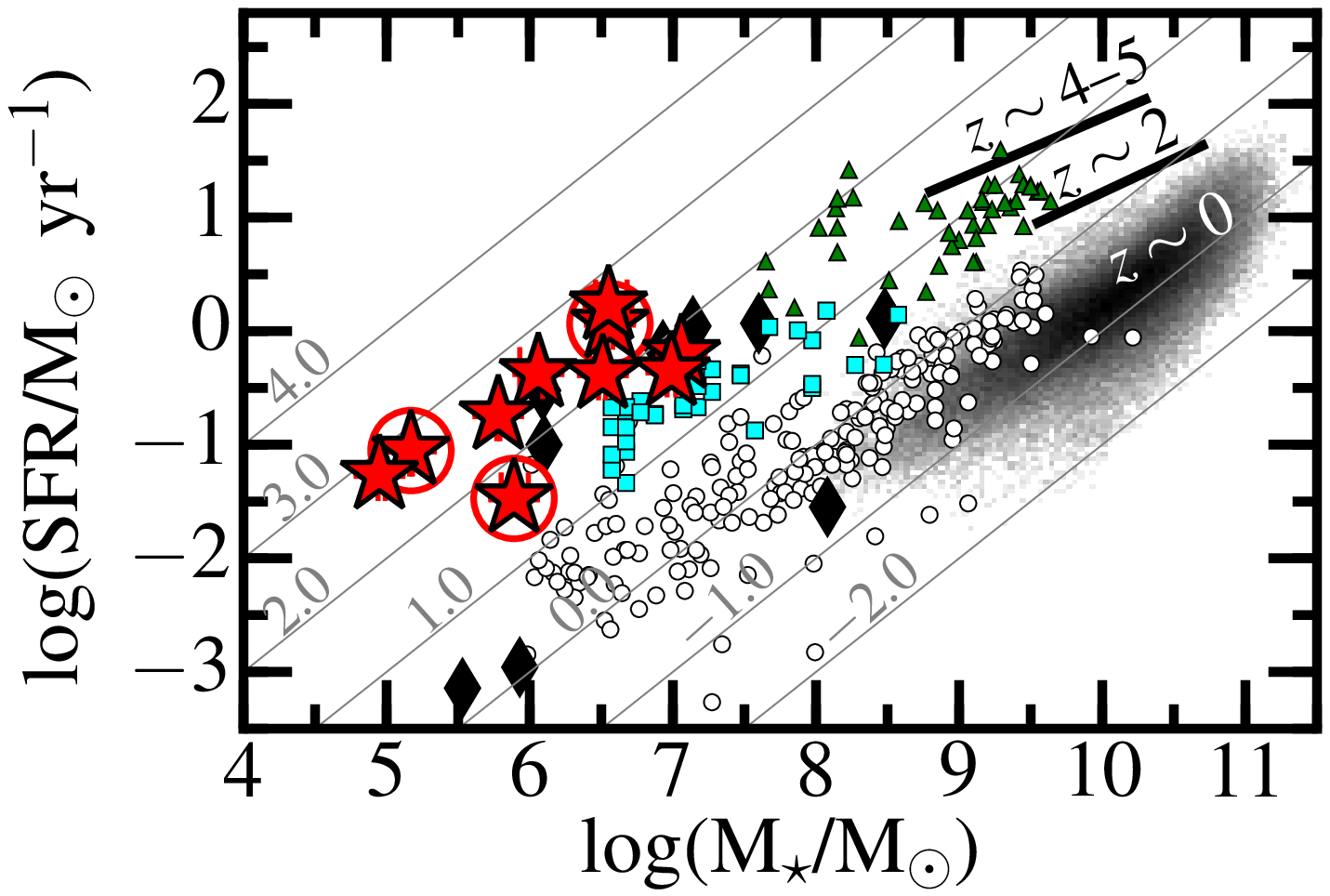}
\caption{Stellar mass and SFR of our metal-poor galaxies with GPs, BBs, and local metal-poor galaxies.
Symbols are the same as in Figure \ref{fig:ew_metal}.
We also show the stellar-mass and SFR distribution of typical $z\sim0$ SFGs (i.e., $z\sim0$ main sequence; black mesh), which we derive from the value-added catalog of SDSS DR7 \citep{Kauffmann2003, Brinchmann2004, Salim2007}.
The solid lines represent the main sequences at $z\sim2$ and $z\sim4$--$5$ \citep[][]{Shivaei2016a, Shim2011}.
The SFRs of \citet[][]{Shivaei2016a} and \citet[][]{Shim2011} are estimated based on the {\Ha} flux.
We convert stellar masses and SFRs derived from the literature into those of the \citet{Chabrier2003} IMF, applying  conversion factors obtained by \citet{Madau2014}.
Gray solid lines and accompanying numbers indicate $\log$(sSFR/Gyr$^{-1}$)=($-$2.0, $-$1.0,..., 4.0).
The stellar masses and SFRs of the representative metal-poor galaxies are derived from the literature \citep{Izotov2018a, Hirschauer2016, Izotov2019a, Hsyu2017, Hunt2015a, Sacchi2016, Rhode2013, Annibali2013}
\label{fig:mass_SFR}}
\end{figure}

Figure \ref{fig:mass_SFR} shows SFRs and stellar masses of our metal-poor galaxies, BBs, GPs, metal-poor galaxies of SA16, and the representative metal-poor galaxies from the literature.
Our metal-poor galaxies, BBs, GPs, and the representative metal-poor galaxies have higher SFRs than typical $z\sim0$ galaxies (i.e., $z\sim0$ star-formation main sequence) for a given stellar mass.
In other words, they have a higher specific SFR (sSFR) than those given by the $z\sim0$ main sequence.
Particularly, our metal-poor galaxies have low stellar mass values in the range of {\mass}$<$6.0, which are lower than BBs, GPs, and metal-poor galaxies of SA16.

The stellar masses of our metal-poor galaxies fall on the typical stellar-mass range of globular clusters, i.e.,  {\mass}$\sim$4--6.
Thus, one may guess that these metal-poor galaxies might be globular clusters that have been formed very recently.
However, further investigation is necessary to understand the association between metal-poor galaxies and globular clusters, which will be discussed in Paper-III (Isobe et al. in prep.).

The solid lines in Figure \ref{fig:mass_SFR} show the main sequences of typical galaxies at $z\sim2$ \citep{Shivaei2016a} and $z\sim4$--$5$ \citep{Shim2011}.
As suggested by solid lines, the main sequence evolves towards high SFR for a given stellar mass with increasing redshift.
Our metal-poor galaxies have higher SFRs for a given $M_{\star}$ than the $z\sim0$ main sequence, falling onto the extrapolation of the $z\sim4$--$5$ main sequence.
Our metal-poor galaxies have as high sSFR values as low-$M_{\star}$ galaxies at $z\gtrsim3$ and local LyC leakers \citep[e.g., $\log$(sSFR/Gyr$^{-1}$)$\sim$1--3,][]{Ono2010b, Vanzella2016a, Izotov2018a, Shim2011}.
Table \ref{table:sSFR} summarizes sSFR values of our metal-poor galaxies and other galaxy populations from the literature for reference.
Based on the high sSFRs, we suggest that our metal-poor galaxies are undergoing intensive star formation comparable to the low-$M_{\star}$ SFGs at $z\gtrsim3$.

Our SFR estimates are obtained under the simple assumption of \citet{Kennicutt1998} because we only have optical observational results for now, and that the simple assumption can be broken in the very young ($\lesssim$10 Myr), metal-poor, low-M$_{\star}$ galaxies because the conversion factor is sensitive to the IMF, the star-formation history, the metallicity, and the escape fraction and dust absorption of ionizing photons \citep{Kennicutt1998}.
Other SFR uncertainties may arise from additional ionizing photon sources, such as a low-luminosity AGN, shock-heated gas, galactic outflows, and X-ray binaries, which are not included in the stellar synthesis models used in the calibration of \citet{Kennicutt1998}.
Further multi-wavelength observations are required to understand the star-formation rate, history, and mechanism of very young, metal-poor, low-M$_{\star}$ galaxies.

\begin{deluxetable}{cccc}
\tablecolumns{4}
\tabletypesize{\scriptsize}
\tablecaption{Typical Values of Specific SFR %
\label{table:sSFR}}
\tablehead{%
\colhead{Population} &
\colhead{$\log$(sSFR)} &
\colhead{redshift} &
\colhead{Ref.} \\
\colhead{} &
\colhead{(Gyr$^{-1}$)} &
\colhead{} &
\colhead{} \\
\colhead{(1)} &
\colhead{(2)} &
\colhead{(3)} &
\colhead{(4)} 
}
\startdata
Our EMPGs & 2.47 & $0.007$--$0.03$ & {\bf This work}\\ 
EMPGs (SDSS DR7) & 0.34 & $\lesssim0.1$ & SA16\\ 
BBs & 1.39 & $\sim0.05$ & Y17b\\ 
GPs & 1.38 & $\sim0.3$ & Y17a\\ 
LyC leaker ($f_\mathrm{esc}^\mathrm{LyC}$=0.46) & 1.29 & 0.37 & I18\\ 
\hline
Main Sequence ($z\sim0$) & $-$0.20 & $\sim0$ & SDSS DR7\\ 
Main Sequence ($z\sim2$) & $\sim0.0$--$0.5$ & $\sim2$ &S16\\ 
Main Sequence ($z\sim4$--$5$) & $\sim1.0$--$1.5$ & $\sim4$--$5$ &S11\\ 
\hline
low-$M_{\star}$ SFG ($z\sim3$) & 1.10/1.80 & 3.12 &V16\\
Little Blue Dots & $\gtrsim$2.0 & $2$--$5$ &E17 \\ 
LAEs ($z=5.7$) & 3.05 & $5.7$ &O10\\ 
LAEs ($z=6.6$) & 3.05 & $6.6$ &O10 
\enddata
\tablecomments{
(1) Galaxy Population.
(2) Average of sSFR in the unit of log(Gyr$^{-1}$). We calculate a linear average of each sample here.
(3) Typical redshift.
(4) References of sSFR---SDSS DR7: \citep{Kauffmann2003, Brinchmann2004, Salim2007}, SA16: \citet{SanchezAlmeida2016}, Y17b: \citet{Yang2017b}, Y17a: \citet{Yang2017a}, I18: \citet{Izotov2018b}, S16: \citet{Shivaei2016a}, S11: \citet{Shim2011}, V16: \citet{Vanzella2016a}, E17: \citet{Elmegreen2017b}, O10: \citet{Ono2010b}.
}
\end{deluxetable}

\subsection{$M_{\star}$-$Z$ Relation} \label{subsec:mz}
\begin{figure}
\epsscale{1.2}
\plotone{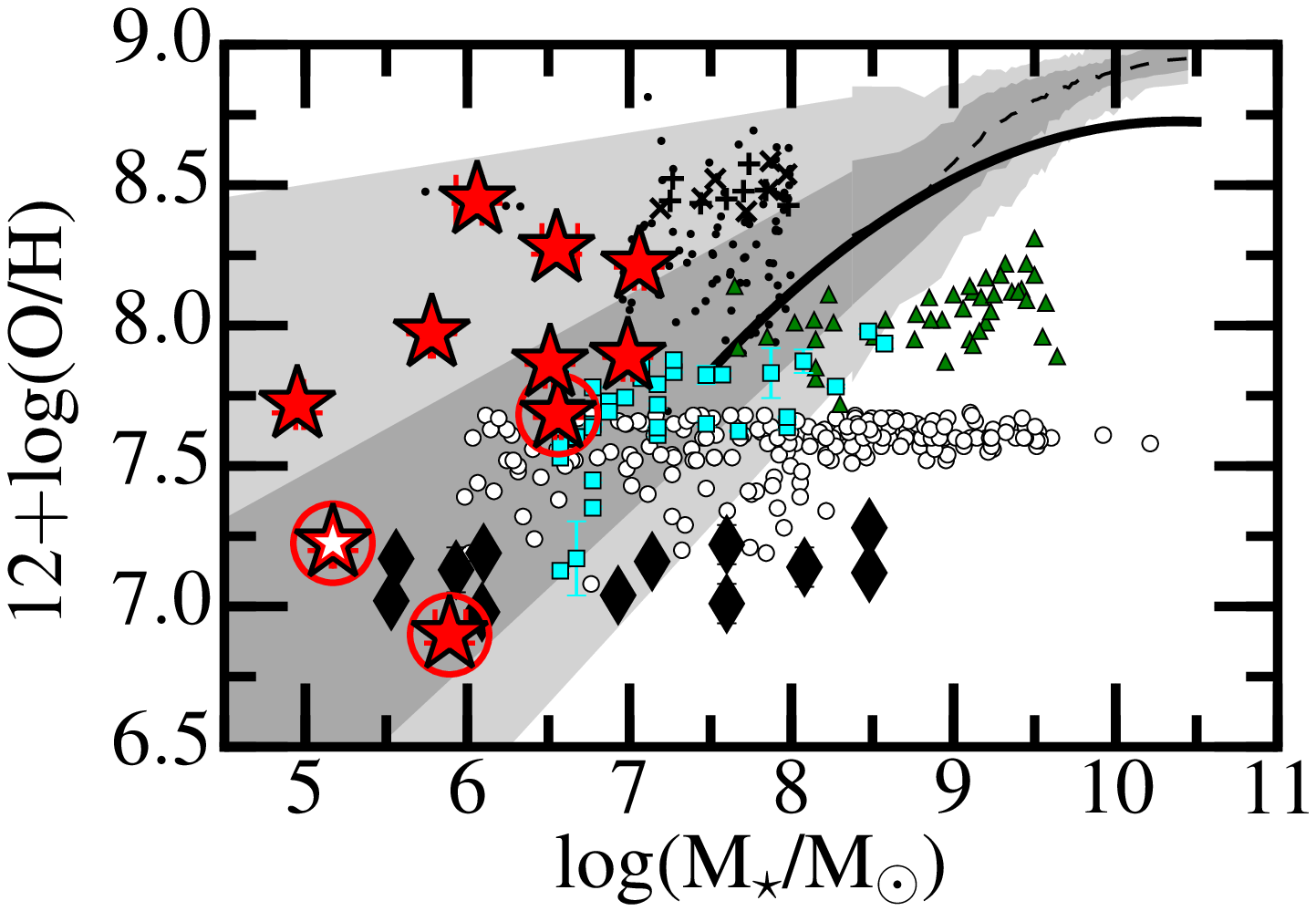}
\caption{Mass-metallicity relation of our metal-poor galaxies.
Symbols are the same as in Figure \ref{fig:ew_metal}.
The solid and dashed lines indicate averaged local SFGs given by \citet{Andrews2013} and Z12 from SDSS data, respectively.
The dark gray and light gray shaded regions represent the 68 and 95-percentile distributions of SFGs of Z12, although the extrapolation is applied below {\mass}$=$8.4.
We also show relatively metal-enriched dwarfs of P08 (cross) and Z12 (plus) from SDSS, as well as DEEP2 galaxies of Z12 (dot) in the stellar mass range of {\mass}$<$8.0.
Typical metallicity error of our metal-poor galaxies is $\Delta$(O/H)$\sim$0.01 dex.
\label{fig:mz}}
\end{figure}

Figure \ref{fig:mz} exhibits a mass-metallicity ($M_{\star}$-$Z$) relation of our metal-poor galaxies.
Our metal-poor galaxies are located around the low-mass end of {\mass}=5--7 among metal-poor galaxy samples of BBs, GPs, the S16 metal-poor galaxies, and the representative metal-poor galaxies in Figure \ref{fig:mz}.
Metallicities of our metal-poor galaxies extend in a relatively wide range, {\metal}$\sim$6.9--8.5. 
The gray shaded regions in Figure \ref{fig:mz} represent the 68 and 95-percentile distributions of local SFGs of \citet[][Z12 hereafter]{Zahid2012}, who have reported that the metallicity scatter of galaxies becomes larger with decreasing metallicity for a given mass.
Although the extrapolation is applied below {\mass}$=$8.4 here, 5 out of our metal-poor galaxies fall in the 68-percentile distribution of the local $M_{\star}$-$Z$ relation.

Interestingly, we find that the other 5 metal-poor galaxies of ours are located above the 68-percentile distribution given by Z12, i.e., higher metallicities for a given stellar mass.
We refer to the 5 metal-poor galaxies located above the 68-percentile distribution as ``above-MZ galaxies'' hereafter.
Our above-MZ galaxies have moderate metallicities of {\metal}$\sim$8.0 in spite of their very low-M$_{\star}$ (i.e.,  {\mass}$=$5--7).
A possible explanation of these above-MZ galaxies has been given by Z12 and \citet[][P08 hereafter]{Peeples2008}.
In Figure \ref{fig:mz}, we also show the low-$z$ galaxy samples of Z12 and P08 in the stellar mass range of {\mass}$<$8.0.
In a sample from the DEEP2 survey, Z12 have found galaxies with a metallicity higher than the local $M_{\star}$-$Z$ relation (Figure \ref{fig:mz}) and a higher SFR for a given stellar mass, which is similar to our above-MZ galaxies.
Z12 have also found counterpart galaxies of their DEEP2 galaxies (i.e., above both the $M_{\star}$--$Z$ and $M_{\star}$--$SFR$ relations) in the SDSS data (Figure \ref{fig:mz}).
Z12 have argued that their DEEP2 and SDSS galaxies may be transitional objects, which have been suggested by P08, from gas-rich dwarf irregulars to gas-poor dwarf spheroidals and ellipticals.
P08 has also investigated local galaxies whose metallicities are higher than the local $M_{\star}$-$Z$ relation, with SDSS data.
Unlike our above-MZ galaxies and the Z12 galaxies, the P08 sample shows redder colors and lower SFRs consistent with the local $M_{\star}$--$SFR$ relation.
P08 have claimed that the P08 galaxies may be in a later stage of the transition from gas-rich dwarf irregulars to gas-poor dwarf spheroidals and ellipticals, and that the gas deficit leads to the low SFRs and high metallicities.
It should be noted that the Z12 and P08 galaxies are located in the relatively isolated environment, similarly to our above-MZ galaxies.
If our above-MZ galaxies are explained by an early stage of the transition, our above-MZ galaxies may be losing (or have lost) gas despite their very recent star formation suggested by the high {\EW}(\Hb) (Section \ref{subsec:identification}).
The gas loss can be caused by the galactic outflow triggered by supernovae (SNe), for example, in young galaxies such as our above-MZ galaxies.
However, to characterize these above-MZ galaxies, more observations are necessary such as far-infrared and radio observations which trace emission from molecular gas, {\HI} gas, and the SNe.

\begin{figure}
\epsscale{1.2}
\plotone{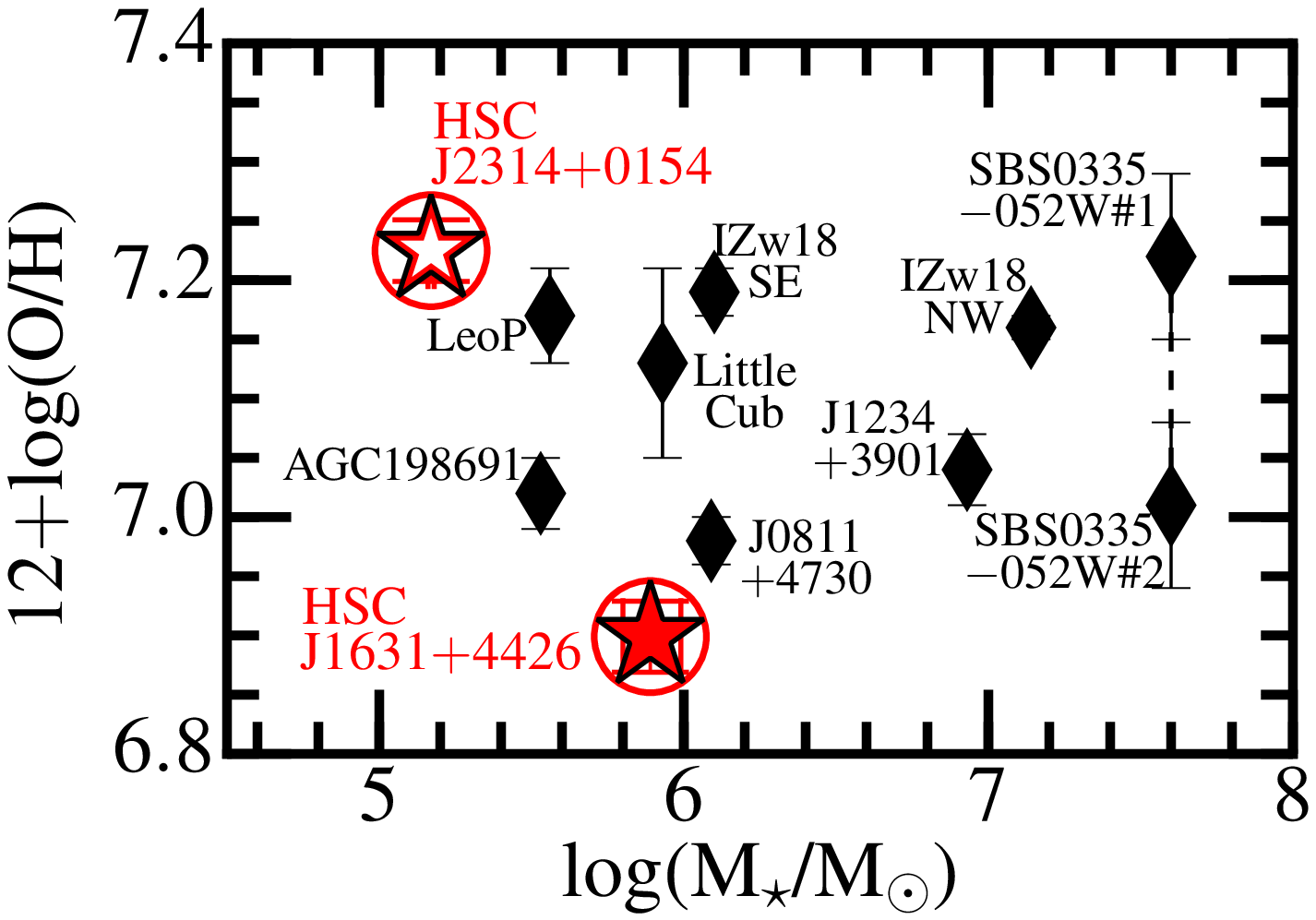}
\caption{Same as top panel of Figure \ref{fig:mz}, but zoom-in around the low-$M_{\star}$, low-metallicity ends.
\label{fig:mz_zoom}}
\end{figure}

Figure \ref{fig:mz_zoom} demonstrates the low-$M_{\star}$, low-metallicity ends of the $M_{\star}$-$Z$ relation.
Here we compare our metal-poor galaxies with the representative metal-poor galaxies.
Among the representative metal-poor galaxies, we find that our {\HSCEMPGd} ({\metal}=6.90, i.e., $Z/Z_{\sun}$$=$0.016) has the lowest metallicity reported ever.
The metallicity of our {\HSCEMPGd} is lower than those of J0811$+$4730 \citep[][]{Izotov2018a}, AGC198691 \citep[][]{Hirschauer2016}, SBS0335$-$052 \citep[e.g.,][]{Izotov2009}, and J1234$+$3901 \citep[][]{Izotov2019a}.
We emphasize that the discovery of the very faint EMPG, {\HSCEMPGd} ($i$$=$22.5 mag) has been enabled by the deep, wide-field HSC-SSP data, which could not be achieved by the previous SDSS surveys.
This paper presents just the first spectroscopic result of 4 out of the 27 HSC-EMPG candidates. 
We expect to discover more EMPGs from our HSC-EMPG candidates in the undergoing spectroscopy.

\subsection{BPT Diagram} \label{subsec:bpt}
\begin{figure}
\epsscale{1.2}
\plotone{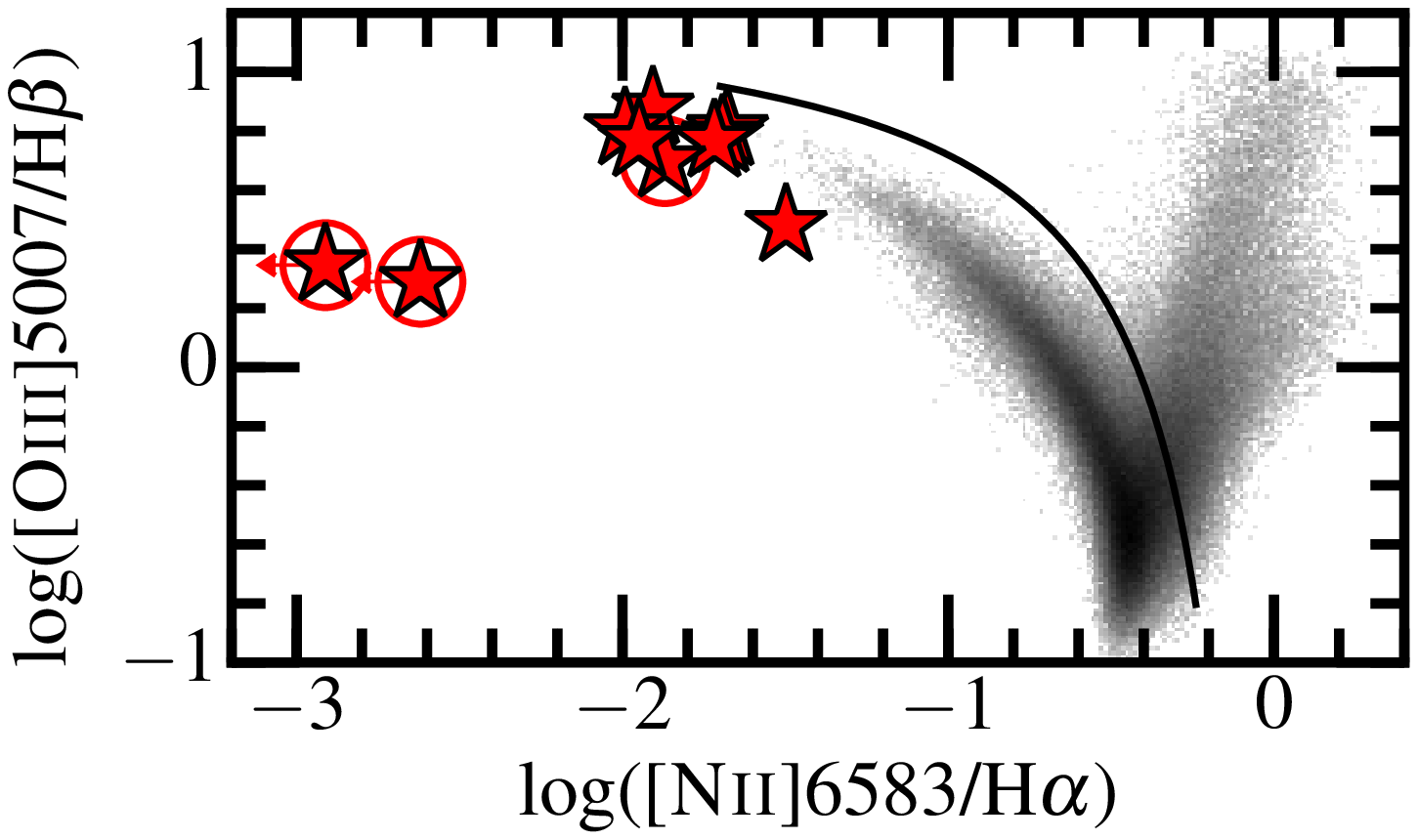}
\caption{Our metal-poor galaxies on the BPT diagram (red stars).
The black mesh represents $z\sim0$ SFGs and AGNs derived from the emission-line catalog of SDSS DR7 \citep{Tremonti2004}.
The solid curve indicates the maximum photoionization models that can be achieved under the assumption of  stellar radiation \citep{Kewley2001}.
The region below the solid curve is defined by the SFG region, while the upper right side is defined by the AGN region. 
\label{fig:bpt}}
\end{figure}

Figure \ref{fig:bpt} is an emission line diagnostic diagram of {\NII}/{\Ha} and {\OIII}/{\Hb} (i.e., BPT diagram) with our metal-poor galaxies.
Our metal-poor galaxies fall on the SFG region defined by the maximum photoionization models under the stellar radiation \citep{Kewley2001}.
We do not find any evidence that our metal-poor galaxies are affected by an AGN or shock heating from the optical emission line ratios.
However, \citet{Kewley2013a} suggest that metal-poor gas heated by the AGN radiation or shock also show emission-line ratios falling on the SFG region defined by \citet{Kewley2001}.
We thus do not exclude the possibility of the existence of a metal-poor AGN or shock heating of metal-poor gas.
We will discuss the ionization state of inter-stellar medium (ISM) and the ionization-photon sources in Paper II.

\subsection{Velocity Dispersion} \label{subsec:velocity}

\begin{figure}
\epsscale{1.2}
\plotone{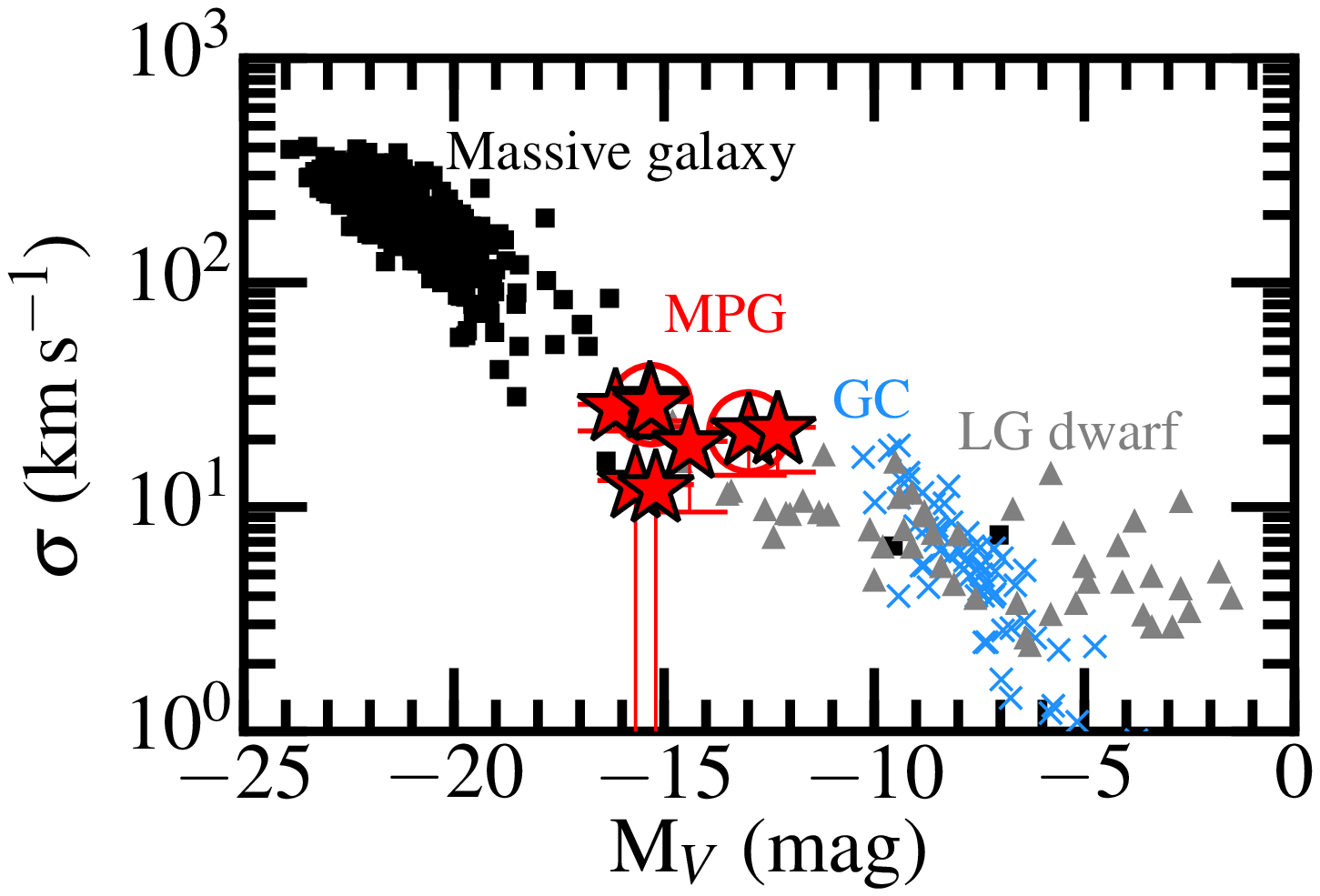}
\plotone{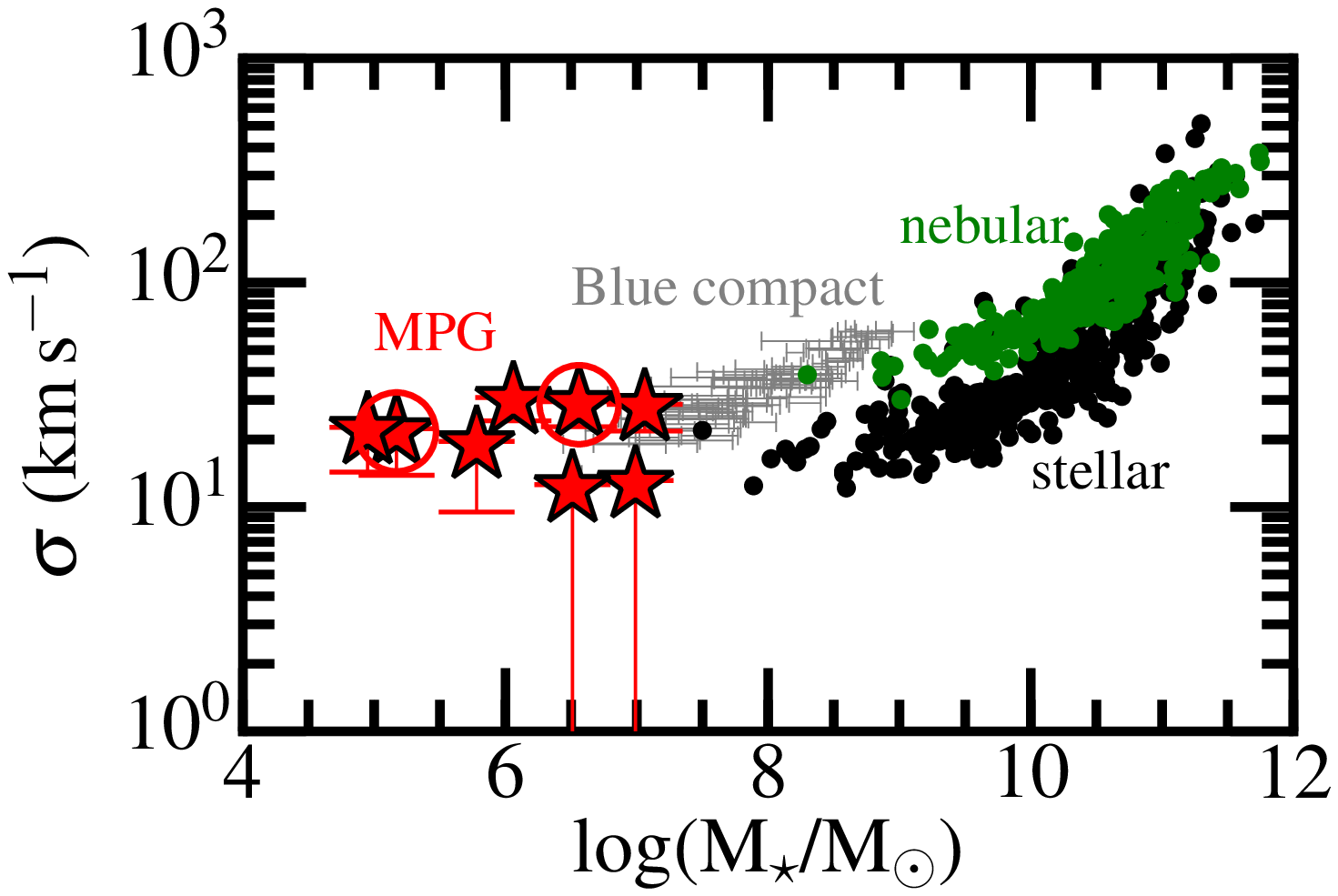}
\caption{Top: Velocity dispersion as a function of optical magnitudes.
Stars are the same as in Figure \ref{fig:ew_metal}.
Squares and triangles represent stellar velocity dispersions of bright galaxies and local faint galaxies from the literature, which are compiled by \citet{Prugniel1996} and \citet{Lin2016}, respectively.
We also show stellar velocity dispersions of globular clusters \citep{Harris1996} with blue crosses.
To estimate the continuum level in $V$-band of our metal poor galaxies, we use an $i$-band magnitude instead of $g$- or $r$-band magnitudes because the $g$ and $r$ bands are strongly affected by strong emission lines.
Here we assume a flat continuum from $V$ to $i$ bands in the unit of {\ergscmHz}.
We confirm that this assumption is correct within $\sim$0.2 mag by looking at a continuum in MagE spectra of our metal-poor galaxies.
Bottom: Same as top panel, but as a function of stellar mass.
Black and green circles represent velocity dispersions obtained with stellar and nebular lines, respectively, from the Sydney-AAO Multi-object Integral field spectrograph (SAMI) galaxy survey \citep{Barat2019}.
Gray bars show blue compact galaxies at $z\sim0.02$--$0.2$ reported by \citet{Chavez2014}.
The upper and lower limits of the gray bars indicate dynamical masses ($M_\textrm{\tiny dyn}$) and stellar masses of ionizing star clusters ($M_\textrm{\tiny cl}$), respectively.
The gray bars represent possible total stellar-mass ranges because an equation, $M_\textrm{\tiny cl}<M_{*}<M_\textrm{\tiny dyn}$ holds by definition.
\label{fig:velocity}}
\end{figure}

We estimate velocity dispersions of our 8 metal-poor galaxies observed with MagE out of our 10 metal-poor galaxies.
We do not estimate velocity dispersions of the other 2 galaxies observed with LDSS-3, DEIMOS, and FOCAS due to spectral resolutions not high enough to resolve emission lines of our very low-mass sample.
We measure emission-line widths of our metal-poor galaxies by a Gaussian fit to an {\Hb} emission line, obtaining $\sigma_\mathrm{obs}$$=$36.5--45.3 km\,s$^{-1}$.
We obtain intrinsic velocity dispersions, $\sigma$ of our metal-poor galaxies with
\begin{eqnarray}
  \sigma=\sqrt{\sigma_\mathrm{obs}^2-\sigma_\mathrm{inst}^2-\sigma_\mathrm{th}^2-\sigma_\mathrm{fs}^2}, \label{eq:sigma}
\end{eqnarray}
where $\sigma_\mathrm{inst}$, $\sigma_\mathrm{th}$, and $\sigma_\mathrm{fs}$ are the instrumental, thermal, and fine-structure line broadening, under the assumption that the broadening can be approximated by a Gaussian profile \citep[e.g.,][]{Chavez2014}.
We measure the instrumental line broadening with arc-lamp frames and find $\sigma_\mathrm{inst}$$=$26.4 and 33.3 km\,s$^{-1}$ with the slit widths of 0.85 and 1.20 arcsec, respectively. 
We calculate $\sigma_\mathrm{th}$ with
\begin{eqnarray}
  \sigma_\mathrm{th}=\sqrt{\frac{kT_{\mathrm{e}}}{m}}, \label{eq:sigmath}
\end{eqnarray}
where $k$ and $m$ represent the Boltzmann constant and the hydrogen mass, respectively.
We use $\sigma_\mathrm{th}$$=$9.1 km\,s$^{-1}$, which is obtained from Equation (\ref{eq:sigmath}) under the assumption of $T_{\mathrm{e}}$=10,000 K.
We adopt the fine-structure line broadening of {\Hb}, $\sigma_\mathrm{fs}=$2.4 km\,s$^{-1}$  \citep{Garcia-Diaz2008}.
Then, we calculate $\sigma$ with Equation (\ref{eq:sigma}), obtaining $\sigma$ values in the range of 12.0--29.8 km\,s$^{-1}$.
The obtained  $\sigma$ values are summarized in Table \ref{table:prop}.
For a careful comparison, we have also checked $\sigma$ values under the assumption of $T_{\mathrm{e}}$=15,000 K in Equation (\ref{eq:sigmath}), which is more consistent with the {\TeOIII} measurements as shown in Table \ref{table:temp}.
However, in this case, two of the $\sigma$ estimates become zero.
Thus, we include the $\sigma$ differences between the two assumptions (i.e., $T_{\mathrm{e}}$=10,000 and 15,000 K) in $\sigma$ errors.
We do not remove an effect of the emission line broadening caused by the dynamical galaxy rotation because the spectral resolution of MagE is still not enough to separate the rotation and the dispersion.
However, previous studies with high spectral-resolution spectroscopy \citep[e.g.,][]{Melnick1988, Chavez2014} find that most of galaxies with $\sigma$$\lesssim$60 km\,s$^{-1}$ are dispersion supported rather than rotation supported, which suggest that our metal-poor galaxies ($\sigma$$=$12.0--29.8 km\,s$^{-1}$) may be also dispersion supported.
Even if this assumption is not true, our $\sigma$ estimates at least provide upper limits on the velocity dispersions.

The top panel of Figure \ref{fig:velocity} demonstrates velocity dispersions of our metal poor galaxies as a function of $V$-band absolute magnitude in comparison with stellar velocity dispersions of massive galaxies \citep{Prugniel1996}, dwarf galaxies \citep[a compiled catalog of][]{Lin2016}, and globular clusters \citep{Harris1996}.
We find that our metal-poor galaxies fall on a velocity-dispersion sequence made of massive galaxies, dwarf galaxies, and globular clusters in the top panel of Figure \ref{fig:velocity}.
The compiled dwarf galaxy catalog of \citet{Lin2016} are derived from literature of dwarf galaxies in the Local Group ($\lesssim$3 Mpc) reported by \citet{McConnachie2012}, \citet{Kirby2015a, Kirby2015b}, \citet{Simon2015}, and \citet{Martin2016}.
On the other hand, our metal-poor galaxies are low-$M_{\star}$ galaxies outside Local Group.
The velocity dispersions of our metal-poor galaxies trace the gas kinematics while those of the massive galaxies, dwarf galaxies, and globular clusters shown here are estimated mainly from the motion of individual stars.
Indeed, as shown in the bottom panel of Figure \ref{fig:velocity}, the velocity dispersions of gas (green circles) and stars (black circles) are different by a factor of 1.0--1.3  in the range of {\mass}$\sim$8.0--11.5 \citep{Barat2019}.
We also show local blue compact galaxies reported by \citet{Chavez2014} in the bottom panel of Figure \ref{fig:velocity}.
\citet{Chavez2014} have performed spectroscopy with high spectral resolutions ($R\sim$10,000--20,000) for the local blue compact galaxies of {\mass}$\sim$6.5--9.0.
They estimate $\sigma$ with an {\Hb} emission line and created a sample of dispersion-supported galaxies.
In the bottom panel of Figure \ref{fig:velocity}, we find that our metal-poor galaxies in the range of {\mass}$\sim$6.5--7.0 overlap with the dispersion-supported galaxies of \citet{Chavez2014}.
Thus, our metal-poor galaxies may also be dispersion supported.


\section{SUMMARY} \label{sec:summary}
We search for extremely metal-poor galaxies (EMPGs) at $z\lesssim0.03$ to construct a local sample whose galaxy properties are similar to those of high-$z$ galaxies in the early star-formation phase (i.e., low $M_{\star}$, high sSFR, low metallicity, and young stellar ages).
We select EMPGs from the wide-field, deep imaging data of the Subaru Strategic Program (SSP) with Hyper Sprime-Cam (HSC) in combination with the wide-field, shallow data of Sloan Digital Sky Survey (SDSS).
This work is the first metal-poor galaxy survey that exploits the wide ($\sim$500 deg$^2$), deep ($i_\mathrm{lim}$$\sim$26 mag) imaging data of HSC SSP, with which we expect to discover faint EMPGs that SDSS could not detect due to magnitude limitations. 
To remove contamination more efficiently than a simple color-color selection from our sample, we develop a new selection technique based on machine learning (ML).
We construct a ML classifier that distinguishes EMPGs from other types of objects, which is well trained by model templates of galaxies, stars, and QSOs.
By testing our ML classifier with the SDSS photometry+spectroscopy data, we confirm that our ML classifier reaches 86\% completeness and 46\% purity.
Then our ML classifier is applied to HSC and SDSS photometry, obtaining 27 and 86 EMPG candidates, respectively.
These EMPG candidates have a wide range of $i$-band magnitudes, $i$=14.8--24.3 mag, thanks to the combination of the SDSS and HSC data.
We have conducted optical spectroscopy with Magellan/LDSS-3, Magellan/MagE, Keck/DEIMOS, and Subaru/FOCAS for 10 out of the 27$+$86 EMPG candidates.
Our main results are summarized below.

\begin{itemize}
\item We confirm that the 10 EMPG candidates are real star-forming galaxies (SFGs) at $z$=0.007--0.03 with strong  emission lines, whose rest-frame {\Hb} equivalent widths ({\EW}) reach 104--265 {\AA}, and a metallicity range of {\metal}=6.90--8.45.
Three out of the 10 EMPG candidates satisfy the EMPG criterion of {\metal}$<$7.69. 
The other 7 galaxies still show low metallicities ($\sim$0.1--0.6$Z_{\odot}$).
We thus conclude that our new selection based on ML successfully selects real EMPGs or metal-poor, strong-line SFGs.

\item The number density of our HSC metal-poor galaxies is 1.5$\times$10$^{-4}$ Mpc$^{-3}$, which is $\times$10 times higher than that of our SDSS metal-poor galaxies (2.8$\times$10$^{-5}$ Mpc$^{-3}$).
This difference is explained by the fact that our HSC metal-poor galaxies (median: $i\sim22.5$ mag) are $\sim30$ times fainter than our SDSS metal-poor galaxies (median: $i\sim18.8$ mag).

\item To characterize the environment of our metal-poor galaxies, we compare nearest neighborhood distances (D$_\mathrm{near}$) of our metal-poor galaxies with those of local, typical SFGs.
The D$_\mathrm{near}$ of our metal-poor galaxies range from 0.49 to 17.69 Mpc with an average of 3.83 Mpc, which is  $\sim$2.5 times larger than that of local, typical SFGs (average 1.52 Mpc).
With a Kolmogorov-Smirnov test ($p=1.9\times10^{-3}$), we significantly confirm that our metal-poor galaxies are located in the relatively isolated environment compared to the local, typical SFGs.

\item We find that our metal-poor galaxy sample encompasses low metallicities, {\metal}=6.90--8.45, low stellar masses, {\mass}=5.0--7.1, and high specific star-formation rates (sSFR$\sim$300\,Gyr$^{-1}$), suggesting the possibility that they are analogs of high-$z$, low-mass SFGs.

\item We find that 5 out of our 10 metal-poor galaxies with the spectroscopically confirmation have moderate metallicities of {\metal}$\sim$8.0 in spite of their very low-M$_{\star}$ (i.e.,  {\mass}$=$5--7), which are located above an extrapolation of the local mass-metallicity relation.
One possible explanation is that the 5 galaxies above the local mass-metallicity relation are in an early stage of the transition from gas-rich dwarf irregulars to gas-poor dwarf spheroidals and ellipticals, which is suggested by \citet{Peeples2008} and \citet{Zahid2012}.

\item We confirm that {\HSCEMPGd} shows the lowest metallicity value {\metal}=6.90$\pm0.03$ (i.e., $Z/Z_{\odot}$=0.016) reported ever. 

\item Our metal-poor galaxies fall on the SFG region of the BPT diagram, and we do not find any evidence that our metal-poor galaxies are affected by an AGN or shock heating from the optical emission line ratios.
However, we do not exclude the possibility of the existence of a metal-poor AGN or shock because little is known about the low-metallicity AGN or shock to date.

\item We roughly measure velocity dispersions of our metal-poor galaxies with an {\Hb} emission line, which may trace the ionized gas kinematics. 
Thanks to a high spectral resolution of MagE ($R$$\sim$4,000), we find that our metal-poor galaxies have small velocity dispersions of $\sigma$$=$12.0--29.8 km\,s$^{-1}$.
The velocity dispersions of our metal-poor galaxies are consistent with a relation between the velocity dispersion and $V$-band magnitude, which is made by a sequence of low-$z$ bright galaxies, dwarf galaxies in Local Group, and globular clusters.

\end{itemize}

\acknowledgments
\begin{acknowledgements}
We are grateful to  
Lennox Cowie,
Akio Inoue,
Taddy Kodama, 
Matthew Malkan, and
Daniel Stark
for their important and useful comments.
We are grateful to John David Silverman and Anne Verhamme for their helpful comments on our survey name. 
We would like to express our special thanks to Daniel Kelson for his great efforts in helping us reduce and calibrate our MagE data.

We also thank staffs of the Las Campanas observatories, the Subaru telescope, and the Keck observatories for helping us with our observations.
The observations were carried out within the framework of Subaru-Keck time exchange program, where the travel expense was supported by the Subaru Telescope, which is operated by the National Astronomical Observatory of Japan.
The authors wish to recognize and acknowledge the very significant cultural role and reverence that the summit of Maunakea has always had within the indigenous Hawaiian community.  
We are most fortunate to have the opportunity to conduct observations from this mountain.

The Hyper Suprime-Cam (HSC) collaboration includes the astronomical communities of Japan and Taiwan, and Princeton University.  The HSC instrumentation and software were developed by the National Astronomical Observatory of Japan (NAOJ), the Kavli Institute for the Physics and Mathematics of the Universe (Kavli IPMU), the University of Tokyo, the High Energy Accelerator Research Organization (KEK), the Academia Sinica Institute for Astronomy and Astrophysics in Taiwan (ASIAA), and Princeton University.  Funding was contributed by the FIRST program from the Japanese Cabinet Office, the Ministry of Education, Culture, Sports, Science and Technology (MEXT), the Japan Society for the Promotion of Science (JSPS), Japan Science and Technology Agency  (JST), the Toray Science  Foundation, NAOJ, Kavli IPMU, KEK, ASIAA, and Princeton University.

This paper makes use of software developed for the Large Synoptic Survey Telescope. We thank the LSST Project for making their code available as free software at \verb|http://dm.lsst.org|.

This paper is based on data collected at the Subaru Telescope and retrieved from the HSC data archive system, which is operated by Subaru Telescope and Astronomy Data Center (ADC) at NAOJ. Data analysis was in part carried out with the cooperation of Center for Computational Astrophysics (CfCA), NAOJ.

The Pan-STARRS1 Surveys (PS1) and the PS1 public science archive have been made possible through contributions by the Institute for Astronomy, the University of Hawaii, the Pan-STARRS Project Office, the Max Planck Society and its participating institutes, the Max Planck Institute for Astronomy, Heidelberg, and the Max Planck Institute for Extraterrestrial Physics, Garching, The Johns Hopkins University, Durham University, the University of Edinburgh, the Queen’s University Belfast, the Harvard-Smithsonian Center for Astrophysics, the Las Cumbres Observatory Global Telescope Network Incorporated, the National Central University of Taiwan, the Space Telescope Science Institute, the National Aeronautics and Space Administration under grant No. NNX08AR22G issued through the Planetary Science Division of the NASA Science Mission Directorate, the National Science Foundation grant No. AST-1238877, the University of Maryland, Eotvos Lorand University (ELTE), the Los Alamos National Laboratory, and the Gordon and Betty Moore Foundation.

This work is supported by World Premier International Research Center Initiative (WPI Initiative), MEXT, Japan, as well as KAKENHI Grant-in-Aid for Scientific Research (A) (15H02064, 17H01110, and 17H01114) through Japan Society for the Promotion of Science (JSPS).
T.K., K.Y., Y.S., and M. Onodera are supported by JSPS KAKENHI Grant Numbers, 18J12840, 18K13578, 18J12727, and 17K14257.
S.F. acknowledges support from the European Research Council (ERC) Consolidator Grant funding scheme (project ConTExt, grant No. 648179). The Cosmic Dawn Center is funded by the Danish National Research Foundation under grant No. 140.

\end{acknowledgements}

\bibliography{library}

\end{document}